# Quantum and Classical Dynamics of Molecular Scale Structures

Eman N Almutib

PhD Thesis in Nanoelectronics

Submitted in part fulfilment of the requirements for
the degree of Doctor of Philosophy

September 30, 2016

Lancaster University

# Declaration

Except where stated otherwise, this thesis is a result of the author's original work and has not been submitted in whole or in part for the award of a higher degree elsewhere. This thesis documents work carried out between January 2012 and August 2016 at Lancaster University, UK, under the supervision of Prof. Colin J. Lambert and funded by Taif University, Saudi Arabia.

Eman Almutib

September 30, 2016

*To my beautiful mother Huda Mahmoud and my deceased father's spirit Nader Almutib, people who had a great influence in my life.*

# Abstract

In this thesis, I investigate the molecular electronic properties of molecular junctions formed from single molecules. Most of my work uses the saturated alkane chain as my reference molecule to study some problems of charge transfer in nanoscale devices. Chapters 2 and 3 present a brief introduction to Density Functional Theory using the SIESTA implementation [2], and the Green's function formalism of electron transport as implemented in the GOLLUM code [3], which is a next-generation equilibrium transport code, born out of the non-equilibrium transport code SMEAGOL [19]. These two techniques are used to study the charge transport through dicarboxylic-acid-terminated alkanes in chapter 4, which are bound to graphene-gold nanogap electrodes. The results are then compared with those using symmetric gold electrodes and reveal that there is a difference between the two situations due to the difference in Fermi energies relative to the frontier orbitals of the molecules. Furthermore, the electrical conductance in the graphene–molecule–Au junction is higher than that in Au–molecule–Au junction, which suggests that graphene offers superior electrode performance when utilizing carboxylic acid anchor groups. In chapter 5, I show that the conductance of the saturated chain is affected by adding oxygen to the chain. Comparisons between my electronic structure calculations on oligoethers such as poly ethylene glycol (PEG) chains and previous work on alkanes shows that the conductance of oligoethers is lower than that of alkane chains with the same length. The calculation of the length dependence of the electrical conductance of alkanes and oligoethers, shows that the beta tunnelling factor $\beta_N$ per methyl unit of the alkanes is lower than the beta factor of oligoethers. In the final chapter, molecular dynamic (MD) simulations using the DLPOLY_4 [31] code, are used to examine the molecular assembly of two candidate molecules for graphene based molecular electronics, one with one pyrene anchor, Pyrene-PEGn-ex-TTF (PPT) and the other with three pyrene anchors, tri-pyrene derivative (TPPT) on a disordered graphene surface. PPT is seen to form flat structures whilst TPPT is seen to form semi-circular cone like micelle structures on the graphene surface. In the presence of water, the PPT tends to aggregate whereas the TPPT micelle expands. The hydrophobic pyrene anchors are firmly attached to the graphene surface in both cases while the hydrophilic dithiol heads groups which allow the water to disperse the micelles.

# Acknowledgments

I would like to express the deepest appreciation to my supervisor, Professor Colin J. Lambert, who has the attitude and the substance of a genius: he continually and convincingly adds a special flavor and spirit of adventure in regard to research by intensive fruitful discussion and excitement in regard to teaching over these years. I would like to thank my co-supervisor Dr. Steve Baily, Dr. Iain Grace for thier encouraged me and continues support.

I would like also to thank my sponsor, the Ministry of Higher Education in Saudi Arabia and Saudi culture mission in London, Al Taif University in Saudi Arabia, for given me this great opportunity to study a Ph.D. in the United Kingdom.

I would like to thank the collaborating experimental groups of Department of Chemistry, Xi'an-Jiaotong Liverpool University, Professor Chzhou Zhao, Dr.Longlong Liu, Dr. Qian Zhang, Dr. Shuhui Tao for their successful experiments. I would like tothank all my friends and colleagues in Colin’s group, especially Dr. Qusiy Al-Galiby, Dr. Hatef Sadeghi, Sara Sangtarash, Ali Ismael.

Last but not the least, I would like to thank my family: my mother Huda and my brothers Khalid, Talal, Ahmed, Rakan and Bader and my other half, my sister Reham, my teachers Kadijah and Abeer, and my best friend and soul sister Amani and all my friends in Saudi Arabia and United Kingdom for supporting me spiritually throughout writing this thesis and my life in general.

# Publications

1. Liu, L, Zhang, G, Tao, S, Zhao, CZ, Almutib, E, Al-Galiby, Q, Bailey, SWD, Grace, IM, Lambert, CJ, Du, J & Yang, L 2016, 'Charge transport through dicarboxylic-acid-terminated alkanes bound to graphene-gold nanogap electrodes' Nanoscale., 10.1039/C6NR03807G.

# Contents

# Chapter 1

# 1. Introduction

## 1.1. Molecular electronics

Molecular electronics, is a branch of nanotechnology used to study the electronic and thermal transport properties of circuits in which single molecules or assemblies of them are used as basic building blocks [20]. In 1947 the first prototype transistor was created by William Shockley, John Bardeen and Walter Brattain at Bell Laboratories. The transistor became the essential building block of all electronic devices through the next six decades, and became recognised as one of the most important inventions of the previous century. During the Silicon revolution in the 1960sthe transistor size was reduced from its initial prototype of a few centimetres to micrometers. This trend has continued into the 21$^{st}$. century and the transistor is now nanometers in size. This downward trend is known as Moore's law [21] which states that the number of transistors on an integrated circuit is approximately doubling every two years leading to a dramatic decrease by several orders of magnitude over six decades. This gives manufacturers the possibility to produce smaller, faster and more energy efficient devices. As the reduction in size reached the nano-scale the electronics industry has been forced to find alternatives to the classic semiconducting materials and one of these possible alternatives is to be found in molecular electronics [22]. The advantages of using molecular electronics are in providing the required reduction in size, flexibility of design and a possibility of exploitation of current silicon technology [23]. The idea of using single molecules as molecular electronic devices, started with theoretical research in the1970s [24], but only recently has it attracted intense scientific interest to explore the unique properties and opportunities. Improvement in the methods used to calculate molecular electronic properties allows theorists to deal with more complicated molecules and to match their calculations more closely to reality. Experimental groups across the world use a variety of measurement techniques to study the molecule's electronic properties. The main problem is the small size of the device leads to uncertainty about exactly what is being measured and how the molecule is orientated or connected to the electrodes. This simple fact causes a disagreement of data between experimental groups and forces them to try to find the best experimental method which can be facilitated by theoretical calculations.

Theoretical and experimental investigators have focused on electrode-molecule-electrode junctions, which will be discussed in this thesis the main experimental technique used to study these systems is the Scanning Tunnelling Microscopy Break Junctions (STM-BJ) [25, 19]. Molecular electronics is a modern technique and a single electron transistor is still to appear on an industrial scale but, a number of other interesting effects have been observed in experiment and theory. This opens the door not just for the construction if nanoscale transistors, but also nano-sensors [26], rectifiers [27], memory [28] and optical devices [29, 30].

## 1.2. Thesis outline

I will begin this thesis by introducing density functional theory (DFT) which is a numerical method I use it to investigate the electrical properties on the molecular scale. The following chapter will introduce a single molecule transport theory by using the retarded Green's function. I will use these two methods to study the charge transport through dicarboxylic-acid-terminated alkanes bound to graphene-gold nanogap electrodes in chapter 4. In chapter 5, I will be compare the conductance of alkane and oligoethylene glycols chains. In chapter 6, I will use classical molecular dynamics, to study micelle formation in selected amphiphilic molecules and how we can combine MD dynamics and DFT to calculate the conductance in the single molecule. The final chapter will give the conclusions of these six chapters, and suggest possible future work which could be carried out.

# Chapter 2

# 2. Density Functional Theory

## 2.1. Introduction

Density functional theory (DFT) is proving to be one of the most successful and promising theories used in physics, chemistry and materials science to compute the electronic structure of the ground state of many body systems, in particular atoms, molecules, and the condensed phases. The name density functional theory comes from the use of functionals of the electron density. DFT has been generalized to treat many different situations such as spin-polarized systems, multicomponent systems, and time-dependent phenomena [1]. In this thesis, I use the SIESTA [2] implementation of DFT to calculate the electrical properties of the system. DFT enables us to extract the Hamiltonian and to determine the optimum geometry for the system. Then I use the equilibrium transport code GOLLUM [3] to compute the transmission coefficient $T(E)$ for electrons of energy $E$ passing from the lower electrode to the upper electrode. Once the $T(E)$ is computed, the zero-bias electrical conductance $G$ using the Landauer formula is calculated.

### a. The Schrödinger equation

Let us start from the solution of the time-independent, non-relativistic Schrödinger equation which is the basic goal of most approaches in solid state physics and quantum chemistry.

$$H\Psi_i(\vec{x}_1, \vec{x}_2, \dots, \vec{x}_N, \vec{R}_1, \vec{R}_2, \dots, \vec{R}_M) = E_i\Psi_i(\vec{x}_1, \vec{x}_2, \dots, \vec{x}_N, \vec{R}_1, \vec{R}_2, \dots, \vec{R}_M) \quad (2.1.1)$$

Where $H$ is the Hamiltonian operator for a system consisting of $M$ nuclei and $N$ electrons in atomic units, and $\Psi_i$ is a set of solutions, or eigenstates of the Hamiltonian. Each solution has an associated eigenvalue $E_i$ which is a real number.

$$H = \sum_{i=1}^{N} -\frac{\nabla_i^2}{2m_i} - \sum_{I=1}^{M} \frac{\nabla_I^2}{2M_I} + \frac{1}{2}\sum_{i \neq j}^{N} \frac{1}{|r_i - r_j|} + \frac{1}{2}\sum_{I \neq J}^{M} \frac{Z_I Z_J}{|R_I - R_J|} - \frac{1}{2}\sum_{i,I=1}^{NM} \frac{Z_i}{|r_i - R_J|} \quad (2.1.2)$$

Here, I and J run over the M nuclei, $i$ and $j$ denote the N electrons in the system. The Hamiltonian equation (2.1.2) describes the kinetic energy of the electrons and nuclei, the attractive electrostatic interaction between the nuclei and the electrons and repulsive potential due to the electron-electron interactions, and nucleus-nucleus interactions respectively. The motion of nuclei is much slower than the electron due to the lightweight of electrons compared with the nuclei. That means the wave function which describe the full system can be separated to a nuclear wave function, and the associated electronic wave function. This separation is the Born-Oppenheimer approximation. Thus, we can write the electronic Hamiltonian in form:

$$H_e = \sum_{i=1}^{N} -\frac{\nabla_i^2}{2m_i} + \frac{1}{2}\sum_{i \neq j}^{N} \frac{1}{|r_i - r_j|} - \frac{1}{2}\sum_{i,I=1}^{NM} \frac{Z_i}{|r_i - R_J|} = T + V_{ee} + V_{ext} \quad (2.1.3)$$

The corresponding time independent Schrödinger equation is:

$$H_e \Psi_e = E_e \Psi_e \tag{2.1.4}$$

Although the Born-Oppenheimer approximation to reduces the system size there is still a hugh difficulty to solve equation (2.1.4), because the diagonalization of the general problem is practically impossible even on a modern supercomputer. The Density functional theory solves this problem by expressing the physical quantities in terms of the ground-state density. The electron density of a general many body state is defined as:

$$n(r) = \int dr_2\, dr_3 \dots dr_i \dots |\Psi(r, r_2 \dots . r_i \dots)|^2 \tag{2.1.5}$$

## 2.2. Kohn-Sham equations and self-consistency

In the density functional theory, the Kohn–Sham equation [4] is the Schrödinger equation of a fictitious system of non-interacting particles, which contains a new effective potential $V_{eff}(r)$. Therefore, the many body interactions in the external potential are modelled as a set of non-interacting particles. This replaces the original Hamiltonian of non-interacting particles by a new effective external potential which has the same ground state density as the original system. The total energy of the non-interacting system is a functional of the charge density:

$$E_{non}[n(r)] = T[n] + \frac{1}{2}\int \frac{n(r')n(r)}{|r_i - r_j|} dr dr' + \int V_{ext}(r) n(r) \mathrm{d}r + E_{xc}[n] \tag{2.2.1}$$

And $E_{xc}[n]$ is the exchange-correlation functional defined as the correction to the Hartree energy functional and the interacting kinetic energies.

$$E_{xc}[n] = T[n] - T_{non}[n] + E_H[n] \qquad (2.2.2)$$

Where

$$E_H[n] = \frac{1}{2}\int \frac{n(r)n(r')}{|r-r'|} dr dr' \qquad (2.2.3)$$

The effective potential given by:

$$V_{eff}[n] = \int \frac{n(r')}{|r-r'|} dr' + \frac{\delta E_{xc}[n]}{\delta n} + V_{ext} = V_H[n] + V_{xc}[n] + V_{ext} \qquad (2.2.4)$$

Where $V_H[n] = \int \frac{n(\acute{r})}{|r-\acute{r}|} d\acute{r}$ is the Hartree potential

And $V_{xc}[n] = \frac{\delta E_{xc}[n]}{\delta n}$ is the exchange-correlation potential. Which results in

$$H_{KS}[n]\psi_{KS} = \varepsilon_{KS}\psi_{KS} \qquad (2.2.5)$$

Where the Kohn-Sham Hamiltonian is

$$H_{KS}[n] = -\frac{1}{2}\nabla^2 + V_{eff}[n] \qquad (2.2.6)$$

There is a similarity between Kohn-Sham equation and the single particle Schrodinger equation, but the Kohn-Sham equation is a non-linear equation because n is dependent on the wavefunction and its solution is found by a self-consistent iteration. Therefore, we can impose some initial charge density on the first iteration and calculate the corresponding effective potential. Then, construct a simple single particle Schrödinger equation. Thus, we can use the solution of this Schrödinger equation to calculate the next generation of the charge density which can be used as initial density in the next iteration loop see Figure 1.1. After numbers of iterations the charge density converges to the ground state charge density. We still have problem to evaluate the exchange-correlation potential, because there is no explicit form for that. But, if one would provide an explicit exact formulation for $E_{xc}[n]$ or $V_{xc}[n]$ then this method would give the exact many-body ground state solution.

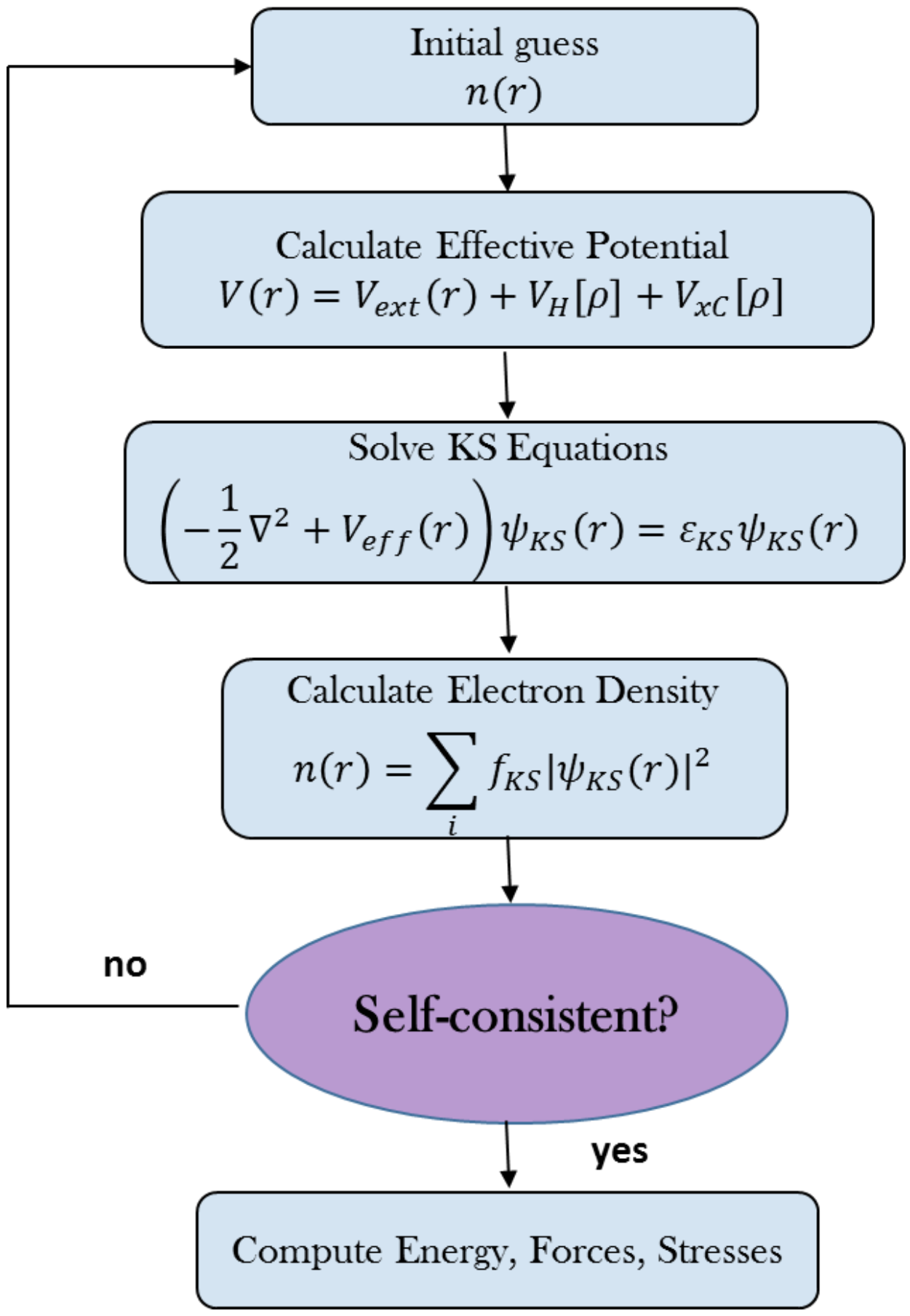

**Figure 1.1.** A typical flow of a DFT self-consistent calculation.

## 2.3. Exchange and correlation

In the previous section we found that, we can define the exchange-correlation potential $V_{xc}[n]$ as functional derivative of the exchange-correlation energy $E_{xc}[n]$. That means if we know the exact forms of $E_{xc}[n]$ and $V_{xc}[n]$, the Kohn-Sham strategy obtains the exact energy. There are numerous approximations in various degrees of accuracy to obtain the agreement results experimental data.

We will start from the basic approximation which is the local density approximation (LDA) [5], where the exchange-correlation functional depends only on the density.

$$\mathrm{E}_{xc}[n(r)] = \int n(r)\epsilon_{xc}[n(r)]d^3r \tag{2.3.1}$$

The other approximations are generalised gradient approximations (GGA) [5], where the exchange-correlation functional depends on both the density and its gradients.

$$E_{xc}[n] \approx E_{xc}^{GGA}[n] = E_x^{GGA}[n] + E_c^{GGA}[n] \tag{2.3.2}$$

Where the exchange term is:

$$E_x^{GGA} = \int n(r)\epsilon_x n(r)F_x\big(n(r), \nabla n(r)\big)dr \tag{2.3.3}$$

In addition to LDA and GGA there are a variety of new approximations such as the Van der Waals approximation which involves the Van der Waals energy functional. These approximations differ in accuracy but each of them gives acceptable results with certain systems.

## 2.4. The pseudopotential approximation

The Kohn-Sham equation simplifies the large interacting problem, but the calculation for the many-body Schrödinger equation for practical purposes is still very large and has the potential to be computationally intensive. Therefore, we need to introduce the pseudopotential approximation to help solve this problem by subtracting out the core electrons from an atom.

Each atom has to kinds of electrons: valence and core electrons. The valence electrons are in partially filled shells but core electrons are those within filled atomic shells. The core electrons are spatially localised in the vicinity of the nucleus and the valence electrons are outside the core region. When the atoms interact only the valence electrons overlap, and the core electrons could be removed and replaced by a pseudopotential. This will decrease the number of electrons in a system and also save the time and memory required to calculate properties of molecules that contain a large number of electrons.

## 2.5. SIESTA basis set

Choosing a suitable type of the basis function is the most important step in calculation by using the SIESTA code. At the self-consistent cycle, the calculation of wavefunctions is required for the diagonalising of the Hamiltonian by the inversion of a large matrix. SIESTA utilises a Linear Combination of Atomic Orbtial (LCAO) basis set to minimize the size of the Hamiltonian, which are constrained to be zero after some defined cut-off radius, and are constructed from the orbitals of the atoms.

The simplest basis set for an atom is a single-$\zeta$ basis (SZ) which corresponds to a single basis function $\psi_{nlm}(r)$ per electron orbital.

$$\psi_{nlm}(r) = \phi_{nl}^{1}(r) Y_{lm}(r) \tag{2.5.1}$$

This type of basis function consists of a product of one radial wavefunction $\phi_{nl}^{1}(r)$ and one spherical harmonic $Y_{lm}(r)$ (2.5.1). The radial part of the wavefunction is found by using Sankey method [17], (CMFinch)where the radial part of the basis functions is obtained by solving a modified version the Schrödinger equation (2.5.2) which solved for the atom placed inside a spherical box and the boundary condition for the radial wavefunction is equal to zero at a cut-off radius $r_{cut}$

$$\left[-\frac{1}{2r}\frac{d^2}{dr^2}+\frac{l(l+1)}{2r^2}+V_{nl}^{ion}(r)\right]r\phi_{nl}^1(r)=(\epsilon_{nl}+\delta E)r\phi_{nl}^1(r) \qquad (2.5.2)$$

Here $V_{nl}^{ion}(r)$ is the corresponding pseudopotential and $\delta E$ is the confinement energy shift.

For more accuracy, multiple- $\zeta$ basis sets that include more than one basis functions to a single orbital. Thus, if there are two basis functions per orbital then it is called a double- $\zeta$ basis set. The two functions are generated by the split valence scheme, which gives some freedom for the core part which is determined by a cut-off radius $r_s$ The first function is simply the single- $\zeta$ function (SZ) and the second function is

$$\phi_{nl}^2(r)=\begin{cases} r^l(a_{nl}-b_{nl}r^2) & r<r_s \\ \phi_{nl}^1 & r<r_s \end{cases} \qquad (2.5.3)$$

Where the $a_{nl}$ and $b_{nl}$ are determined by the continuity and derivative continuity conditions at $r_s$. If the real orbitals polarised due to the surrounding external electric field, the double- $\zeta$ basis set with polarisation, yields the so called double- $\zeta$ -polarised (DZP) basis set. In this thesis the DZ and DZP basis set will be used as standard and " = 0:02Ry. Table (2.5.1) shows the number of basis orbitals for a selected number of atoms for single- $\zeta$, single- $\zeta$ polarised, double- $\zeta$, double- $\zeta$ polarised.

| Atom | Valence configuration | SZ | SZP | DZ | DZP |
|---|---|---|---|---|---|
| **H** | (1s) | 1 | 4 | 2 | 5 |
| **C** | ($2s^2$ $2P^2$) | 4 | 9 | 8 | 13 |
| **N** | ($2S^2$ $2P^3$) | 4 | 9 | 8 | 13 |
| **S** | ($3S^2$ $3P^4$) | 4 | 9 | 8 | 13 |
| **Au** | ($6S^1$ $5d^{10}$) | 6 | 9 | 12 | 15 |

**Table 2.5.1**: Example of the number of radial basis functions per atom as used within the SIESTA for different degrees of precision.

molecular orbitals can be represented as Linear Combinations of Atomic Orbitals (LCAO) given by:

$$\psi_i = \sum_{\mu=1}^{L} c_{\mu i}\phi_\mu \tag{2.5.4}$$

Where $\psi_i$ represents the molecular orbitals (basis function), $\phi_\mu$ are basis functions, $c_{\mu i}$ are numerical coefficients and L is the total number of the basis functions.

## 2.6. Calculation in practice using SIESTA

To start a calculation of transport by using SIESTA method [2], we need to construct the atomic configuration of the system and chose suitable pseudopotentials [5] for each element, which can be different for every exchange-correlational functional. Furthermore, we need appropriate basis set which has to be made for each element present in the calculation. There are other computational input parameters we need it in our calculation, such as the grid fineness and density or energy convergence tolerances, and for periodic systems the Brillouin zone sampling for the k-space integral. These parameters control the accuracy of the numerical procedures and there is a trade-off between the speed of the computational and numerical accuracy. Another type of computational parameters are the convergence controlling parameters, such as the Pulay parameters, which are responsible for accelerating or maintaining the stability of the convergence of the charge density.

Then we need to generate the initial charge density for non-interacting systems from the pseudopotential. Then, the self-consistent calculation starts by solving the Poisson equation to calculating the Hartree potential $V_H$ which obtained by with the multigrid [6, 7] or fast Fourier-transform [6, 8] method, and also calculating the exchange correlation potential which obtained by performing the integrals given in Sec. 2

After solving the Kohn-Sham equations and obtained a new density $n(r)$ we start the next iteration. The iteration continues until we get the necessary convergence criteria. As a result, we obtain the ground state Kohn-Sham orbitals and the ground state energy for a given atomic configuration. For structural optimisation the procedure described above is in another loop, which is controlled by the conjugate gradient [9, 6] method for finding the minimal ground state energy and the corresponding atomic configuration.

# Chapter 3

# 3. Transport Theory

## 3.1. Introduction

In the previous chapter I have displayed the DFT method of calculating the electronic structure of an isolated molecule or molecular wire. and in this chapter we go beyond to the next step which is to connect the isolated molecule to metallic electrodes and investigate its electronic properties such as transmission and reflection. This is done by using the Green's function scattering formalism.

I will start this chapter with a brief overview of the Landauer formula. Then, I will introduce the simplest form of a retarded Green' function for one-dimensional tight binding lattice. Next, I will show how the Green's function is related to the transmission coefficient across the scattering region. This method which we used it in the simple structure is the same method we will used it to derive the transmission coefficient of mesoscopic conductors of arbitrarily complex geometry.

## 3.2. The Landauer formula

In the late 1950's, Rolf Landauer related the electrical resistance of a conductor to its scattering properties in a new formula called Landauer Formula. This formula is the most popular way to describe coherent transport in nanodevices. The main idea of this process is that, a single wave function is sufficient to describe the electronic flow. That means the transport properties of a mesoscopic system like the conductance are related to the transmission probability for an electron passing through this system [1]. To derive the Landauer Formula, we consider the system shown in Figure 3.2.1. The system consisted of a mesoscopic scatterer connected to the two leads which in turn connect to external reservoirs. The chemical potentials ($\mu_L$ and $\mu_R$), for the reservoirs are slightly different ($\mu_L - \mu_R > 0$) that will drive electrons from the left to the right reservoir.

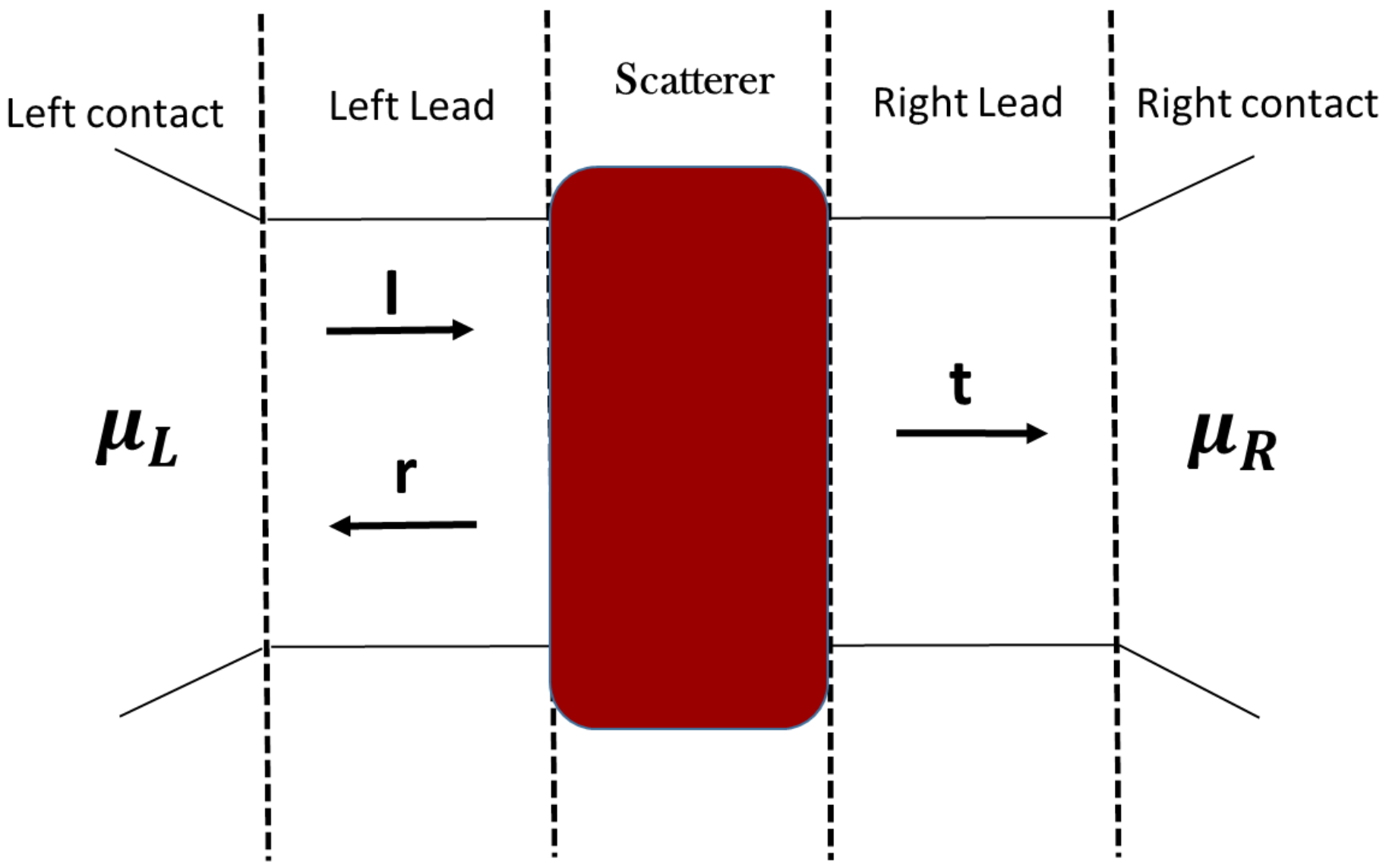

**Figure 3.2.1**: A mesoscopic scatterer connected to contacts by ballistic leads. Where $\mu_R$ (right) and $\mu_L$ (Left) are the chemical potential in the contacts (Figure is taken from (10)).

The incident current passing through this system from the left to the write reservoir is:

$$\delta I = ev\left(\frac{\partial n}{\partial E}\right)(\mu_L - \mu_R) \quad (3.1.1)$$

Where $e$ is the electronic charge, $v$ is the group velocity in the lead at $\mu$ energy, and $\left(\frac{\partial n}{\partial E}\right)$ is the density of states. For on-dimensional system:

$$\frac{\partial n}{\partial E} = \frac{\partial n}{\partial k}\frac{\partial k}{\partial E} = \frac{\partial n}{\partial k}\frac{1}{v\hbar} \quad (3.1.2)$$

As in one-dimension, $\partial n/\partial k = 1/\pi$ and $\partial n/\partial E = 1/\hbar v$, since the group velocity is $v = \frac{1}{\hbar}\frac{dE}{dk}$, thus we can rewrite equation (3.1.1)

$$\delta \mathrm{I} = \frac{2e}{h}(\mu_L - \mu_R) = \frac{2e^2}{h}\delta \mathrm{V} \quad (3.1.3)$$

Where $\delta V$ is the voltage which corresponds to the chemical potential difference, and number 2 is a factor for spin dependency. Equation (3.1.3) is important, because it reveals that the conductance for one open channel in the absence of a scattering region is $\left(\frac{e^2}{h}\right)$ which is $\approx 77.5\ \mu S$, this it corresponds to the universal resistance $h/e^2 \approx 12.9 k\Omega$. On the other hand, if the system has a scattering region, the current is partially reflected with a probability $R = |r|^2$ and partially transmitted with a probability $T = |t|^2$. The current passing through the scatterer to the right lead will be:

$$\delta \mathrm{I} = \frac{2e^2}{h}T\delta \mathrm{V} \Rightarrow \frac{\delta \mathrm{I}}{\delta \mathrm{V}} = G = \frac{2e^2}{h}T \quad (3.1.4)$$

This is the Landauer formula, were the conductance is $G = I/V = (2e^2/h)T$, and transmission is evaluated at the Fermi energy [1]. At zero voltage and finite temperature the conductance is:

$$G = \frac{I}{V} = G_0 \int_{-\infty}^{\infty} dE T(E)\left(-\frac{df(E)}{dE}\right) \quad (3.1.5)$$

Where $G_0$ is the quantum of conductance $G_0 = \left(\frac{2e^2}{h}\right)$, $f(E)$ is Fermi distribution function which have this form at left and right reservoir:

$$f_{left}(E) = \frac{1}{\left[e^{\beta\left(E-E_F^{left}\right)}+1\right]} \quad , \quad f_{left}(E) = \frac{1}{\left[e^{\beta\left(E-E_F^{right}\right)}+1\right]}$$

Where $E_F$ is the Fermi energy of the reservoir, $E_F^{left} = E_F + \frac{eV}{2}$ , $E_F^{rigth} = E_F - \frac{eV}{2}$ , and $\beta = \frac{1}{k_B T}$ where $T$ here is the temperature, and $k_B$ is Boltzmann constant$k_B = 8.62 \times 10^{-5} eV/k$. Since the quantity $-\frac{df(E)}{dE}$ is a normalised probability distribution of width approximately equal to $k_BT$, centred on the Fermi energy $E_F$, the integral in equation (3.1.5) represents a thermal average of the transmission function *T(E)* over an energy window of the width $k_BT$ (= 25 meV at room temperature) [11].

In zero voltage and zero temperature,

$$G = G_0 T(E_F) \tag{3.1.6}$$

If we have multiple open channels the Landauer formula is:

$$\frac{\delta I}{\delta V} = G = \frac{2e^2}{h}\sum_{i,j}\left|t_{i,j}\right|^2 = \frac{2e^2}{h}Tr(tt^{\dagger}) \tag{3.1.7}$$

The transmission amplitude $t_{i,j}$ passing through scattering region from the $j-th$ channel on the left side to the $i-th$ channel on the right side, and the reflection amplitude $r_{i,j}$ passing also through scattering region but in opposite way. Therefore, transmission and reflection together will make the S matrix:

$$S = \begin{pmatrix} r & t' \\ t & r' \end{pmatrix} \tag{3.1.8}$$

The S matrix described the complete scattering process, where $t, t'$ are the transmission amplitude matrices from left to right and right to left respectably, and $r, r'$ are the reflection amplitude matrices from left to right and right to left respectably. From equation (3.1.7), $r, t$ and $r', t'$ are matrices for more than one channel, and could be complex (e.g. presence of a magnetic field). Furthermore, the S matrix is unitary $SS^{\dagger} = I$ due to charge conservation. The S matrix is a essential topic of scattering theory. It is useful not just in describing transport in the linear response regime, but also in other problems, such as to describe adiabatic pumping [12].

## 3.3. Scattering Theory

### 3.3.1. One dimensional (1-D) linear crystalline lattice.

To understand the electronic scattering problem, I will start with a simple one-dimensional crystalline system as shown in Figure 3.2.2, where each atom has site energy $\varepsilon_0$ and is coupled with the next atom by hopping parameters $-\gamma$. We can describe this system by using Schrodinger equation (3.2.1).

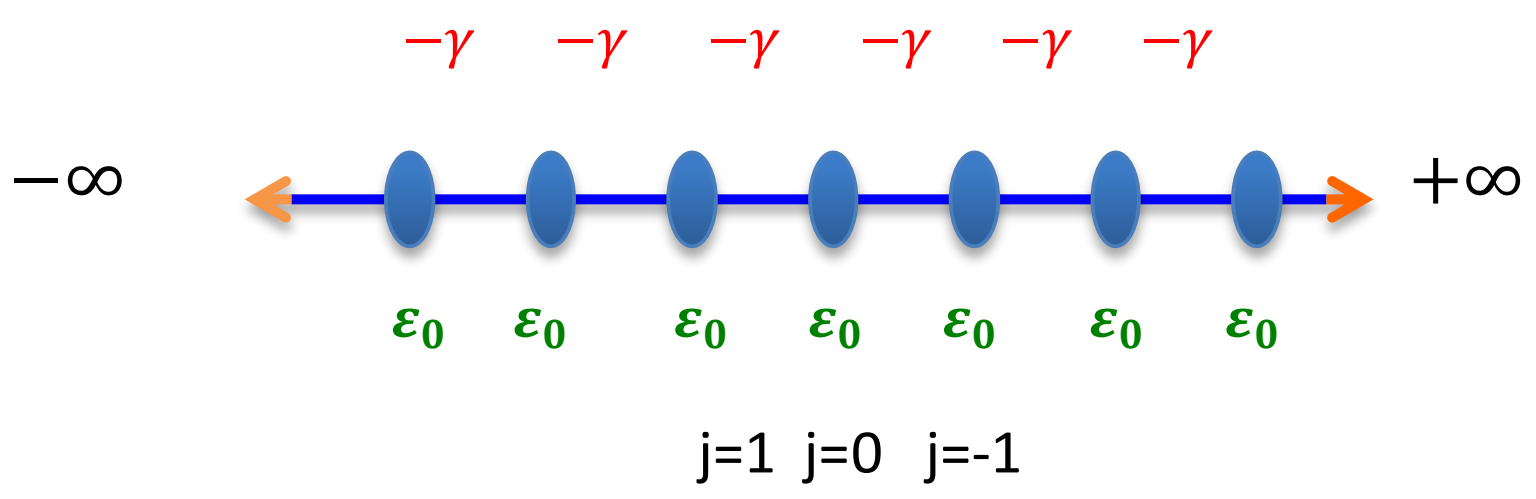

**Figure 3.2.2** Tight-binding model of a one dimensional crystalline chain, $\varepsilon_0$ are theon-site energies $\gamma$ the inter-site couplings, and j is just a label.

$$H|\psi\rangle = E|\psi\rangle \qquad (3.2.1)$$

To solve Schrödinger equation in (3.2.1) for the system in Figure 3.2.2, we need to define the Hamiltonian for this system first:

$$H = \begin{pmatrix} \ddots & -\gamma & & \\ -\gamma & \varepsilon_0 & -\gamma & \\ & -\gamma & \varepsilon_0 & -\gamma \\ & & -\gamma & \ddots \end{pmatrix} \qquad (3.2.2)$$

The Hamiltonian matrix $H$ is real symmetric matrix, thus the Hamiltonian is Hermitian ($H = H^{\dagger}$), and eigenvalue $E_n$ is a real, thus we have N solutions for the matrix($N \times N$). The Schrödinger equation for infinite crystalline chain at point $j$ is:

$$\varepsilon_0\psi_j - \gamma\psi_{j-1} - \psi_{j+1} = E\psi_j \qquad (3.2.3)$$

Where $\psi$ is the wavefunction, and equation (3.2.2) satisfied for all j going from$+\infty\ to - \infty$. After we factored out $\psi_{j+1}$ we get Recurrent Relation

$$\psi_{j+1} = \left(\frac{\varepsilon_0 - E}{\gamma}\right)\psi_j - \psi_{j-1} \qquad (3.2.4)$$

And by using Bloch's wave we can define the wave function $\psi_j = e^{ikj}$ (for electron in a perfect crystal and substituting in (3.2.3) we get the dispersion relation in equation (3.2.5).

$$E(k) = \ \varepsilon_0 - 2\gamma cosk \qquad (3.2.5)$$

Dispersion relation is very important relation in electron transport because study of transport properties of solids is related to study the dispersion relation of electrons. As well as, it is connecting between the waves properties such as $k$ (wavenumber) and $E$ (wave energy).

The speed of the current passing through the chain is the group velocity $v$, which is defined as the derivative of the energy (dispersion relation) with respect the wave number $E(k)$:

$$v = \frac{1}{\hbar}\frac{\partial E}{\partial k} \tag{3.2.6}$$

The normalised wave function with a factor $1/\sqrt{v}$, will takes this form:

$$\psi_j = \frac{1}{\sqrt{v}} e^{ikj} \tag{3.2.7}$$

### 3.3.2. Retarded Green's Function

To obtain the transmission and reflection amplitudes we need to calculate the Green's function of the system. Suppose that we have a source $l$ in the one-dimensional chain as shown in Figure 3.2.3.

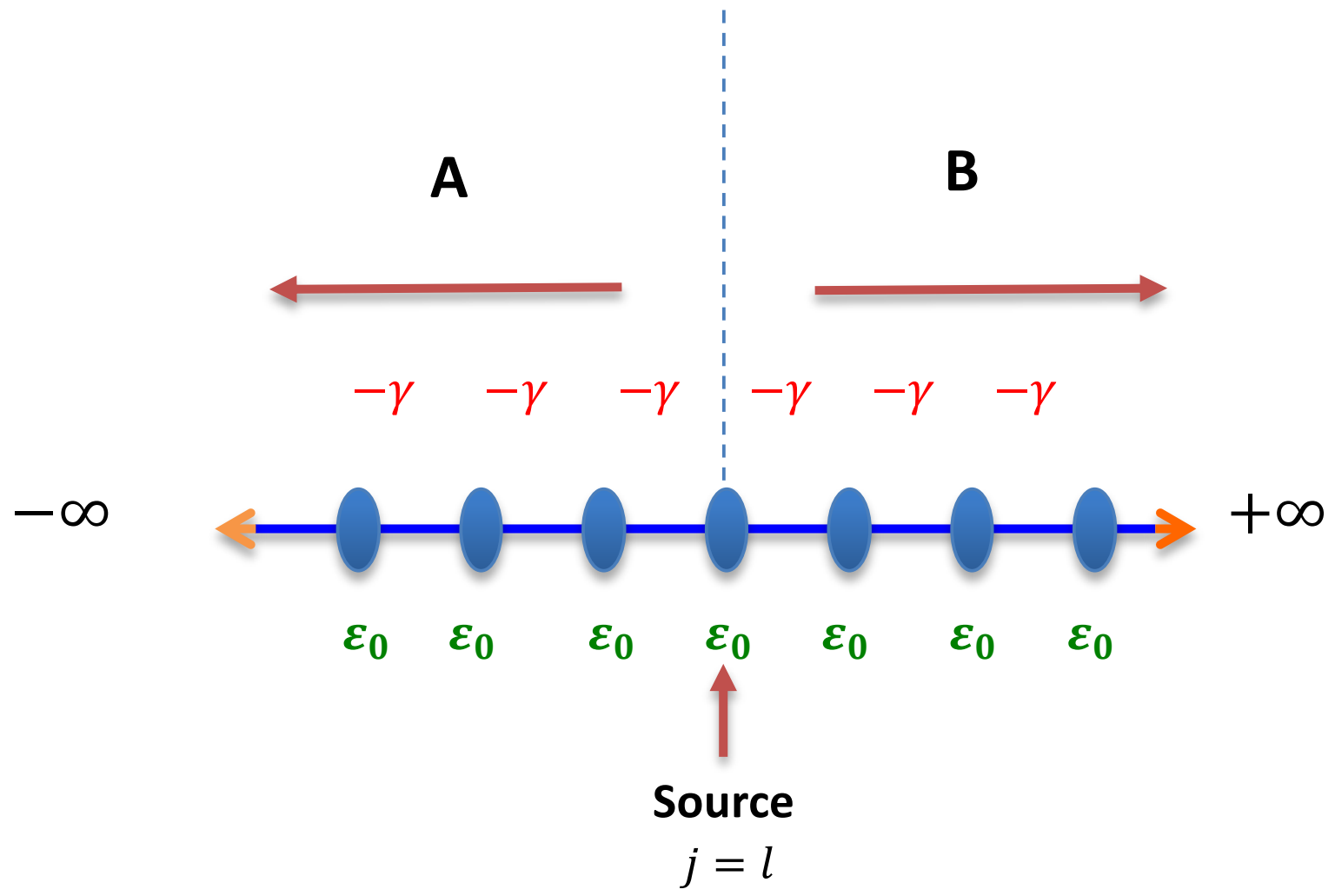

**Figure 3.2.3**. Retarded Green's Function of an infinite one-dimensional chain. The wave emitted from the source $l$ propagate left and right with amplitudes A and B respectively.

We can rewrite Schrödinger equation (3.2.1) in this form:

$$(E - H)|\psi\rangle = 0 \tag{3.2.8}$$

So, what is the Green's function? Green function is satisfying this equation:

$$(E - H)G = I \tag{3.2.9}$$

Thus, the difference between Green's function and wavefunction that the Green's function $G$ is a matrix and wavefunction $|\psi\rangle$ is a vector. $I$ is the identity matrix, and $H$ is the Hamiltonian matrix of infinity system. The solution of equation (3.2.9) can be written as:

$$G = (E - H)^{-1} \tag{3.2.10}$$

Equation (3.2.10) is singular if the energy $E$ is equal to an eigenvalue of the Hamiltonian $H$. To avoid this singularity, considers in practice the limit to be the solution of (3.2.9)

$$G_{\pm} = \lim_{\eta \to 0} (E - H \pm i\eta)^{-1} \tag{3.2.11}$$

Where $\eta$ is a positive number, and $G_{+}(G_{-})$ is the retarded (advanced) Green's function, respectively. I use retarded Green's functions in all the calculation done of this thesis. The retarded Greens function $g_{jl}$, describes the response of a system at a point $j$ due to a source $l$ . This source causes excitation which rise two waves, travelling to the left and right as shown in Figure 3.2.4.

$$g_{jl} = \begin{cases} Ae^{-ikj} & , \quad j \leq l \\ Be^{+ikj} & , \quad j \geq l \end{cases} \tag{3.2.12}$$

Where, A and B are the amplitude of two outgoing waves travelling to the left (A) and right (B) as shown in Figure 3.2.4. These expressions at (3.2.12) satisfy two conditions:

First condition: The Green' function must be continuous at $j = l$:

$$Ae^{-ikl} = Be^{ikl} = C \tag{3.2.13}$$

And from (3.2.12) we can find that:

$$g_{jl|_{j=l}} = \begin{cases} Ae^{-ikl}, & A = C\ e^{+ikl} \\ Be^{+ikl}, & B = C\ e^{-ikl} \end{cases} \tag{3.2.14}$$

So, from (3.2.13) and (3.2.14) we found the Green' function at point:

$$g_{il} = Ce^{ik|j-l|} \tag{3.2.15}$$

Second condition: the expression (3.2.15) should satisfy Green's Equation, $(E - H)g_{jl} = \delta_{jl}$:

Where $\delta_{jl}$ is Kronecker delta. For infinite system:

$$Eg_{jl} - \sum_{j=-\infty}^{\infty} H_{ji}\, g_{il} = \delta_{jl}$$

$$\varepsilon_0 g_{jl} - \gamma g_{j+1,l} - \gamma g_{j-1,l} = Eg_{jl} - \delta_{jl} \tag{3.2.16}$$

And to find the constant C, we use equation (3.2.15) into (3.2.16) at $j = l$:

$$(\varepsilon_0 - E)C - \gamma Ce^{ik} - \gamma Ce^{ik} = -1$$

$$2\gamma C cosk - 2\gamma C e^{ik} = -1$$

$$C = \frac{1}{2i\gamma sink} = \frac{1}{i\hbar v} \quad (3.2.17)$$

By substituted the value of C into equation (3.2.15), we found the retarded Green's function of an infinite one-dimensional chain (3.2.18), which describe outgoing waves from the source at point $l$.

$$g_{jl} = \frac{e^{ik|j-l|}}{i\hbar v} \quad (3.2.18)$$

If the two waves incoming from left and right into point, so we can imagine point $l$ as a sink not a source, then we can found another solution to this problem which is the advanced Green's function:

$$g_{jl} = \frac{e^{ik|j-l|}}{i\hbar v} \quad (3.2.19)$$

In this thesis, I will focus in the retarded Green' function.

### 3.3.3. Semi-Infinite One-Dimensional chain

After we found the Green's function of an infinite one-dimensional chain, in this section we want to define the Green's function for a semi-infinite one-dimensional chain see Figure 3.2.5, with site energies, $\varepsilon_0$, and hopping elements, $\gamma$, terminates at a given point $M - 1$, where the Green's function for site $M$ is zero. Therefore, we expected that the Green's function for the semi-infinite chain is the Green's function of an infinite one-dimensional chain adding a wave function which reflected from the boundary.

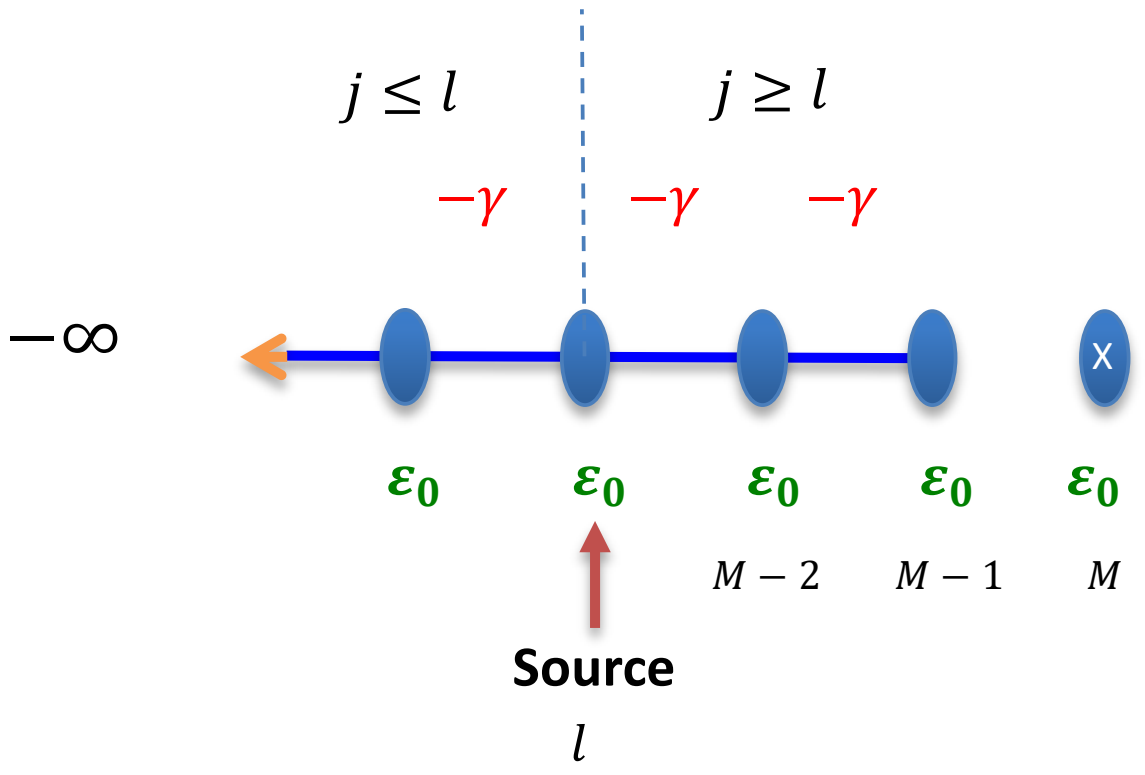

**Figure 3.2.5** Tight-binding approximation of a semi-infinite one-dimensional chain with on-site energies $\varepsilon_0$ and couplings $\gamma$.

$$g_{j,l} = \frac{e^{ik|j-l|}}{i\hbar v} + De^{-ik|j-l|} \tag{3.2.20}$$

In order to find the constant D, we will use the boundary condition for semi-infinity chain. If the source $l$ is at point $(M = l)$, then there is no effect on the chain. That means the $g_{j,M} = 0 \; for \; j \leq l$ . apply this condition in equation (3.2.20) we get:

$$g_{j,M} = \frac{e^{ik|j-M|}}{i\hbar v} + De^{-ik|j-M|} = 0$$

$$g_{j,M} = \frac{e^{ik(M-j)}}{i\hbar v} + De^{-ik(M-j)} = 0$$

$$D = -\frac{e^{2ik(M-j)}}{i\hbar v} \tag{3.2.21}$$

Substituting this back into the Green's function (3.2.20), we find:

$$g_{j,l} = \frac{1}{i\hbar v}\left(e^{ik(l-j)} - e^{ik(2M-j-l)}\right) \qquad (3.2.22)$$

The second condition is, there is no bond between site (M-1) and M, so the source dos not effect of any point beyond (M-1). Thus if $j \geq l \;\; and \;\; j = M \;\; \Rightarrow \;\; g_{M,l} = 0$ :

$$g_{M,l} = \frac{e^{ik(M-l)}}{i\hbar v} + De^{-ik(M-l)} = 0$$

$$D = -\frac{e^{2ik(M-l)}}{i\hbar v} \qquad (3.2.23)$$

Substituting D into the Green's function, obtained:

$$g_{j,l} = \frac{1}{i\hbar v}\left(e^{ik(j-l)} - e^{ik(2M-l-j)}\right)$$

$$g_{j,l} = \begin{cases} \frac{1}{i\hbar v}\left(e^{ik(j-l)} - e^{ik(2M-l-j)}\right), \; \geq l \\ \frac{1}{i\hbar v}\left(e^{ik(l-j)} - e^{ik(2M-j-l)}\right), \; j \leq l \end{cases} \qquad (3.2.24)$$

$$g_{j,l} = g_{j,l}^{\infty} + \psi_{j,l}^{M} \qquad (3.2.25)$$

Where $g_{j,l}^{\infty}$ is the Green's function for the infinite chain and $\psi_{j,l}^{M}$ is the wave function represent the boundary condition:

$$\psi_{j,l}^{M} = -\frac{e^{ik(2M-j-l)}}{i\hbar v} \qquad (3.2.26)$$

The Green's function in the end of the chain will have the following form at the boundary $= l = M - 1$ :

$$g_{M-1,M-1} = -\frac{e^{ik}}{\gamma} \qquad (3.2.27)$$

This is the surface Green's function.

### 3.3.4. One dimensional (1-D) scattering

Now we can study the scattering matrix of a simple one dimensional scatterer by connecting the two semi-infinite electrodes coupled by hopping $\propto$, and the end of the left and right semi-infinite leads is $\varepsilon_L$ and $\varepsilon_R$ respectively, see Figure 3.2.6.

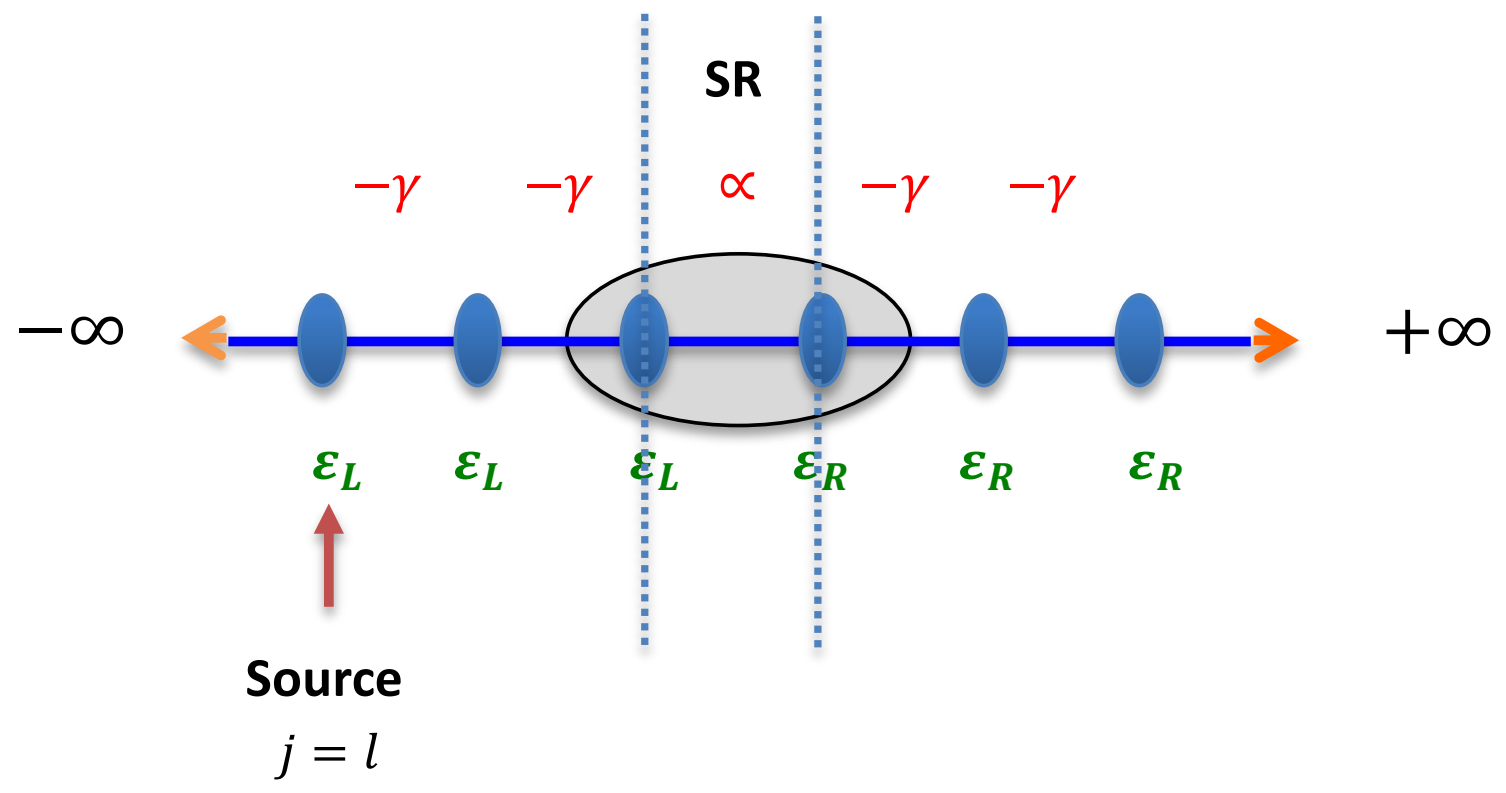

**Figure 3.2.6.** Tight-binding model of two semi-infinite one-dimensional leads, with on-site energies $\boldsymbol{\varepsilon_0}$ and couplings $\gamma$, they coupled by a hopping parameter $\propto$.

From the surface green's function of the semi-infinite leads (3.2.27), we can write the Green's function form of the decoupled system or $\propto = 0$:

$$g = \begin{pmatrix} g_L & 0 \\ 0 & g_R \end{pmatrix} = \begin{pmatrix} -\frac{e^{ik}}{\gamma} & 0 \\ 0 & -\frac{e^{ik}}{\gamma} \end{pmatrix} \qquad (3.2.28)$$

Where $g$ is the decupled Green's function.

If we the two leads connected then, we need to use the Dyson's equation to get the Green's Function of the coupled system $G$:

$$G = (g^{-1} - V) \qquad (3.2.29)$$

This equation is Dyson Equation, and the operator $V$ define scattering region,

$$V = \begin{pmatrix} 0 & V_c \\ V_c^{\dagger} & 0 \end{pmatrix} = \begin{pmatrix} 0 & \propto \\ \propto^{\dagger} & 0 \end{pmatrix} \qquad (3.2.30)$$

Substituted equation (3.2.28) and (3.2.30) into (3.2.29) to get the solution to the Dyson equation:

$$G = \frac{1}{\gamma^2 e^{-2ik} - \propto^2} \begin{pmatrix} -\gamma e^{-ik} & \propto \\ \propto^{\dagger} & -\gamma e^{-ik} \end{pmatrix} \qquad (3.2.31)$$

## 3.3.5. Transmission and Reflection:

Now we can calculate the transmission $t$, and the reflection $r$ amplitudes by using Green's function (3.2.31). This is done by using the Fisher-Lee relation [13, 14] which relates the scattering amplitudes of a scattering problem to its Green's function. Suppose that the source $l$ is in the left lead and emits two waves travelling to the left and to the right, and point $j$ is in the right lead. The right-going wave to effect point $j$, it will travel through the scatterer. Thus, the Green's function effected by two waves; left moving wave $A(e^{-ik|j-l|} + re^{+ik|j-l|})$, and the transmitted wave $(Ate^{+ik|j-l|})$, where $A = {}^{1}/_{i\hbar v}$ . The Fisher-Lee relations at the point before the scatterer $M - 1$ is:

$$j = l = M - 1$$

$$r = G_{M-1,M-1}\, i\hbar v - 1 \qquad (3.2.32)$$

$$t = G_{M-1,M}\, i\hbar v e^{ik} \qquad (3.2.33)$$

If the wave sending through right to left lead The Fisher-Lee relations will be:

$$r' = G_{M,M}\, i\hbar v - 1 \qquad (3.2.34)$$

$$t' = G_{M,M-1}\, i\hbar v e^{-ik} \qquad (3.2.35)$$

Then by using Green's function (3.2.31) we can calculate the transmission and reflection cofficients:

$$r = -\frac{(i\hbar v)\gamma e^{-ik}}{\gamma^2 e^{-2ik} - \propto^2} - 1$$

$$t = \frac{(i\hbar v) \propto e^{ik}}{\gamma^2 e^{-2ik} - \propto^2}$$

$$r' = -\frac{(i\hbar v)\gamma \quad {}^{-ik}}{\gamma^2 e^{-2ik} - \propto^2} - 1$$

$$t' = \frac{(i\hbar v) \propto e^{-ik}}{\gamma^2 e^{-2ik} - \propto^2}$$

Where the probabilities of transmission and reflection are:

$$T = |t|^2, R = |r|^2, T' = |t'|^2, R' = |r'|^2$$

We have now the full scattering matrix so, by using the the Landauer formula, we can calculate the zero bias conductance of the system.

## 3.4. Generalization of the scattering formalism

In this section, we generalize the Green's function formal in previous section to we can use it to calculate the conductance of any an arbitrarily complex conductor. I will follow the derivation of Lambert, which is presented in [15]. we started by computed the Green's function of crystalline leads, then we used the decimation method to reduce the dimensionality of the scattering region. Finally, we got the form of the scattering amplitudes by means of generalization of the Fisher-Lee relation.

### 3.4.1. Hamiltonian and Green's function of the leads

Consider the system in Figure 3.3.1 that shows a given number of semi-infinite and crystalline leads connected to a scattering region of arbitrary geometry and the transport is in the *z* direction. Since the leads are semi-infinite and crystalline, the Hamiltonian describing this system can be written as a generalization of a one dimensional lead:

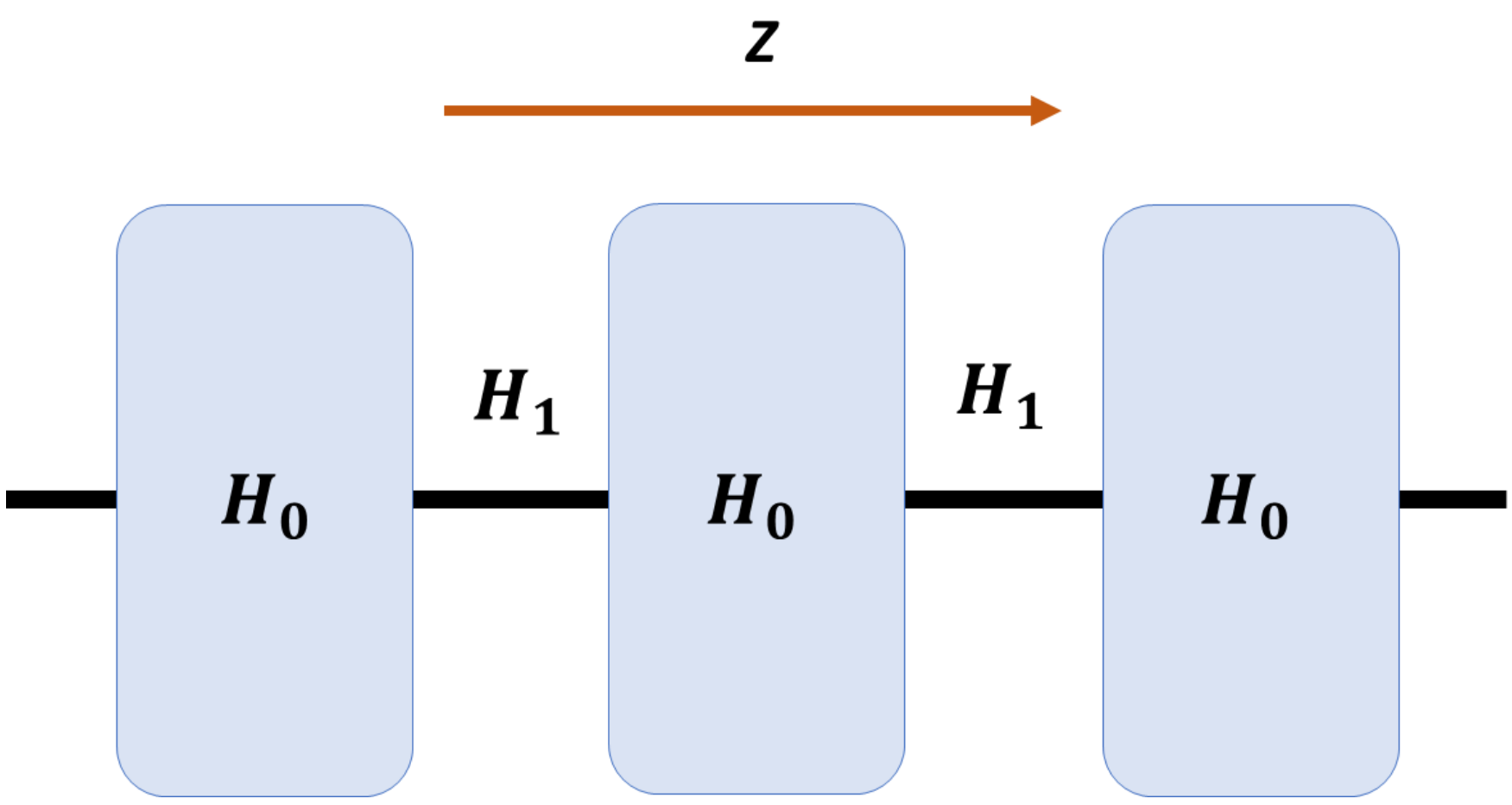

**Figure 3.3.1**. An infinite generalized lead. States described by the Hamiltonian $H_0$ are connected via generalized hopping $H_1$. The direction z is defined to be parallel to the axis of the chain. One can assign for each slice an individual z value. [15]

$$H = \begin{pmatrix} \ddots & \ddots & \ddots & & \\ \ddots & H_0 & H_1 & 0 & \\ \ddots & H_1^\dagger & H_0 & H_1 & \ddots \\ & 0 & H_1^\dagger & H_0 & \ddots \\ & & \ddots & \ddots & \ddots \end{pmatrix}_{\infty \times \infty} \tag{3.3.1}$$

Where $H_0$ are the hopping component and $H_1$ are the unit cells of the electrode repeated in the z-direction. $H_0 and\, H_1$ are in general complex matrices and the total Hamiltonian $H$ is Hermitian matrix and has real eigenvalues and complete set of orthonormal eigenvector. In order to calculate the Green's function of any lead, we have to calculate the spectrum of its Hamiltonian by solving the Schrödinger equation of this lead:

$$H_0\Psi_z + H_1\Psi_{z+1} + H_1^\dagger\Psi_{z-1} = E\Psi_z \tag{3.3.2}$$

The wave function $\Psi_z$ is describing the slice located at position $z$ along the $z$ axis. The general system is infinitely periodic only in the z direction and. This means we can be solving the Schrödinger equation by using the Bloch form which consisting of a product of a propagating plane wave, and a wavefunction $\emptyset_k$ in the perpendicular direction to transport and it have $M$ degrees of freedom ($1 \times M$ dimensional vector).

$$\Psi_z = \frac{1}{\sqrt{v_k}} e^{ikz} \emptyset_k \tag{3.3.3}$$

Where $v_k$ is the group velocity, $k$ is the wave number. Substituting the wave function (3.3.3) into the Schrödinger equation (3.3.2) gives:

$$(H_0 + e^{ik}H_1 + e^{-ik}H_1^{\dagger} - E)\emptyset_k = 0 \quad (3.3.4)$$

To compute the Green's function $g$ of such a structure, for all real energies, we need to solve for the eigenvalue $E(k)$, so we have to determine the allowed values of the wavevector $k$ by solving the secular equation.

$$\det(H_0 + e^{ik}H_1 + e^{-ik}H_1^{\dagger} - E) = 0 \quad (3.3.5)$$

The conventional band-theory for real $k$, yields M energy bands, $E_n(k)$, $n = 1, \ldots\ldots, M$, with $E_n(k + 2\pi) = E_n(k)$. If we define $\chi_k = e^{-ik}\emptyset_k$, then we can write equation (3.3.4) in its matrix form:

$$\begin{pmatrix} -H_1^{-1}(H_0 - E) & -H_1^{-1}H_1^{\dagger} \\ I & 0 \end{pmatrix}\begin{pmatrix} \emptyset_k \\ \chi_k \end{pmatrix} = e^{ik}\begin{pmatrix} \emptyset_k \\ \chi_k \end{pmatrix} \quad (3.3.6)$$

$H_0$ is the slice Hamiltonian with size $M \times M$, $I$ is the $M$ dimensional identity matrix. Equation (3.3.6) have two states, one on the left and one on the right. if wave number $k_l$ is a real wave number, a state is propagating and if the state has an imaginary part, then the state is decaying. If the imaginary part of the wave number is positive then it is a left decaying state, if it has a negative imaginary part it is a right decaying state. The propagating states are sorted according to the group velocity of the state defined by:

$$v_{k_l} = \frac{1}{\hbar}\frac{\partial E_{k,l}}{\partial k} \quad (3.3.7)$$

If the state has positive group velocity $v_{k_l}$, means it is a right propagating state, otherwise it is a left propagating state. To summarise:

| | Left | Right |
|---|---|---|
| **Decaying** | $Im(k_l) > 0$ | $Im(k_l) < 0$ |
| **Propagating** | $Im(k_l) = 0, v_{k_l} < 0$ | $Im(k_l) = 0, v_{k_l} > 0$ |

**Table 1**: Sorting the eigenstates into left and right propagating or decaying states according to the wave number and group velocity.

I will indices for the right propagating/decaying wave numbers by $k_l$and for the left propagating/decaying wave numbers by $\bar{k}_l$. Thus $\emptyset_{k_l}$ is a wave function associated to a right state and $\emptyset_{\bar{k}_l}$ is associated to a left state.

As the wavefunctions for different values of k need not be orthogonal equation (3.3.4), so we have to deal with the nonorthogonality when constructing the Green's function. Therefore, we will introduce the duals to the $\emptyset_{k_l}$ and $\emptyset_{\bar{k}_l}$ in a way that they should obey:

$$\widetilde{\emptyset}_{k_j}\emptyset_{k_j} = \widetilde{\emptyset}_{\bar{k}_l}\emptyset_{\bar{k}_l} = \delta_{ij} \qquad (3.3.8)$$

This yields the generalized completeness relation:

$$\sum_{l=1}^{M} \widetilde{\emptyset}_{k_l}\emptyset_{k_l} = \sum_{l=1}^{M} \widetilde{\emptyset}_{\bar{k}_l}\emptyset_{\bar{k}_l} = I \qquad (3.3.9)$$

Now we can start by calculating the Green's function first for the infinite system, and then, by satisfying the appropriate boundary conditions, for the semi-infinite leads at their surface. Since the Green's function satisfies the Schrödinger equation when $z \neq z'$, so the Green's function will have consisted from the mixture of the eigenstates $\emptyset_{k_l}$ and $\emptyset_{\bar{k}_l}$:

$$g_{z,z'} = \begin{cases} \sum_{l=1}^{M} \emptyset_{k_l} e^{ik_l(z-z')} w_{k_l}, & z \geq z' \\ \sum_{l=1}^{M} \emptyset_{\bar{k}_l} e^{i\bar{k}_l(z-z')} w_{\bar{k}_l}, & z \leq z' \end{cases} \tag{3.3.10}$$

Now we want to find the $w$ vectors by using tow condition for The Green's function (3.3.10):

1. Must be continuous at $z = z'$:

$$\sum_{l=1}^{M} \emptyset_{k_l} w_{k_l} = \sum_{l=1}^{M} \emptyset_{\bar{k}_l} w_{\bar{k}_l} \tag{3.3.11}$$

2. Satisfy Green's function Equation: $(E - H)g_{z,z'} = \delta_{z,z'}$

$$\sum_{l=1}^{M} [(E - H_0)\emptyset_{k_l} w_{k_l} + H_1 \emptyset_{k_l} w_{k_l} e^{ik_l} + H_1^{\dagger} \emptyset_{\bar{k}_l} w_{\bar{k}_l} e^{-i\bar{k}_l}] = I \tag{3.3.12}$$

After some algebra we find:

$$\sum_{l=1}^{M} H_1^{\dagger} (\emptyset_{\bar{k}_l} e^{-i\bar{k}_l} w_{\bar{k}_l} - \emptyset_{k_l} e^{-ik_l} w_{k_l}) = I \tag{3.3.13}$$

Now we will use the dual vectors defined in (3.3.8) and multiplying equation (3.3.11) by $\widetilde{\emptyset}_{k_p}$we get:

$$\sum_{l=1}^{M} \widetilde{\emptyset}_{k_p} \phi_{\bar{k}_l} w_{\bar{k}_l} = \sum_{l=1}^{M} \widetilde{\emptyset}_{k_p} \phi_{k_l} w_{k_l}$$

$$\Rightarrow \sum_{l=1}^{M} \widetilde{\emptyset}_{k_p} \phi_{\bar{k}_l} w_{\bar{k}_l} = w_{k_l} \tag{3.3.14}$$

And multiplying equation (3.3.11) by $\widetilde{\emptyset}_{\bar{k}_p}$we get:

$$\sum_{l=1}^{M} \widetilde{\emptyset}_{\bar{k}_p} \phi_{k_l} w_{k_l} = \sum_{l=1}^{M} \widetilde{\emptyset}_{\bar{k}_p} \phi_{\bar{k}_l} w_{\bar{k}_l}$$

$$\Rightarrow \sum_{l=1}^{M} \widetilde{\emptyset}_{\bar{k}_p} \phi_{k_l} w_{k_l} = w_{\bar{k}_l} \qquad (3.3.15)$$

Substitute equations (3.3.14), (3.3.15) into equation (3.3.13), and using (3.3.11) we get:

$$\sum_{l,p=1}^{M} H_1^{\dagger} (\emptyset_{k_l} e^{-ik_l} \widetilde{\emptyset}_{k_l} - \emptyset_{\bar{k}_l} e^{-i\bar{k}_l} \widetilde{\emptyset}_{\bar{k}_l}) \phi_{\bar{k}_p} w_{\bar{k}_p} = I \qquad (3.3.16)$$

Which gives us an expression for $w_k$:

$$w_k = \widetilde{\emptyset}_{k_l} v^{-1} \qquad (3.3.18)$$

$$\mathcal{V} = \sum_{l=1}^{M} H_1^{\dagger} (\emptyset_{k_l} e^{-ik_l} \widetilde{\emptyset}_{k_l} - \emptyset_{-_l} e^{-i\bar{k}_l} \widetilde{\emptyset}_{\bar{k}_l}) \qquad (3.3.19)$$

The wave number $k$ in (3.3.18) refers to left and right states. The Green’s function of an infinite system is:

$$g_{z,z'}^{\infty} = \begin{cases} \sum_{l=1}^{M} \emptyset_{k_l} e^{ik_l(z-z')} \widetilde{\emptyset}_{k_l} \mathcal{V}^{-1}, & z \geq z' \\ \sum_{l=1}^{M} \emptyset_{\bar{k}_l} e^{i\bar{k}_l(z-z')} \widetilde{\emptyset}_{\bar{k}_l} \mathcal{V}^{-1}, & z \leq z' \end{cases} \qquad (3.3.20)$$

To get the Green's function for a semi-infinite lead we need to add a wave function to the Green's function in satisfy the boundary conditions at the end of the lead, as what we did with the one dimensional case.

$$g_{z,z'} = g_{z,z'}^{\infty} + \sum_{l=1}^{M} \emptyset_{\bar{k}_l} e^{i\bar{k}_l z} \Lambda_l(z, z', z_0) \qquad (3.3.21)$$

Where $\Lambda_l(z, z', z_0)$ is unknown amplitude depend in $z, z', and\ z_0$. The wave cannot propagate beyond the end of the chain so, $g_{z_0,z'} = 0$

$$g_{z_0,z'} = g_{z_0,z'}^{\infty} + \sum_{l=1}^{M} \emptyset_{\bar{k}_l} e^{i\bar{k}_l z_0} \Lambda_l(z', z_0) = 0$$

$$\sum_{l=1}^{M} \emptyset_{\bar{k}_l} e^{i\bar{k}_l z_0} \Lambda_l(z', z_0) = -g_{z_0,z'}^{\infty}$$

$$\Lambda_l(z', z_0) = -g_{z_0,z'}^{\infty} \sum_{l=1}^{M} \left(\emptyset_{\bar{k}_l} e^{i\bar{k}_l z_0}\right)^{-1}$$

$$= -\sum_{p=1}^{M} \emptyset_{k_p} e^{ik_p(z_0 - z')} \tilde{\emptyset}_{k_p} \mathcal{V}^{-1} \sum_{l=1}^{M} \tilde{\emptyset}_{\bar{k}_l} e^{-i\bar{k}_l z_0}$$

$$= -\sum_{l,p}^{M} e^{-i\bar{k}_l z_0} \tilde{\emptyset}_{\bar{k}_l} \emptyset_{k_p} e^{ik_p(z_0 - z')} \tilde{\emptyset}_{k_p} \mathcal{V}^{-1}$$

$$g_{z,z'} = g_{z,z'}^{\infty} - \sum_{l,p}^{M} \emptyset_{\bar{k}_l} e^{i\bar{k}_l z} e^{-i\bar{k}_l z_0} \tilde{\emptyset}_{\bar{k}_l} \emptyset_{k_p} e^{ik_p(z_0 - z')} \tilde{\emptyset}_{k_p} \mathcal{V}^{-1}$$

$$\Rightarrow g_{z,z'} = g_{z,z'}^{\infty} - \sum_{l,p}^{M} \emptyset_{\bar{k}_l} e^{i\bar{k}_l(z - z_0)}\ \tilde{\emptyset}_{\bar{k}_l} \emptyset_{k_p} e^{ik_p(z_0 - z')} \tilde{\emptyset}_{k_p} \mathcal{V}^{-1} \qquad (3.3.22)$$

We can rewrite equation (3.3.22) in this form:

$$g_{z,z'} = g_{z,z'}^{\infty} + \Delta$$

Where

$$\Delta = \sum_{l,p}^{M} \emptyset_{\bar{k}_l} e^{i\bar{k}_l(z-z_0)} \,\widetilde{\emptyset}_{\bar{k}_l} \emptyset_{k_p} e^{ik_p(z_0-z')} \widetilde{\emptyset}_{k_l} \mathcal{V}^{-1}$$

And $\Delta$ is the green's function for a semi-infinite lead terminating at $z_0 - 1$, where the chain is to the left of the boundary condition. And we can also calculate a semi-infinite lead terminating at $z_0 + 1$ so that the chain exists to the right of this boundary condition. The Green's Function for this situation can be written:

$$g_{z,z'} = g_{z,z'}^{\infty} - \sum_{l,p}^{M} \emptyset_{\bar{k}_l} e^{i\bar{k}_l(z-z_0)} \,\widetilde{\emptyset}_{\bar{k}_l} \emptyset_{k_p} e^{ik_p(z_0-z')} \widetilde{\emptyset}_{k_l} \mathcal{V}^{-1} \quad (3.3.23)$$

The surface Green's function calculated at each end point. For the left lead, $z = z' = z_0 - 1$ and for the right lead $z = z' = z_0 + 1$.

$$g_L = \left[ \sum_{l=1}^{M} \emptyset_{\bar{k}_l} e^{i\bar{k}_l(z_0-1-z_0+1)} \widetilde{\emptyset}_{\bar{k}_l} - \sum_{l,p}^{M} \emptyset_{\bar{k}_l} e^{i\bar{k}_l(z_0-1-z_0)} \,\widetilde{\emptyset}_{\bar{k}_l} \emptyset_{k_p} e^{ik_p(z_0-z_0+1)} \widetilde{\emptyset}_{k_p} \right] \mathcal{V}^{-1}$$

$$g_L = \left[ I - \sum_{l,p}^{M} \emptyset_{\bar{k}_l} e^{-i\bar{k}_l} \,\widetilde{\emptyset}_{\bar{k}_l} \emptyset_{k_p} e^{ik_p} \widetilde{\emptyset}_{k_p} \right] \mathcal{V}^{-1} \quad (3.3.24)$$

$$g_R = \left[ \sum_{l=1}^{M} \emptyset_{k_l} e^{ik_l(z_0+1-z_0-1)} \widetilde{\emptyset}_{k_l} - \sum_{l,p}^{M} \emptyset_{k_l} e^{ik_l(z_0+1-z_0)} \,\widetilde{\emptyset}_{k_l} \emptyset_{\bar{k}_p} e^{i\bar{k}_p(z_0-z_0-1)} \widetilde{\emptyset}_{\bar{k}_p} \right] \mathcal{V}^{-1}$$

$$g_R = \left[ I - \sum_{l,p}^{M} \emptyset_{k_l} e^{ik_l} \; \widetilde{\emptyset}_{k_l} \emptyset_{\bar{k}_p} e^{-i\bar{k}_p} \widetilde{\emptyset}_{\bar{k}_p} \right] \mathcal{V}^{-1} \qquad (3.3.25)$$

So now we can represent the two semi-infinite crystalline electrode by using equations (3.3.24)) and (3.3.25). The next step is to bringing these two together and apply them to a scattering problem to produce the green's function for an infinite system.

### 3.4.2. Effective Hamiltonian of the scattering region

If we want to solving a scattering problem, we just need to describe the surface of the system. The Fisher-Lee relation [14] gives use the transmission and reflection amplitudes as a function of the Green's function of the surface sites of the system, which in turn gives us the conductance through this interface. Then, by using the Dyson Equation, we can couple the leads and the scatterer with a matrix $V$, which contains the hopping parameters. The scattering region is a complicated matrix, so we cannot describe it as a coupling matrix between the surfaces. Therefore, the decimation method is used to reduce the Hamiltonian down to such a structure.

Let us consider again the Schrödinger equation

$$\sum_j H_{ij} \psi_j = E \psi_j \qquad (3.3.26)$$

$$H_{il} \psi_l + \sum_{j \neq l} H_{ij} \psi_j = E \psi_i \qquad (i \neq l), \qquad (3.3.27)$$

$$H_{ll}\psi_l + \sum_{j\neq l} H_{lj}\psi_j = E\psi_l \qquad (i = l), \qquad (3.3.28)$$

From equation (3.3.28) $\psi_l$ is:

$$\psi_l = \sum_{j\neq l} \frac{H_{lj}\psi_j}{E - H_{ll}} \qquad (3.3.29)$$

Then by using (3.3.27) into (3.3.29) we get:

$$\sum_{j\neq l} \left[H_{ij}\psi_j + \frac{H_{il}H_{lj}\psi_j}{E - H_{ll}}\right] = E\psi_i \qquad (i \neq l), \qquad (3.3.30)$$

This equation is the effective Schrödinger equation, and we can introduce a new effective Hamiltonian $H'$ as:

$$H'_{ij} = H_{ij} + \frac{H_{il}H_{lj}}{E - H_{ll}} \qquad (3.3.31)$$

The Hamiltonian $H'_{ij}$ is the decimated Hamiltonian produced by simple Gaussian elimination. This Hamiltonian is a function of the energy $E$, which it suits to the method presented in the previous section. By using the decimation method, we can reduce the complicated scattering problem to the simple one, which was described by the Hamiltonian

$$H = \begin{pmatrix} H_L & V_L & 0 \\ V_L^\dagger & H_{SR} & V_R \\ 0 & V_R^\dagger & H_R \end{pmatrix} \qquad (3.3.32)$$

Where $H_L$ *and* $H_R$ are the semi-infinite leads, and the Hamiltonian of the scatterer is $H_{SR}$. $V_L$ *and* $V_R$ are denoted for the couple the original scattering region to the leads. The scattering region disappears after decimation and we are left with the effective Hamiltonian.

$$H = \begin{pmatrix} H_L & V_c \\ V_c^\dagger & H_R \end{pmatrix} \tag{3.3.33}$$

Where $V_c$ is the effective coupling Hamiltonian, which describes the whole scattering process.

Now we can be applying the same steps as we did in the case of the one-dimensional scattering, and calculate the Green's function of the whole system. The surface Green's function is given by the Dyson's equation:

$$G = \begin{pmatrix} g_L^{-1} & V_c \\ V_c & g_R^{-1} \end{pmatrix}^{-1} = \begin{pmatrix} G_{00} & G_{01} \\ G_{10} & G_{11} \end{pmatrix} \tag{3.3.34}$$

$g_L$ $and$ $g_R$ are the surface Green's functions of the leads which they defined in equations (3.3.24) and (3.3.25).

### 3.4.3. Scattering Matrix

Now, we can be calculating the scattering amplitudes by using the Fisher-Lee relation. A generalization of the Fisher-Lee relation [13, 15,16], assuming that states are normalized to carry unit flux, will give the transmission amplitude from the left lead to the right lead as:

$$t_{hl} = \widetilde{\emptyset}_{k_h}^{\dagger} G_{01} \mathcal{V}_L \emptyset_{k_l} \sqrt{\left|\frac{v_h}{v_l}\right|} \tag{3.3.35}$$

Where $\emptyset_{k_h}$ and $\emptyset_{k_l}$ are a right moving state vector in the right and left lead, respectively. $v_h$ and $v_L$ are the corresponding group velocities. The reflection amplitudes in the left lead is:

$$r_{hl} = \widetilde{\emptyset}^{\dagger}_{\bar{k}_h}(G_{00}\mathcal{V}_L - I)\emptyset_{k_l}\sqrt{\left|\frac{v_h}{v_l}\right|} \quad (3.3.36)$$

State $\emptyset_{\bar{k}_h}$ is a left moving state, $\emptyset_{k_l}$ is a right moving state and $\mathcal{V}_L$ is the $\mathcal{V}$ operator defined before in equation (3.3.19) for the left lead.

The same strategy we use it to define the scattering amplitude for particles coming from the right:

$$t'_{hl} = \widetilde{\emptyset}^{\dagger}_{\bar{k}_h}G_{10}\mathcal{V}_R\emptyset_{\bar{k}_l}\sqrt{\left|\frac{v_h}{v_l}\right|} \quad (3.3.37)$$

$$r'_{hl} = \widetilde{\emptyset}^{\dagger}_{\bar{k}_h}(G_{11}\mathcal{V}_R - I)\emptyset_{\bar{k}_l}\sqrt{\left|\frac{v_h}{v_l}\right|} \quad (3.3.38)$$

Then, we can use these values into the Landauer-Buttiker Formula to calculate the conductance.

## Reference

1. Scheer, E., 2010. Molecular electronics: an introduction to theory and experiment (Vol. 1). World Scientific.
2. Soler, J.M., Artacho, E., Gale, J.D., García, A., Junquera, J., Ordejón, P. and Sánchez-Portal, D., 2002. The SIESTA method for ab initio order-N materials simulation. *Journal of Physics: Condensed Matter*, *14*(11), p.2745.

3. Ferrer, J., Lambert, C.J., García-Suárez, V.M., Manrique, D.Z., Visontai, D., Oroszlany, L., Rodríguez-Ferradás, R., Grace, I., Bailey, S.W.D., Gillemot, K. and Sadeghi, H., 2014. GOLLUM: a next-generation simulation tool for electron, thermal and spin transport. *New Journal of Physics*, *16*(9), p.093029.

4. Kohn, W. and Sham, L.J., 1965. Self-consistent equations including exchange and correlation effects. *Physical review*, *140*(4A), p.A1133.

5. Harrison, N.M., 2003. An introduction to density functional theory. NATO SCIENCE SERIES SUB SERIES III COMPUTER AND AYSTEMS SCIENCES, 187, pp.45-70

6. Fermi, E., 1934. Sopra lo spostamento per pressione delle righe elevate delle serie spettrali. *Il Nuovo Cimento (1924-1942)*, *11*(3), pp.157-166.

7. Press, W.H., 2007. *Numerical recipes 3rd edition: The art of scientific computing*. Cambridge university press.

8. Astrakhantsev, G.P., 1971. An iterative method of solving elliptic net problems. *USSR Computational Mathematics and Mathematical Physics*,*11*(2), pp.171-182.

9. Cooley, J.W. and Tukey, J.W., 1965. An algorithm for the machine calculation of complex Fourier series. *Mathematics of computation*, *19*(90), pp.297-301.

10. Payne, M.C., Teter, M.P., Allan, D.C., Arias, T.A. and Joannopoulos, J.D., 1992. Iterative minimization techniques for ab initio total-energy calculations: molecular dynamics and conjugate gradients. *Reviews of Modern Physics*, *64*(4), p.1045.

11. L. Oroszlány. Carbon based nanodevices. PhD thesis, Lancaster University, 2009.

12. Brouwer, P.W., 1998. Scattering approach to parametric pumping. Physical Review B, 58(16), p.R10135.

13. Lambert, C.J., 2015. Basic concepts of quantum interference and electron transport in single-molecule electronics. *Chemical Society Reviews*, *44*(4), pp.875-888.

14. Fisher, D.S. and Lee, P.A., 1981. Relation between conductivity and transmission matrix. *Physical Review B*, *23*(12), p.6851.

15. Datta, S., 1997. *Electronic transport in mesoscopic systems*. Cambridge university press.

16. S. Sanvito, Giant Magnetoresistance and Quantum Transport in Magnetic Hybrid Nanostructures. PhD thesis, Lancaster University, 1999.

17. Schneider, G.F. and Dekker, C., 2012. DNA sequencing with nanopores. Nature biotechnology, 30(4), pp.326-328.

18. Sankey, O.F. and Niklewski, D.J., 1989. Ab initio multicenter tight-binding model for molecular-dynamics simulations and other applications in covalent systems. Physical Review B, 40(6), p.3979.

19. Xu, B. and Tao, N.J., 2003. Measurement of single-molecule resistance by repeated formation of molecular junctions. Science, 301(5637), pp.1221-1223.

20. Zotti, L.A., Kirchner, T., Cuevas, J.C., Pauly, F., Huhn, T., Scheer, E. and Erbe, A., 2010. Revealing the Role of Anchoring Groups in the Electrical Conduction Through Single-Molecule Junctions. small, 6(14), pp.1529-1535.

21. Moore, G.E., 2006. Cramming more components onto integrated circuits, Reprinted from Electronics, volume 38, number 8, April 19, 1965, pp. 114 ff. IEEE Solid-State Circuits Newsletter, 3(20), pp.33-35.

22. Pignedoli, C.A., Curioni, A. and Andreoni, W., 2007. Anomalous behavior of the dielectric constant of hafnium silicates: a first principles study. Physical review letters, 98(3), p.037602.

23. Piva, P.G., DiLabio, G.A., Pitters, J.L., Zikovsky, J., Rezeq, M.D., Dogel, S., Hofer, W.A. and Wolkow, R.A., 2005. Field regulation of single-molecule conductivity by a charged surface atom. Nature, 435(7042), pp.658-661.

24. Aviram, A. and Ratner, M.A., 1974. Molecular rectifiers. Chemical Physics Letters, 29(2), pp.277-283.

25. Li, C., Pobelov, I., Wandlowski, T., Bagrets, A., Arnold, A. and Evers, F., 2008. Charge transport in single Au| alkanedithiol| Au junctions: coordination geometries and conformational degrees of freedom. Journal of the American Chemical Society, 130(1), pp.318-326.

26. E. Leary, S. J. Higgins, H. van Zalinge, W. Haiss, R. J. Nichols, H. Hbenreich, I. Grace, C. M. Finch, C. J. Lambert, R. McGrath and J. Smerdon, Submitted to Nature, 2008.

27. Ashwell, G.J., Urasinska, B. and Tyrrell, W.D., 2006. Molecules that mimic Schottky diodes. Physical Chemistry Chemical Physics, 8(28), pp.3314-3319.

28. He, J., Chen, B.O., Flatt, A.K., Stephenson, J.J., Doyle, C.D. and Tour, J.M., 2006. Metal-free silicon–molecule–nanotube testbed and memory device. Nature materials, 5(1), pp.63-68.

29. J. He, F. Chen, P. A. Liddell, J. Andreasson, S. D. Straight, D. Gust, T. A. Moore, A. L. Moore, J. Li, O. F. Sankey and S.M. Lindsay, Nanotechnology, 16, 695, (2005).

30. Türksoy, F., Hughes, G., Batsanov, A.S. and Bryce, M.R., 2003. Phenylene–2, 5-dimethylpyrazine co-oligomers: synthesis by Suzuki couplings, X-ray structures of neutral and diprotonated teraryl species and efficient blue emission. Journal of Materials Chemistry, 13(7), pp.1554-1557.

31. Todorov, I.T., Smith, W., Trachenko, K. and Dove, M.T., 2006. DL_POLY_3: new dimensions in molecular dynamics simulations via massive parallelism. Journal of Materials Chemistry, 16(20), pp.1911-1918.

# Chapter 4

## 4. Charge transport through dicarboxylic-acid-terminated alkanes bound to graphene-gold nanogap electrodes

The following work was carried out in collaboration with the experimental groups of Department of Chemistry, Xi'an-Jiaotong Liverpool University, Suzhou, China. I will present my theoretical work on a single-molecule dicarboxylic-acid-terminated alkanes bound to graphene-gold electrodes. The results presented in this chapter were published in: Charge transport through dicarboxylic-acid-terminated alkanes bound to graphene–gold nanogap electrodes. Nanoscale, 2016, Advance Article, DOI: 10.1039/C6NR03807G

‘Graphene-based electrodes are attractive for single-molecule electronics due to their high stability and conductivity and reduced screening compared with metals. In this chapter, I describe a joint project in which the STM-based matrix isolation I(s) method to measure the performance of graphene in single-molecule junctions with one graphene electrode and one gold electrode. By measuring the length dependence of the electrical conductance of dicarboxylic-acid-terminated alkanes, it is found that the transport is consistent with phase coherent tunnelling, but with an attenuation factor of $\beta_N$ = 0.69 per methyl unit, which is lower than the value measured for Au–molecule–Au junctions. Comparison with my density-functional-theory calculations of electron transport through graphene–molecule–Au junctions and Au–molecule–Au junctions reveals that this difference is due to the difference in Fermi energies of the two types of junction, relative to the frontier orbitals of the molecules. For most molecules, their electrical conductance in graphene–molecule–Au junctions is higher than that in Au–molecule–Au junctions, which suggests that graphene offers superior electrode performance, when utilizing carboxylic acid anchor groups.’

Most experimental work to date has used gold to form electrical contacts to single molecules due to its chemical stability, lack of oxidation, high conductivity and ease of junction fabrication. However, there are a number of drawbacks of working with gold electrodes for example their non-compatibility with complementary metal–oxide–semiconductor (CMOS) technology with high atomic mobility. To overcome some of these drawbacks experimentalists have used other metals such as Pt, Ag, Pd [1] to form electrodes in single-molecule junctions together with a range of different anchor groups such as amine [2], pyridine [3], carboxylic acids [4,5] and thiol. In recent years, researchers have used carbon-based materials as non-metallic electrodes for the investigation of molecular junctions [6]. The essential studies such as in [7,8, 9], encouraged me to start using carbon-based materials as the alternative electrode materials for molecular electronics in next generation nanostructured devices. In this chapter, I demonstrate the use of graphene as a bottom electrode in place of the more-commonly used gold. The conductance through such a system has been experimentally measured for various lengths of bicarboxylic alkanes in Au-dicarboxylic acid-graphene junctions using the STM-based matrix isolation I(s) method in which the STM tip is brought close to the graphene surface without making contact. Carboxylic acid group contact is achievable when connected to gold by electronic coupling between the carboxylate group and the gold surface. The conductance histograms for each molecule reveal that the conductance values are dependent on the alkane length and decrease exponentially with increasing molecular length. Figure. 4.1a shows four typical I(s) curves which display current plateaus, and Figure. 1b shows the 2 dimensional conductance histogram for succinic acid constructed where the first peak at 15.6 nS observed in the conductance histogram can be ascribed to the single molecule conductance with a second peak about two times the single molecule conductance.

To understand the mechanism of charge transport in the molecule and to investigate the length dependence of conductance, the collaborating group measured the conductance of HOOC-(CH2)n-COOH (n=3~6) molecules in contact with gold and graphene electrodes. They used the same parameters in all measurements, except when they measuring the rather small conductance of octanedioic acid, where the bias voltage was increased to 500 mV to minimise the impact due to the instrumental errors.

Eman it would be a good idea to include the chemical formulae for the names below eg HOOC-(CH2)n where n= 2,4,…etc

Figure 4.2 presents the conductance histograms of the alkanedicarboxylic acids, from which the most probable conductances are found to be 10.3±2.8, 5.1±1.2, 2.4±0.5 and 1.08±0.34 nS for pentanedioic acid (a), hexanedioic acid (b), heptanedioic acid (c), octanedioic acid (d), respectively. Theses Figuer proved that the conductance values decrease with increasing molecular length.

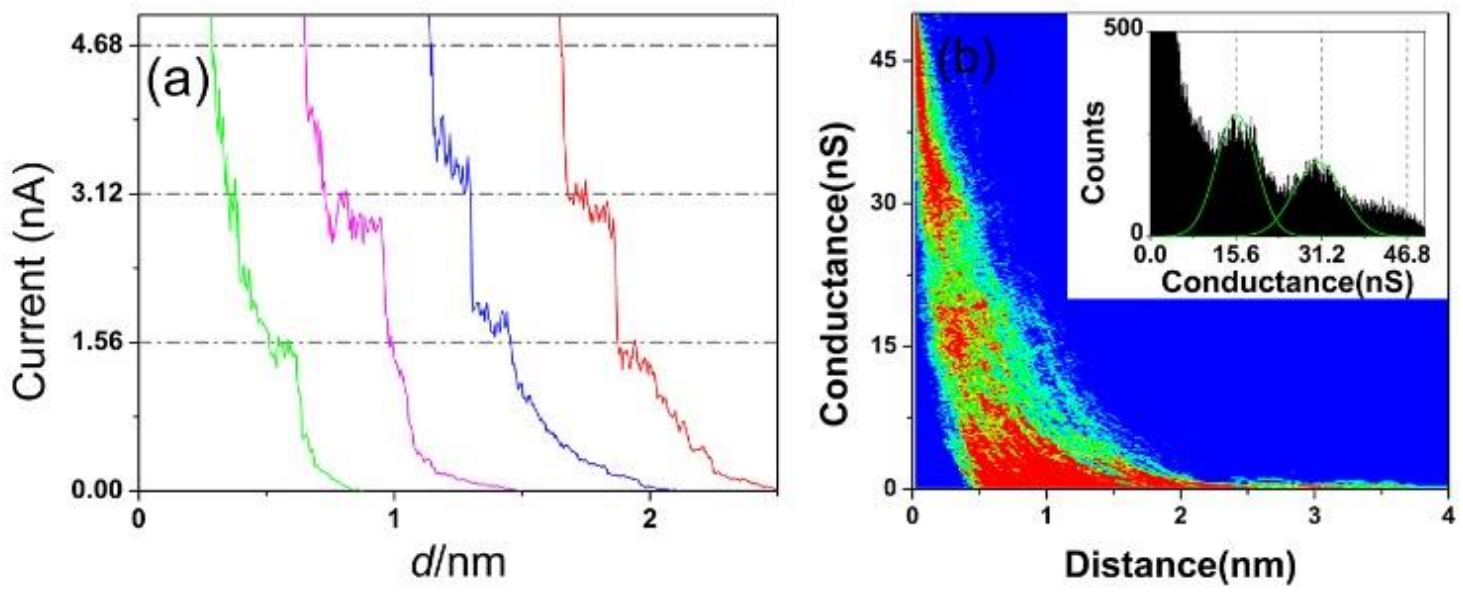

**Figure**. **4.1** (a) Typical I–s curves collected by the I(s) method at 100 mV. (b)The two dimensional (2D) histograms of single molecule conductance of the Au–HOOC–(CH2)2 COOH–graphene constructed from 400 curves. The inset is the corresponding conductance histogram.

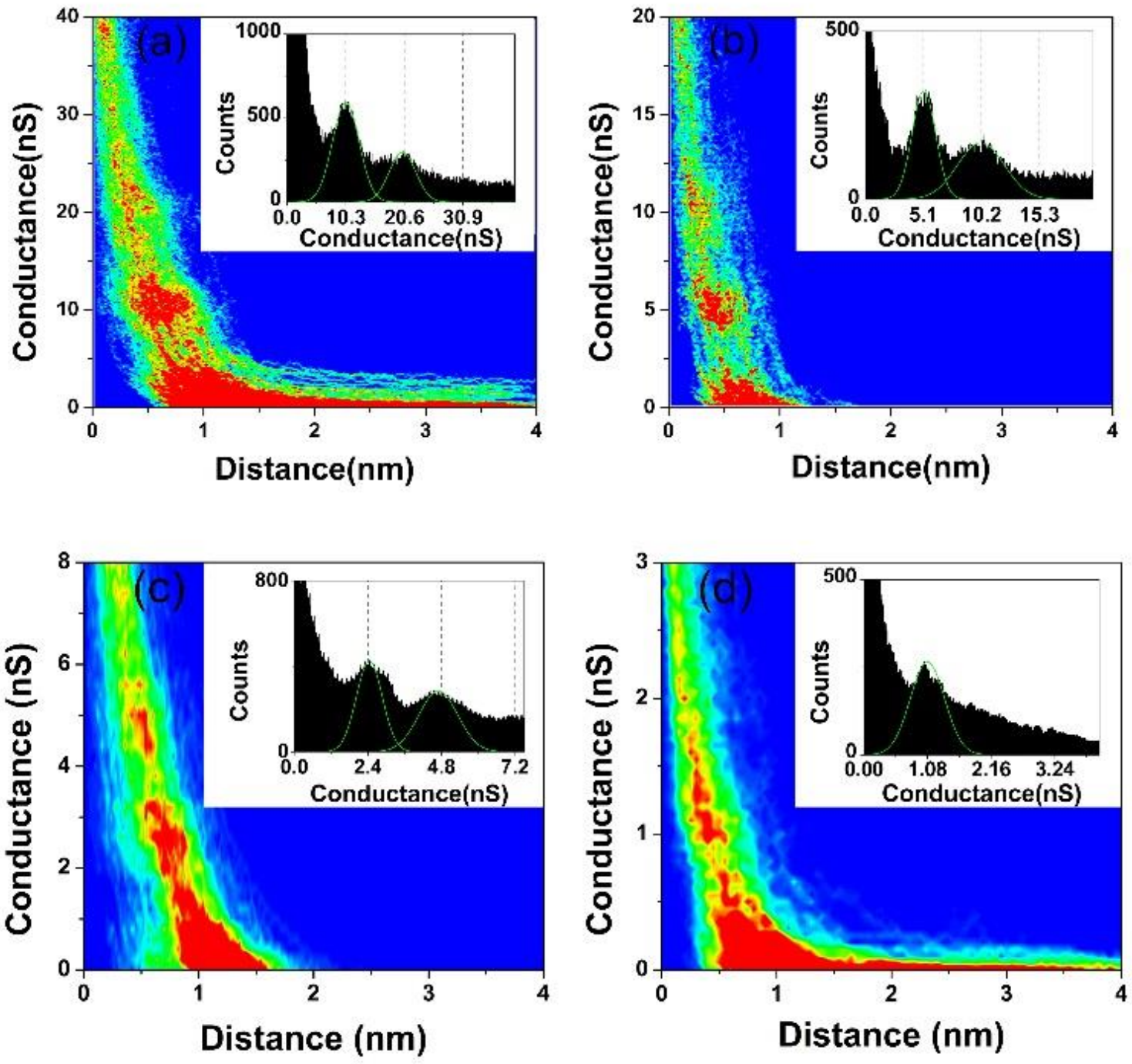

**Figure 4.2**. The 2D histograms of single molecule conductance of the Au-HOOC-(CH2)n-COOH-graphene with (a) n=3, (b) n=4, (c) n=5 and (d) n=6. Insets are the corresponding conductance histograms. All the histograms were constructed from more than 400 curves.

Table 4.1. summarizes the comparison of the dicarboxylic acid conductance in different junctions with literature results. As expected for phase-coherent tunnelling [13] the conductance decreases exponentially with increasing molecular chain length and can be described as Equation (4.1), where G is the single molecular conductance, A is a constant influenced by the coupling between the contact of molecule and electrode, $\beta_N$ is the tunneling decay constant that reflects the efficiency of electron transport, N is the number of methylene units. Values for $\beta_N$ are also presented in Table 1.

$$G = Ae^{-\beta_N N} \tag{4.1}$$

$$\log G = -\beta_N N$$

$$\beta_N = -\frac{logG}{N} \tag{4.2}$$

From equation (4.2), a prefactor, beta, is introduced as the relationship between natural logarithmic of single-molecule conductance and the molecule length or number of (–CH2) units as in our case. The measured conductance decreases with the molecular length and the linear fit yields a tunneling decay constant of ~0.69 per (–CH2) unit, which as shown in Table 4.1 is slightly smaller than that obtained using Au, Cu and Ag electrodes. Furthermore, the conductance of n = 3, 4, 5 and 6 molecules is slightly bigger than those obtained using Au, Cu and Ag electrodes, while collaborating group measured value for the single-molecule conductance of succinic acid is slightly smaller.

| Molecular junction | Conductance (Ns, HC) | | | | | Tunneling decay constant ($\beta_N$) |
|---|---|---|---|---|---|---|
| | n=2 | n=3 | n=4 | n=5 | n=6 | |
| **Au-HOOC-($CH_2$)n-COOH-Graphene** | 15.6 | 10.3 | 5.1 | 2.4 | 1.08 | 0.69±0.04 |
| **Au-HOOC-(CH2)n-COOH-Au[15]** | 20.9 | ….. | 3.7 | ….. | 0.77 | 0.81±0.01 |
| **Ag-HOOC-(C2)nCOO-Ag[16]** | 13.2 | 8 | 3.7 | 1.7 | ….. | 0.71±0.03 |
| **Cu-HOOC(CH2)n-COOH-Cu[16]** | 18.2 | 7.5 | 2.9 | 1.2 | ….. | 0.95±0.02 |
| **Au-molecule-Au[17]** | ….. | ….. | ….. | ….. | ….. | 0.78 |
| **Theory:perfect graphene Au-HOOC-(CH2)n-COOH- Graphene** | 38.5 | 13 | 5.3 | 4 | 2.2 | 0.69 |
| **Theory: defective graphene Au-HOOC-(CH2)n-COOH-Graphene** | 27.1 | 5.19 | 2.96 | 1.09 | 0.68 | 0.89 |

**Table 4.1**. Comparison of the conductance of dicarboxylic acid in different junctions. Note:"……."represents that the data is unavailable or unadopted from the references.

# 4.1. Theoretical calculations

To calculate electrical properties of the molecules in Figure 4.3 (Graphene-Molecule-Au junctions), with different lengths of molecules (n=2, 4, and 6) the relaxed geometry of each isolated molecule was found using the density functional theory (DFT) code SIESTA [10] which employs Troullier-Martins pseudopotentials to represent the potentials of the atomic cores and a local atomic-orbital basis set. I used a double-zeta polarized basis set for all atoms and the local density functional approximation (LDA-CA) by Ceperley and Adler [11]. The Hamiltonian and overlap matrices are calculated on a real -space grid defined by a plane-wave cut off of 150 Ry. Each molecule was relaxed to the optimum geometry until the forces on the atoms are smaller than 0.02 eV/Å and in case of the isolated molecules, a sufficiently-large unit cell was used to avoid spurious steric effects.

After obtaining the relaxed geometry of an isolated molecule, the molecule was then placed between graphene and (111) gold electrodes and the molecules plus electrodes were allowed to further relax to yield the optimized structures shown in Figures 1a-c. The same strategy was followed for the cases of two (111) gold electrodes Figures 3a-c and defected-graphene and (111) gold electrodes in Figures 6a-c. For each structure, Gollum was used [12] to compute the transmission coefficient $T(E)$ for electrons of energy $E$ passing from the lower electrode to the upper gold electrode. Once the $T(E)$ is computed, the zero-bias electrical conductance $G$ using the Landauer formula was calculated from:

$$G = \frac{I}{V} = G_0 \int_{-\infty}^{\infty} dE T(E) \left( -\frac{df(E)}{dE} \right) \quad (4.3)$$

where $G_0 = \left( \frac{2e^2}{h} \right)$ is the quantum of conductance, $f(E)$ is Fermi distribution function defined as $f(E) = e^{(E-E_F)k_B T}$ where $k_B$ is Boltzmann constant and $T$ is the temperature. Since the quantity $-\frac{df(E)}{dE}$ is a normalised probability distribution of width approximately equal to $k_B T$, centred on the Fermi energy $E_F$, the above integral represents a thermal average of the transmission function $T(E)$ over an energy window of the width $k_BT$ (equal to 25 meV at room temperature) [13].

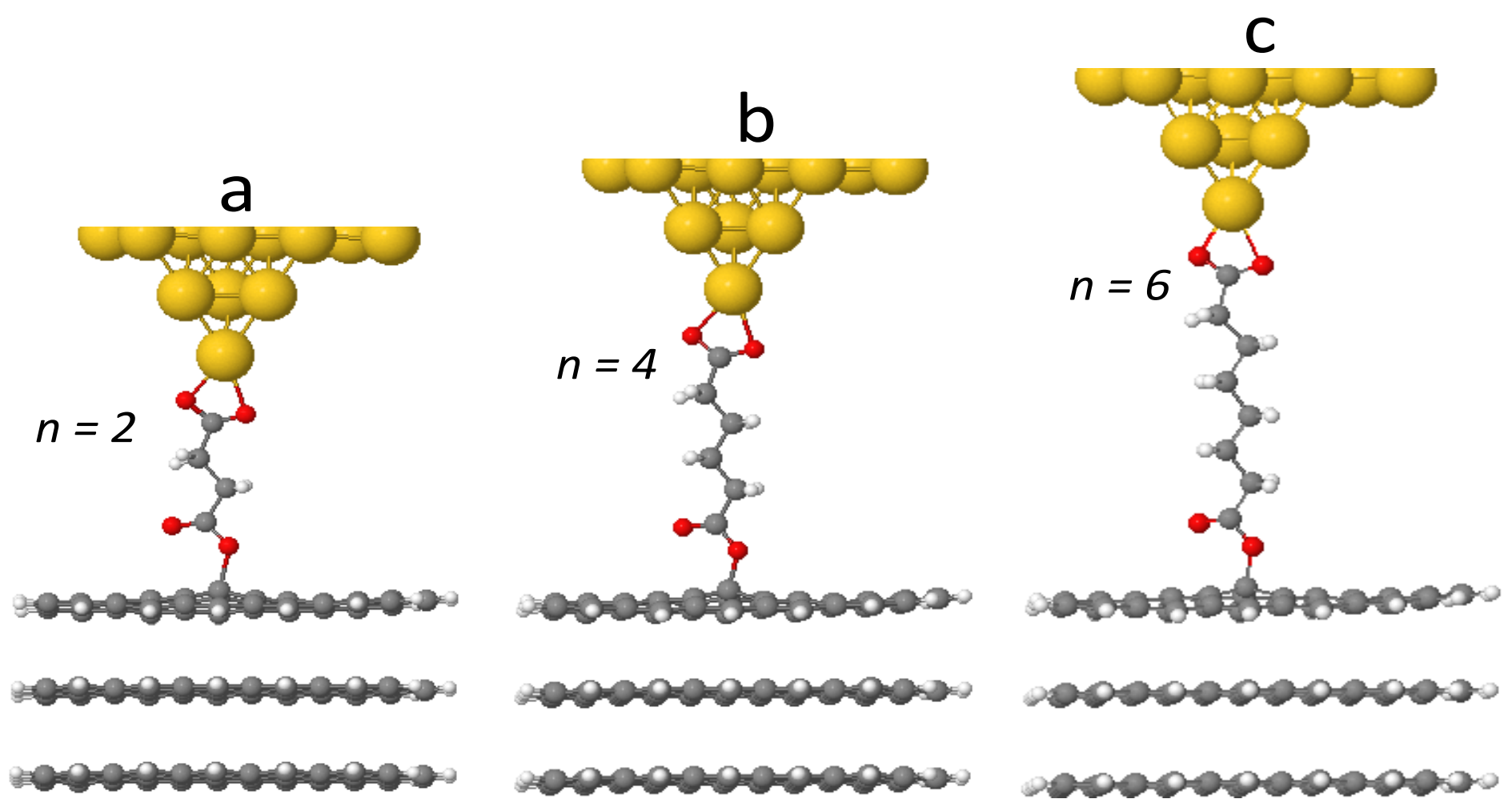

**Figure 4. 3**. (a, b and c) show the optimized geometries of systems containing the dicarboxylic-acid-terminated alkane molecule at lengths (n=2, 4 and 6) connected to the graphene-gold electrodes.

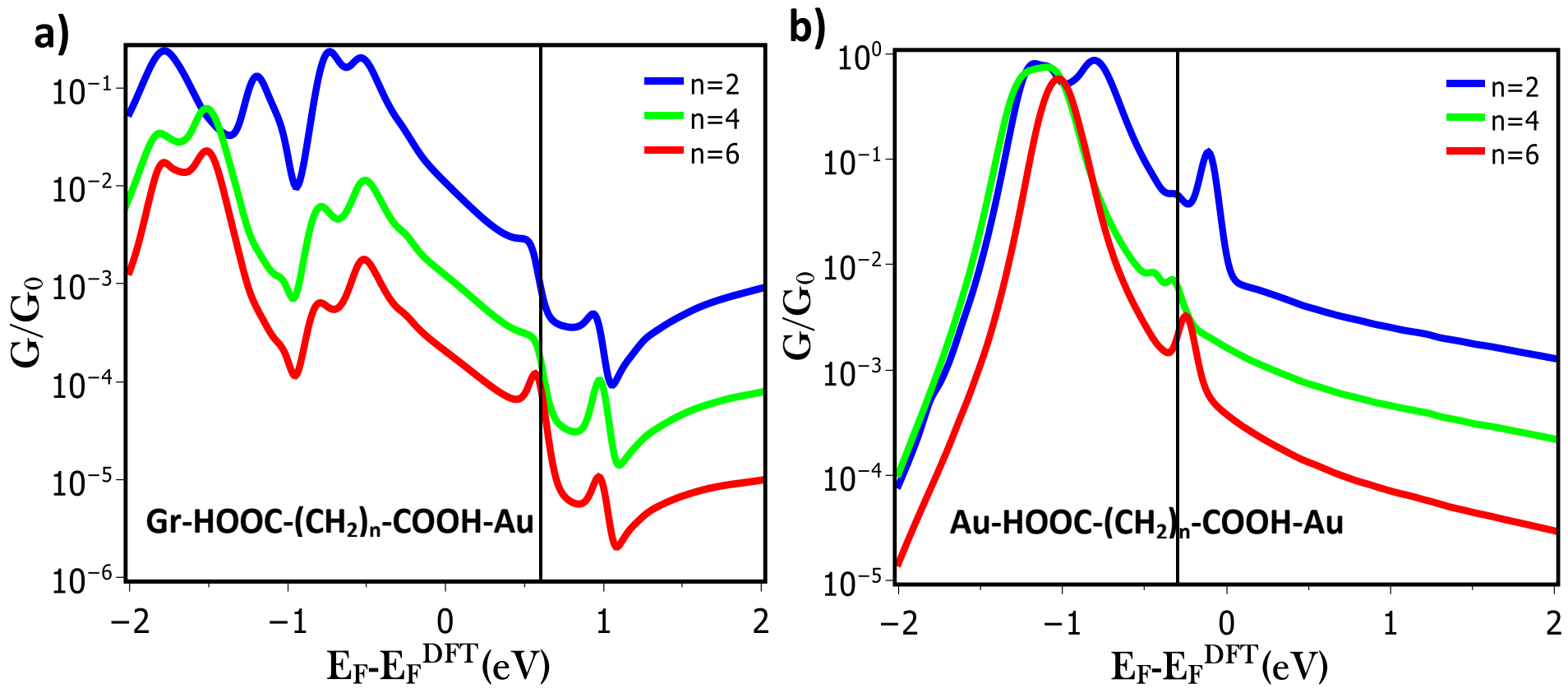

**Figure 4.2**. The room-temperature electrical conductance over a range of Fermi energies of the systems containing: a) the dicarboxylic-acid terminated alkane molecule with the length n=2, n=4 and n=6 of CH2 attached to the graphene-gold electrodes, b) the molecule with the same lengths attached to two gold electrodes.

For the structures in the Figures 4.1a-c, Figure 4.2a shows the room-temperature electrical conductance over a range of Fermi energies $E_F$ in the vicinity of the DFT-predicted Fermi energy $E_F^{DFT}$ . For a wide range of values of $E_F$, the conductance decreases with molecular length, in agreement with experimental measurements. Similarly, Figure 4.2b shows that for the Au-molecule-Au structures in the Figures 4.3a-c, their room-temperature electrical conductances decrease with length.

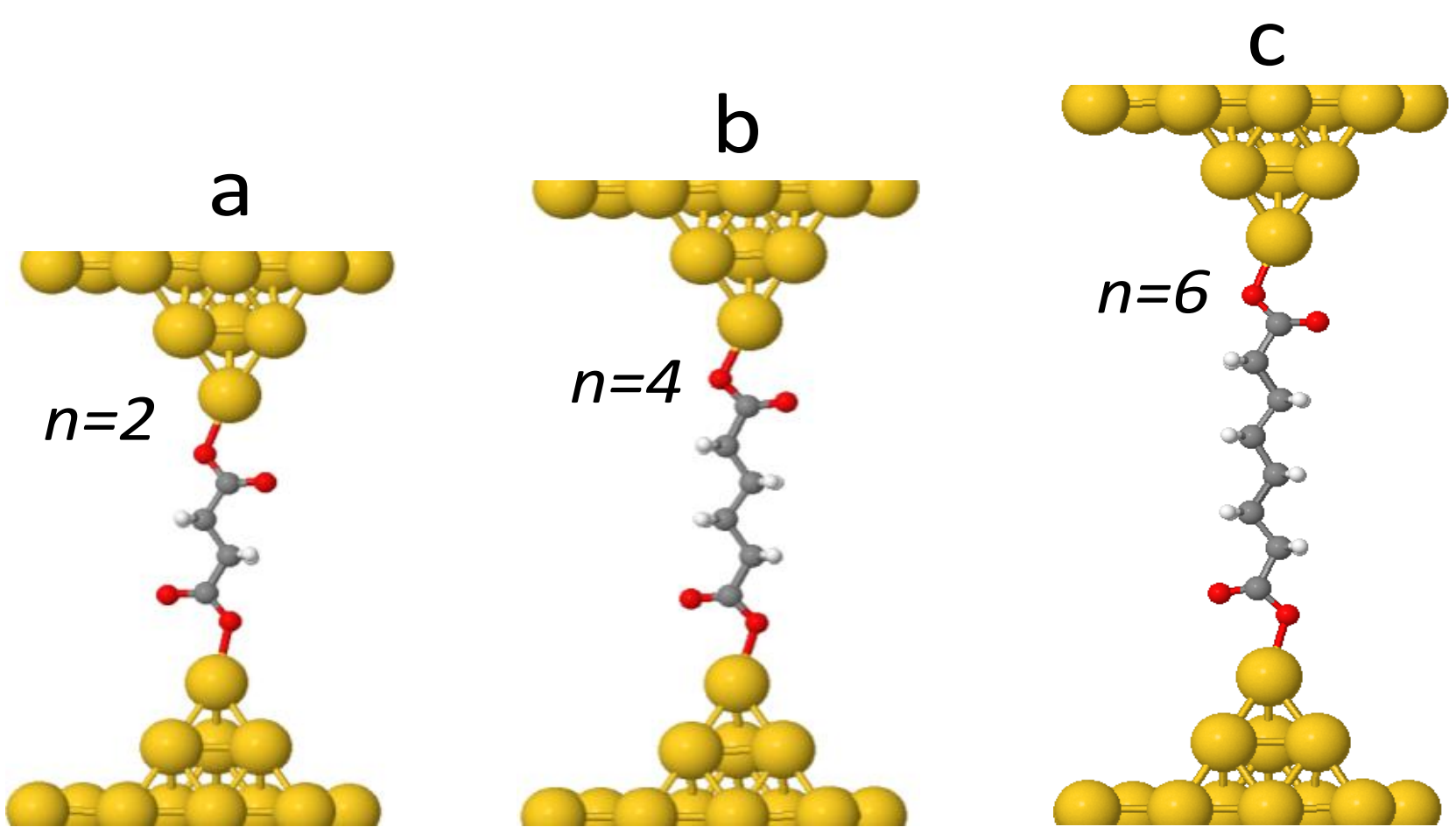

**Figure 4.3**. (a, b and c) show the optimized geometries of systems containing the dicarboxylic-acid-terminated alkanes of lengths (n=2, 4 and 6) connected to two gold electrodes.

The predicted value of the attenuation coefficient $\beta_N$ depends on the precise value of $E_F$ and therefore we computed $\beta_N$ for a range of Fermi energies. For Au-molecule-Au junctions, the closest fit with experiment was found for $E_F = -0.3\,eV$ relative to $E_F^{DFT}$, whereas for graphene-molecule-Au junctions, the closest fit was found for $E_F - E_F^{DFT} = 0.65\,eV$. Figure 4.5 shows a logarithmic plot of predicted single-molecule conductances versus the number of $(-CH_2)$ units in the alkane chain, along with a comparison with experiment. The close agreement between theory and experiment for $\beta_N$ suggests that the difference between the attenuation coefficients of graphene-molecule-Au and Au-molecule-Au junctions arises from a difference in the positions of their frontier orbitals relative to $E_F$. Figure4. 5 also shows that the theoretical conductance values are slightly higher than measured ones for all molecular lengths, which can be attributed to the tendency of LDA to underestimate the HOMO-LUMO gap, which results in an overestimated of the conductance [14].

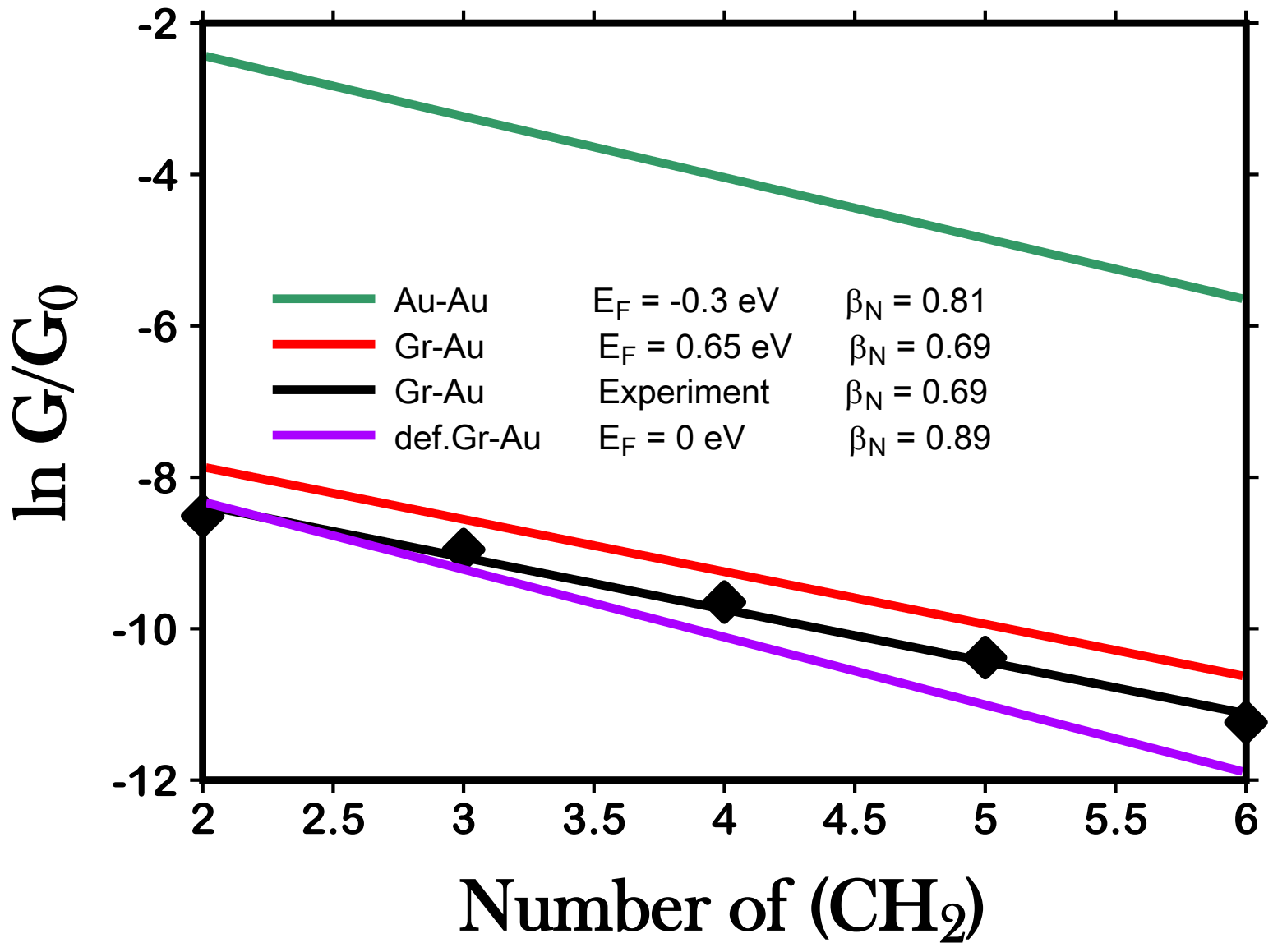

**Figure 4.5**. Comparison between theory and experiment of the logarithm of single – molecule conductance versus number of (-$CH_2$) units in dicarboxylic-acid-terminated alkanes. The green, red and purple lines represent the theoretical results obtained using two gold electrodes (green curve) and perfect graphene-gold electrodes (red curve) and defected graphene-gold electrodes (purple curve), respectively. The black line shows the experimental measurements. The values of the Fermi energy $E_F$ (relative to $E_F^{DFT}$) giving the closest fit to experiment depends on the nature of the contact. For Au-Au, the best fit is found at EF= -0.3 eV and yields a decay constant of $\beta_N$ = 0.81. For Gr-Au, the closest fit is found at $E_F$ = 0.65 eV and yields $\beta_N$ = 0.69, For defect graphene contact (def. Gr-Au) we found $E_F$= 0.0 eV, $\beta_N$ = 0.89. These compare with the experimental decay constant of $\beta_N$ = 0.69.

To examine the role of defects in the graphene substrate, we computed electrical conductances when the lower oxygen of the anchor group binds to a defective site formed by removing a carbon from the graphene sheet and passivating the dangling bonds with hydrogen as show in Figure 4.6a-c. The resulting conductances are shown in the bottom row of table 1 and a comparison with experiment shown in Figure 4.5. These show that defects lower the conductance and increase the attenuation coefficient and that the measured results lie between the defective of defect-free theoretical values.

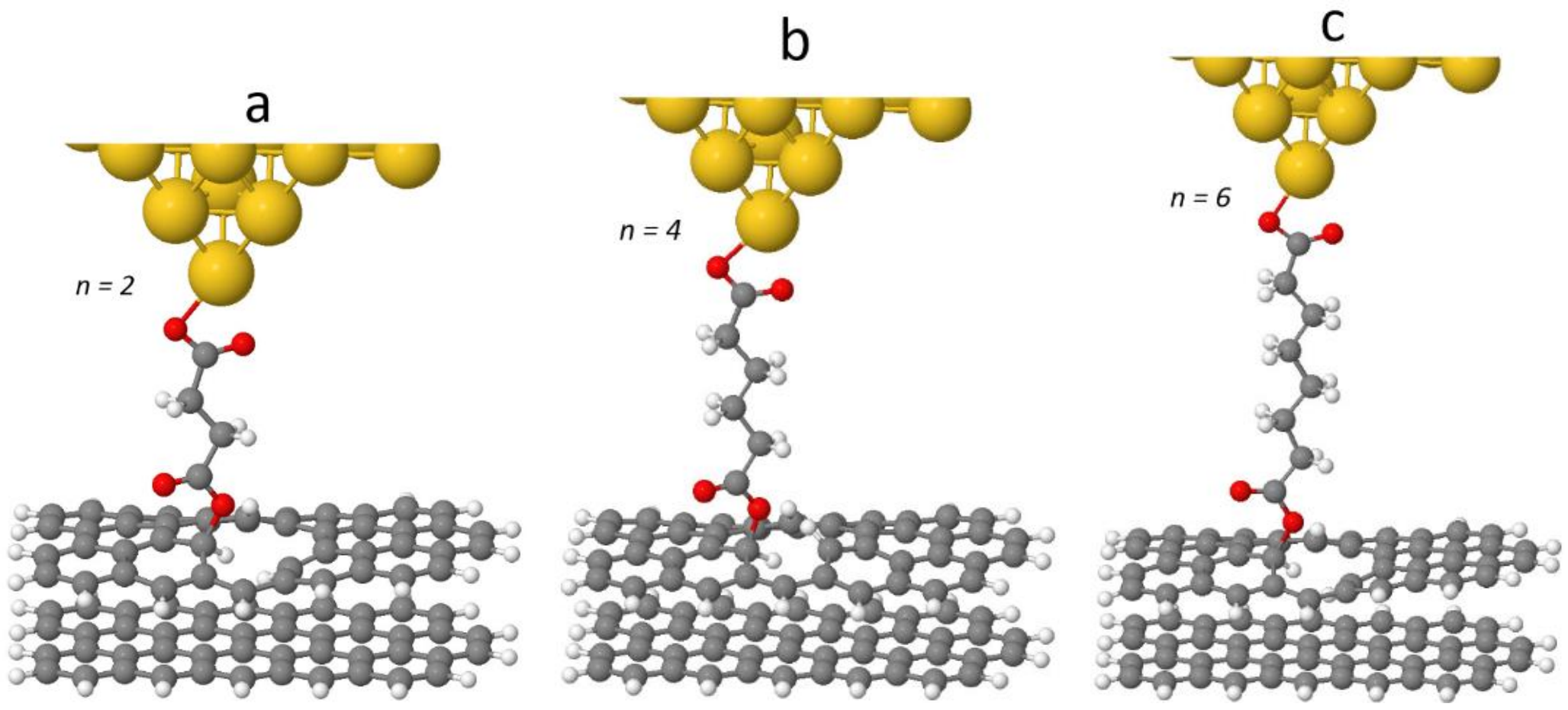

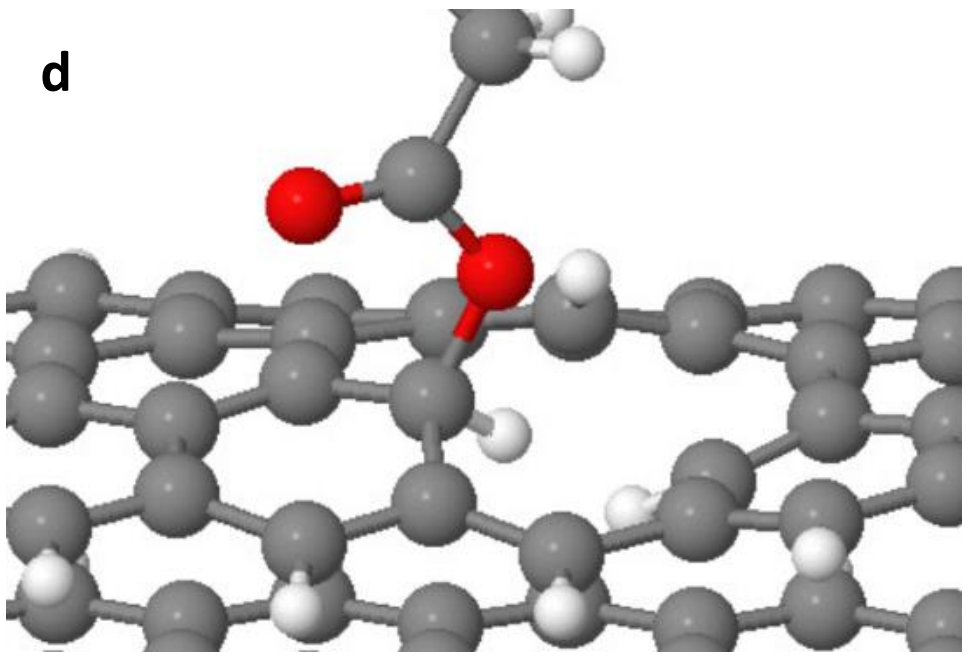

**Figure 4.6**. (a, b and c) show the optimized geometries of systems containing the dicarboxylic-acid-terminated alkanes of lengths (n=2, 4, and 6) connected to a defected graphene sheet. (d) Shows the detail of the defected graphene sheet.

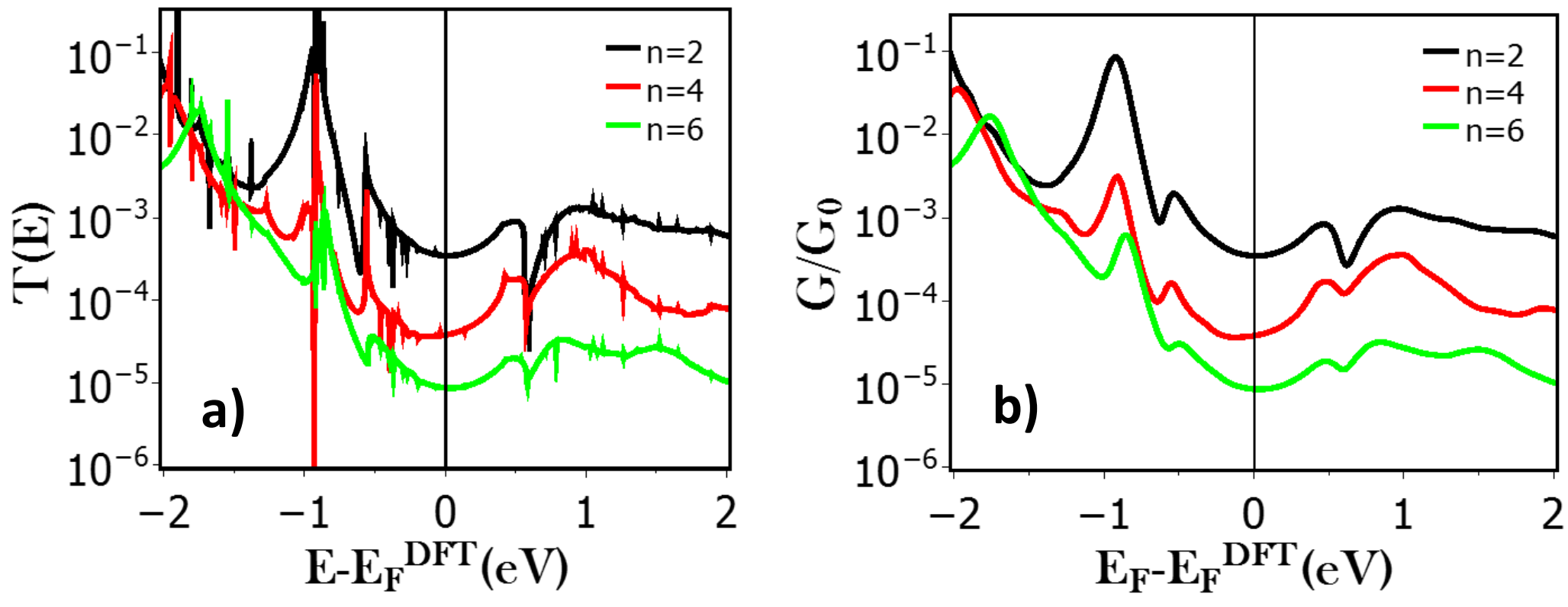

**Figure 4.7.** For the structures in Figure 4.6a-c, this Figure shows a) the transmission coefficients as a function of energy of the systems containing the dicarboxylic-acid-terminated alkane molecule with the lengths n=2, n=4 and n=6 of $CH_2$ attached to the defected graphene-gold electrodes, b) demonstrates the room temperature electrical conductances over a range of Fermi energies.

## 4.2. Effect of Anchoring Groups: Thiol-, and Carboxylic-Acid, Amine-

In this section I have collected results for different anchor group to investigate the effect of these anchors and how that will be different in symmetric and asymmetric electrodes.

### 4.2.1. Symmetric electrodes

For symmetric electrodes I used gold-gold electrodes, and to perform conductance calculations we use the same theoretical techniques in the previous section to calculate electrical properties of the molecules in Figure 4.8, with different lengths of molecules (n= 4 -10), and different anchor group. The conductance varying in the three anchors group as you see in Table 4.2. The alkane's conductance which terminate by thiol and carboxylic-acid, decreases with increased the molecule length as we expected and it is agreement with the result report in [15], and the conductance with thiol is greater than the conductance with carboxylic-acid terminated, that due to the oxygen effect to reduce the charge transport which we will discuss in the next chapter. The lowest conductance is the conductance of the alkane anchors by amine. The amine anchors have different affect in alkane conductance, where the conductance values varying from up and down with increased molecule length. The conductance of alkane has odd-even affected. (that is clear in the conductance curves in Figure 4.9.

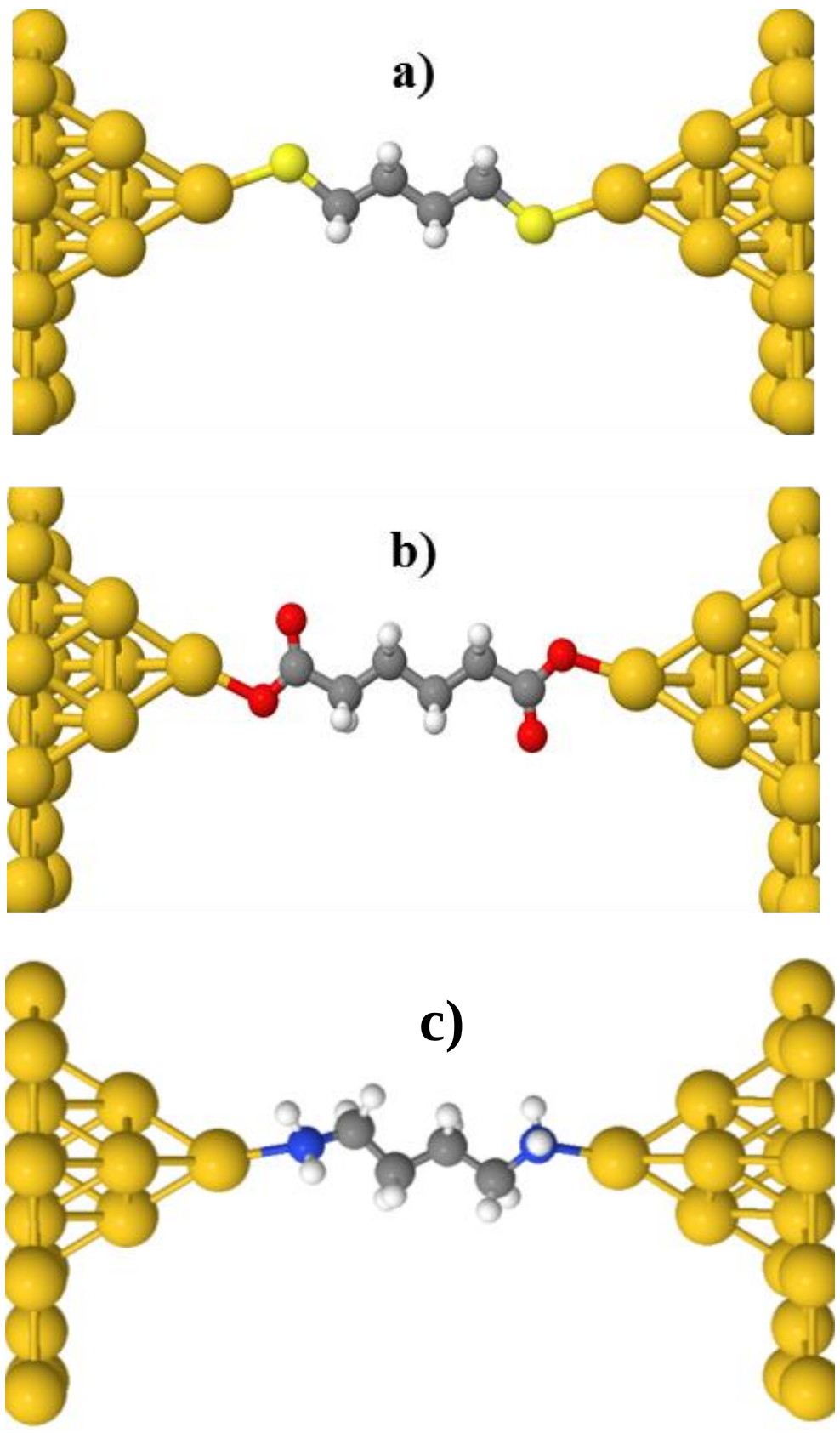

**Figure 4.8.** Example of optimized configuration of Alkane molecule attached to gold electrodes with different ancho group, where a) Thiol (-SH), b) carboxylic-acid (-COOH), and c) amine (-$NH_2$).

| | Conductance, $G_0$ | | |
|---|---|---|---|
| **n** | $HS(CH_2)_nSH$ 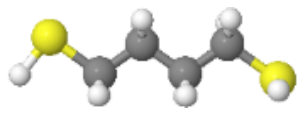 | $HOOC(CH_2)_nCOOH$ 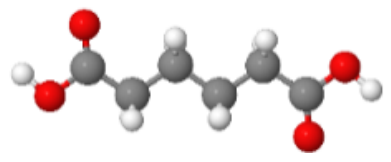 | $NH_2(CH_2)_nNH_2$ 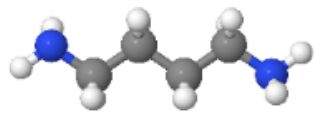 |
| **4** | 951.7 | 125.5 | 5.4 |
| **5** | 321.4 | 84 | 11.6 |
| **6** | 180.7 | 29.1 | 7.7 |
| **7** | 50.2 | 8.4 | 0.6 |
| **8** | 19.6 | 2.4 | 1.4 |
| **9** | 7.2 | 1.1 | 0.57 |
| **10** | 2.5 | 0.29 | 0.1 |

**Table 4.2**. This table show the effect of different anchor group: thiol (-SH), carboxylic-acid (-COOH), and amine (-NH2), of the conductance values with different length (4-10), and symmetric electrodes.

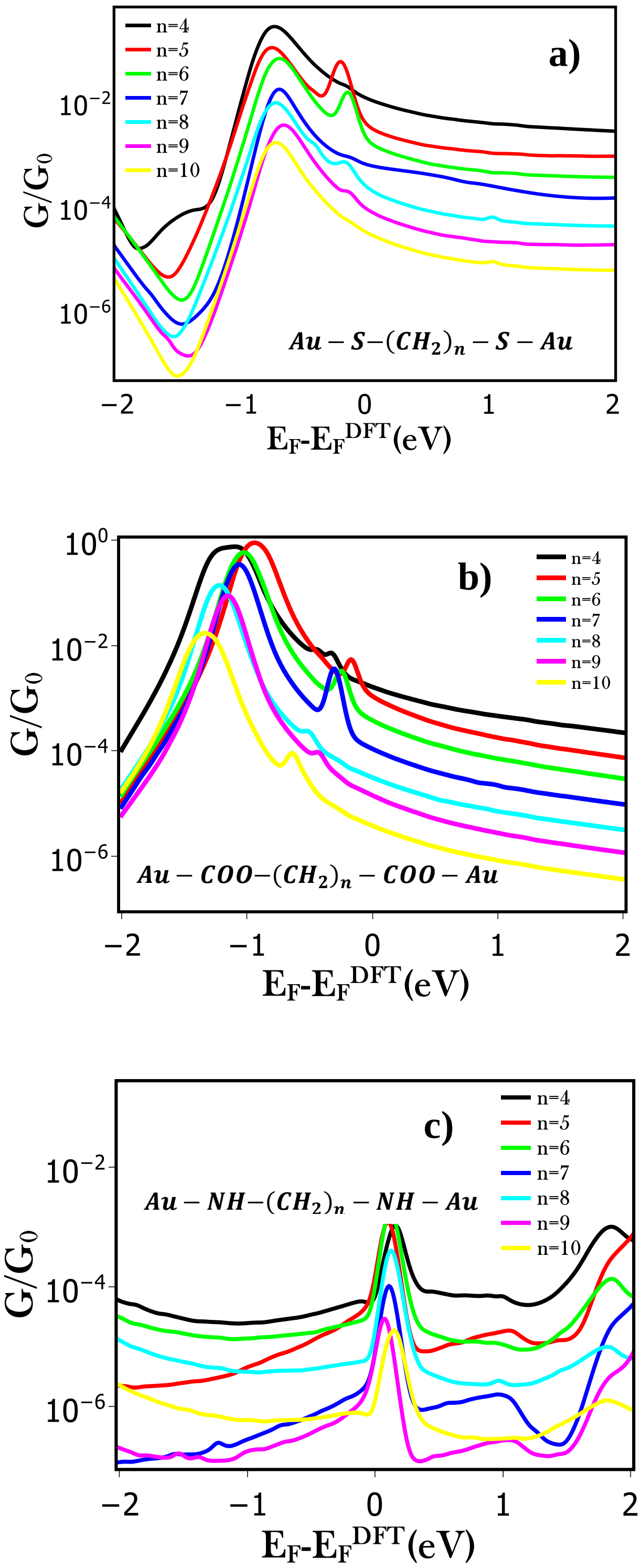

**Figure 4.9.** The room-temperature electrical conductance over a range of Fermi energies of the alkane with different anchor group: a) thiol (-SH), b) carboxylic-acid (-COOH), and c) amine (-NH2)), with different length (4-10), and symmetric electrodes.

As we see the conductance depended on the molecule length, and it is sensitive to the type of anchoring group [15], the relation between the conductance and molecule length describe by beta tunnelling factor ( $\beta_N = -logG/n$), where $n$ is unit length and in our case $(-CH_2)$. Beta factor is important to describe the efficiency of electron transport through the molecules. Figure 10 show the beta factor for the alkane terminated by thiol (-SH) in green line, carboxylic-acid (-COOH) red line, and amine (-NH2) blue line. I found the difference between the beta in the three anchors is (-COOH) is βN = 1.03 > (-SH) is βN = 0.99 > (-NH2) is βN = 0.71, this compering for the length (4-10) and it is very close to the experiment measurement report it in ref. [15].

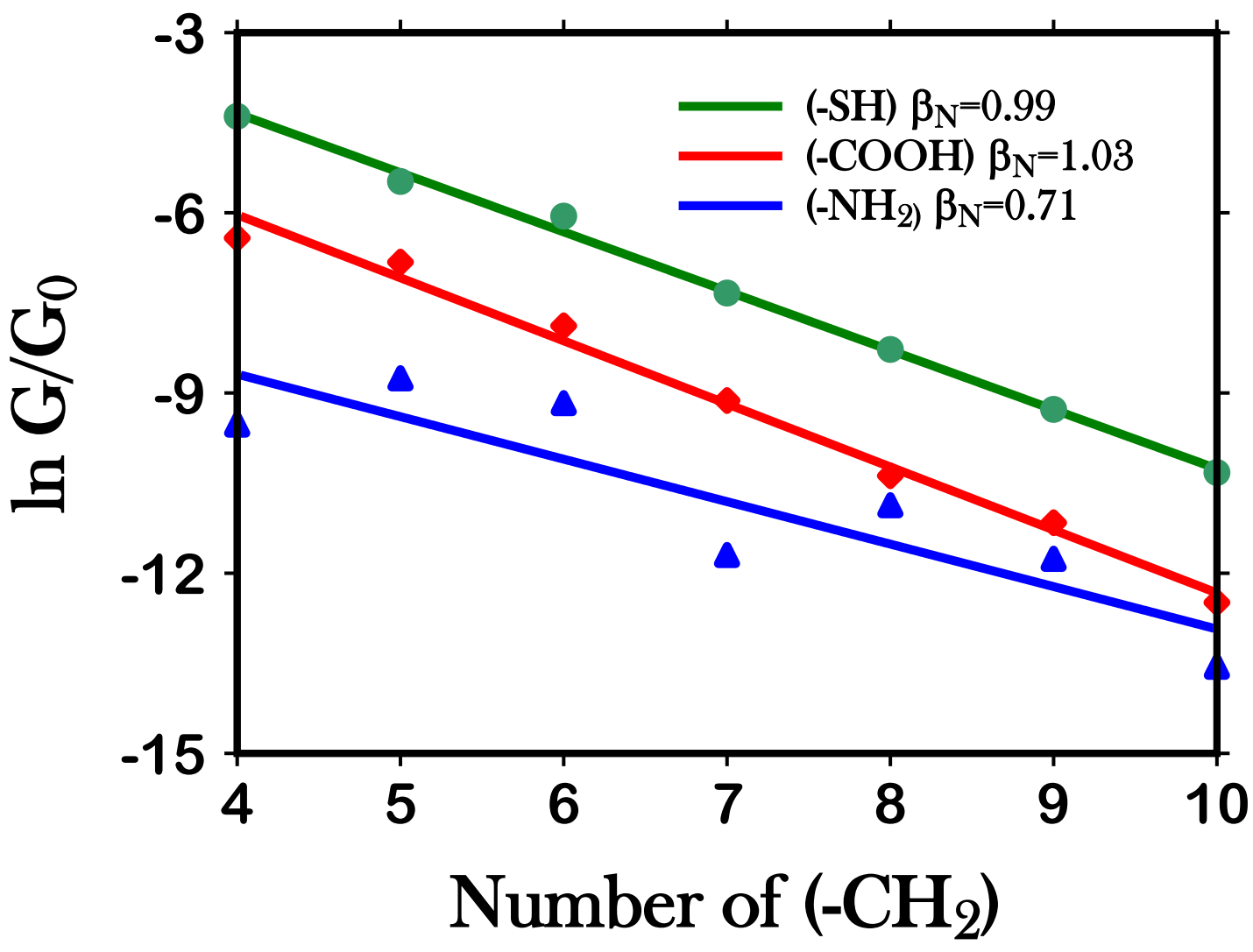

**Figure 4.10**. Comparison the logarithm of single –molecule conductance versus number of (-CH2) units of alkanes with three anchor group: thiol (-SH) in red line, carboxylic-acid (-COOH) in blue line, and amine (-NH2) in green line, with different length (4-10), and symmetric electrodes. The values of the Fermi energy $E_F$ = 0 eV (relative to $E_F^{DFT}$) For Au-Au. A decay constant of(-COOH) is $\beta_N$ = 1.03 > (-SH) is $\beta_N$ = 0.99 > (-$NH_2$) is $\beta_N$ = 0.71.

These differences in the $\beta_N$ values can be attributed to the different anchoring groups which effect the long-distance electron transport in the molecules because of the exponential dependence of the conductance on the length of the molecule. $\beta_N$ therefore depends upon the alignment of the molecular energy levels relative to the Fermi energy of the electrodes and the different anchoring groups can change this energy level alignment [15].

### 4.2.2. Asymmetric electrodes(graphene-gold)

In this section I use graphene-gold electrodes, and to perform conductance calculations we use the same theoretical techniques in the previous section to calculate electrical properties of the molecules in Figure 11, with different lengths of molecules (n= 4, and 10), and different anchor group. Table 4.3 shows the conductance in the three anchor group. The alkane conductance which terminate by thiol decreases with increased the molecule length as we saw in the symmetric electrodes, but in the carboxylic-acid the conductance varing up and down in length n=4-7, and it is constant in last three length n=8, 9, and 10. The amino anchor group have the different story because we are not sure about how he amine bond to the GS. For that matter, we will study three different ways to connect amine to GS. First way, I connect the amine ($NH_2$) to the prefect GS, and after relaxing the geometry I found there is repulsive between the amine and GS see Figure 4.11. Thus we have to try another way to connect the amine to GS, I removed one hydrogen from amine then connected again to GS and let them relaxed. The amine bond to CS (HN-C) and the relaxing geometry is shown in Figure 4.12(a-c). The conductance of alkane terminated by amine with asymmetric electrodes decreased with increased the molecule length.

| | Conductance, $G_0$ |
|---|---|

| n | $Gr(S(CH_2)_nS)Au$ 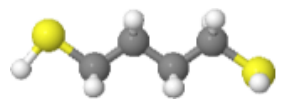 | $Gr(OOC(CH_2)_nCOO)Au$ 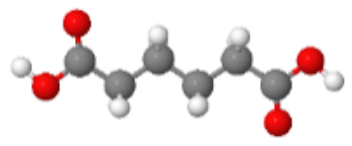 | $Gr(HN(CH_2)_nNH_2)Au$ 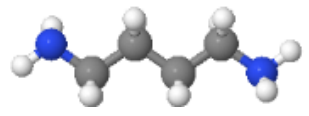 |
|---|---|---|---|
| **4** | 1711 | 94.5 | 255.3<br>Ex:38.33 |
| **5** | 351.17 | 63.19 | 104.8 |
| **6** | 163.89 | 5.48 | 32.5 |
| **7** | 76.08 | 13.26 | 11 |
| **8** | 22.29 | 1.09 | 4.4<br>Ex:4.26 |
| **9** | 11.75 | 1.48 | 1.1 |
| **10** | 3.13 | 1.6 | 0.69 |

**Table 4.3**. This table show the effect of different anchor group: thiol (-SH), carboxylic-acid (-COOH), and amine (-NH2), of the conductance values with different length (4-10), and asymmetric electrodes (graphene-gold).

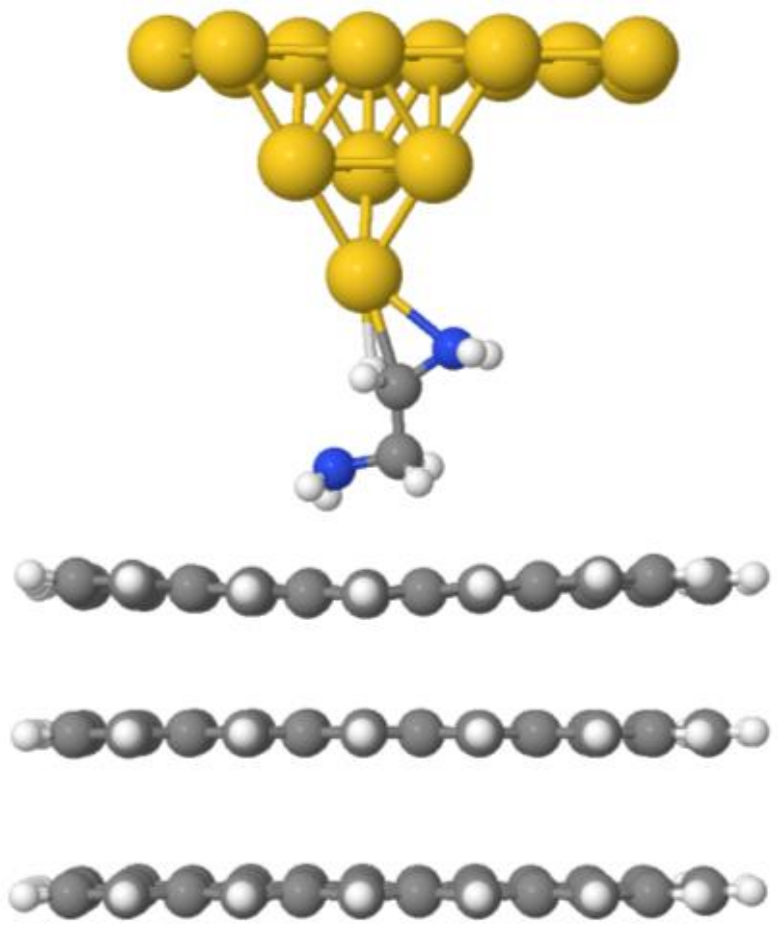

**Figure 4.11.** show the optimized geometries of systems containing the alkanes of lengths (n=2) with amine ($NH_2$) terminated, connected to graphene-gold electrodes.

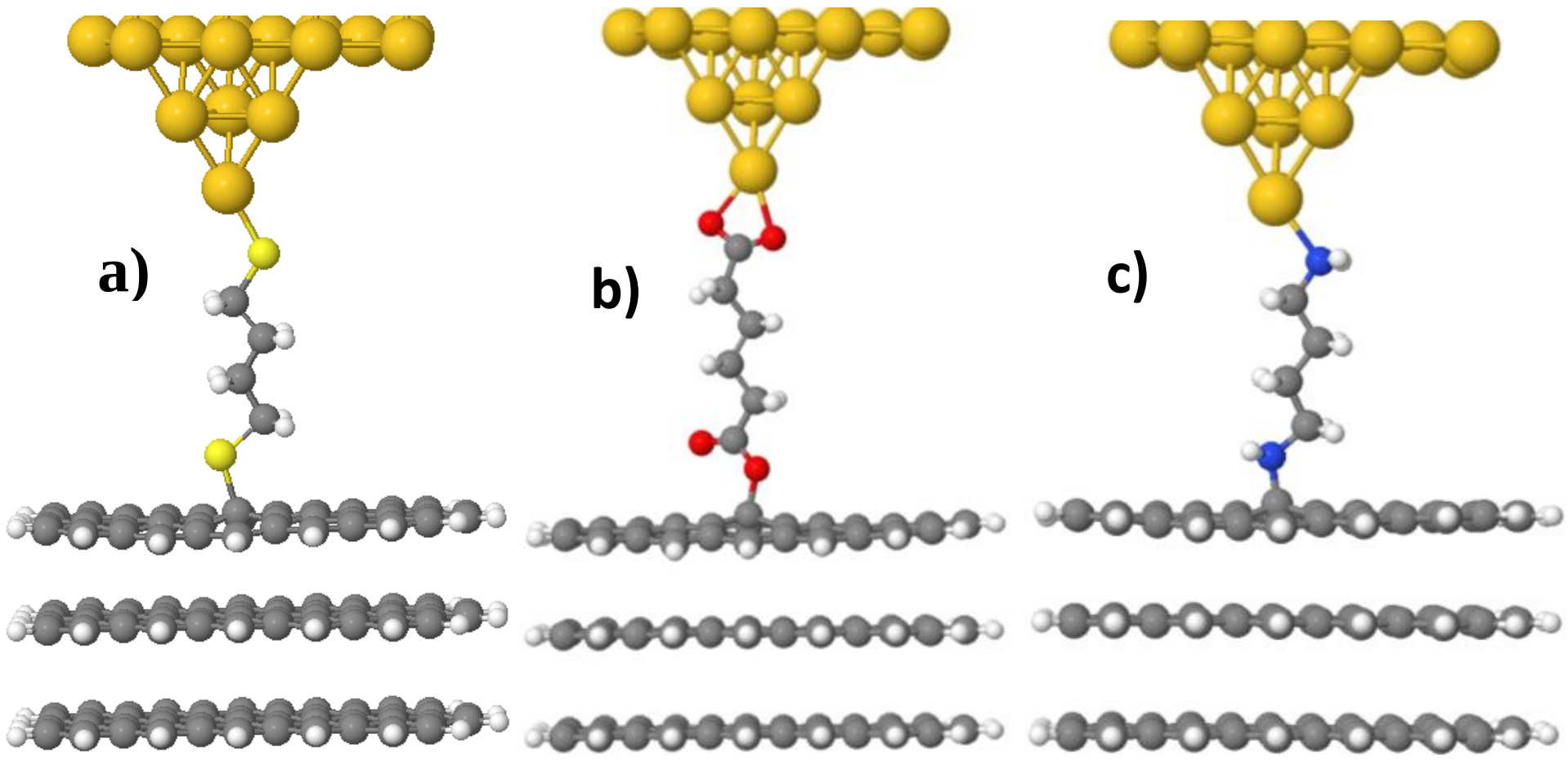

**Figure 4.12**. (a, b and c) show the optimized geometries of systems containing the alkanes of lengths (n=4) with different terminated: a) thiol, b) carboxylic acid, and c) amine, connected to graphene-gold electrodes. I removed one hydrogen from amine to connected with graphene sheet.

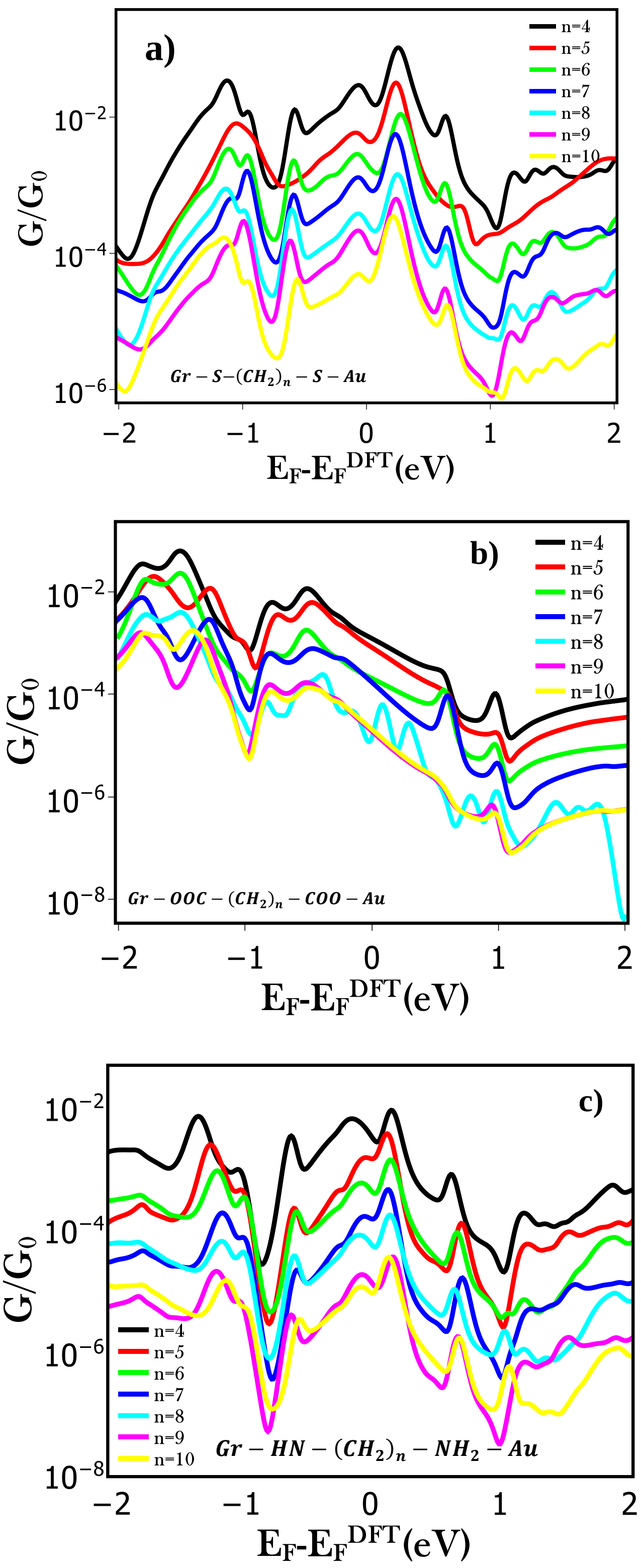

**Figure 4.13.** The room-temperature electrical conductance over a range of Fermi energies of the alkane with different anchor group: a) thiol (-SH), b) carboxylic-acid (-COOH), and c) amine (-NH2)), with different length (4-10), and asymmetric electrodes (graphene-gold).

Figure 14 shows the beta factor for the alkane terminated by thiol (-SH) in green line, carboxylic-acid (-COOH) red line, and amine (-NH2) in blue line. I found that the difference between the beta in the three anchors is (-NH2) is $\beta_N$ = 1.03 > (-SH) is $\beta_N$ = 0.99 > (-COOH) is $\beta_N$ = 0.77, and compering to the trend in symmetric electrodes they are different. These differences proofed that $\beta_N$ depends on the alignment of the molecular energy levels relative to the Fermi energy level of the electrodes [15].

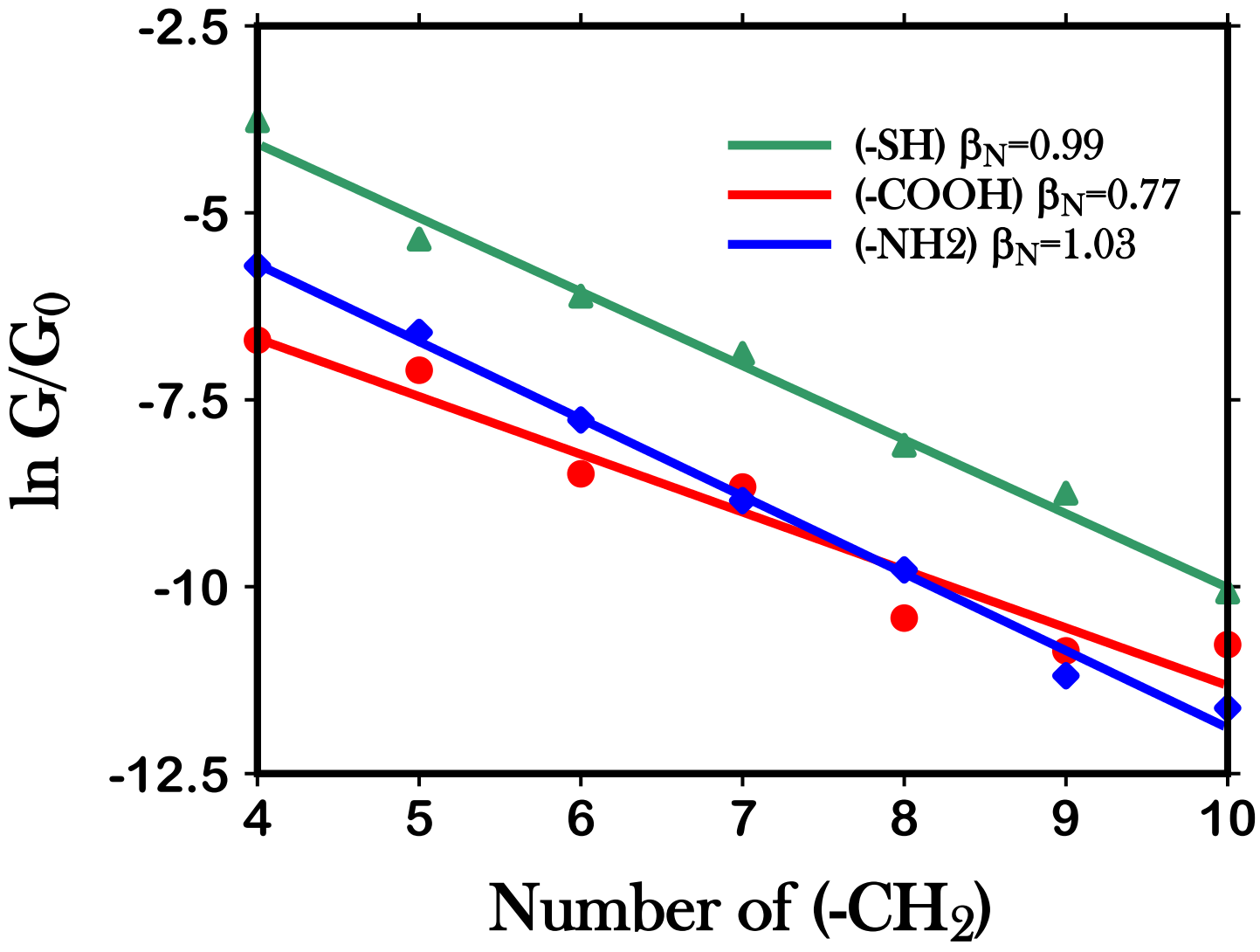

**Figure 4.14.** Comparison the logarithm of single –molecule conductance versus number of (-CH2) units of alkanes with three anchor group: thiol (-SH) in red line, carboxylic-acid (-COOH) in blue line, and amine (-NH2) in green line, with different length (4-10), and asymmetric electrodes (graphene-gold). The values of the Fermi energy $E_F$ = 0 eV (relative to $E_F^{DFT}$) For Gr-Au. A decay constant of (-$NH_2$) is $\beta_N$ = 1.03 > (-SH) is $\beta_N$ = 0.99 > (-COOH) is $\beta_N$ = 0.77.

In conclusion, using graphene as the electrode (Gr-molecule-Au) for single-molecule electronics and for molecules terminated by carboxylic-acid anchor groups mainly leads to an increase in the electrical conductance compared with gold lead (Au-molecule-Au junctions). Density functional theory was combined with a Greens function scattering approach to calculate the electrical conductance of different molecule length through graphene molecule-Au junctions and Au-molecule-Au junctions. The transport is consistent with phase-coherent tunneling, but with an attenuation factor $\beta_N = 0.69$ per methyl unit, which is lower than the value measured for Au-molecule-Au junctions. This difference is due the difference in Fermi energies of the two types of junction, relative to the frontier orbitals of the molecules. This result is agreement with the collaborating experimental groups measurement using the STM-based matrix isolation.

In the second section, I demonstrated and compared single-molecule conductance of alkanes terminated with three different anchoring groups dicarboxylic-acid, diamine, and dithiol anchoring groups and connected with symmetric electrodes (gold-gold) and asymmetric electrodes (graphene-gold). In symmetric electrodes case, at Fermi energy $E_F = 0$ eV (relative to $E_F^{DFT}$) For Au-Au, the factor $\beta_N$ decays (-COOH > -SH > -NH2) is given small different values for each anchoring group. Asymmetric electrodes (graphene-gold), do not affect the thiol anchor effects which are observed in others anchors such as (-NH2 > -SH > -COOH). This is explained by observing that graphene has a weak bond with COOH ref. [15], the amine loses one hydrogen to bond with GS and difference in Fermi energies of the two types of junction.

## 4.3. Acknowledgment

This work was supported by the National Natural Science Foundation of China (NNSFC Grants 21503169), the Jiangsu Science and Technology programme (BK 20140405), Suzhou Industrial Park Initiative Platform Development for Suzhou Municipal Key Lab for New Energy and Environmental Protection Techniques (RR0140), the XJTLU Research Development Fund (PGRS-13-01-03, RDF-14-02-42), EPSRC project UK EPSRC (grant nos. EP/ M014452/1 and EP/N017188/1), by the EU ITN MOLESCO

## 4.4. Reference

1. García-Suárez, V.M., Rocha, A.R., Bailey, S.W., Lambert, C.J., Sanvito, S. and Ferrer, J., 2005. Single-channel conductance of H 2 molecules attached to platinum or palladium electrodes. Physical Review B, 72(4), p.045437.

2. Leff, D.V., Brandt, L. and Heath, J.R., 1996. Synthesis and characterization of hydrophobic, organically-soluble gold nanocrystals functionalized with primary amines. *Langmuir*, *12*(20), pp.4723-4730.

3. Xu, B., Xiao, X. and Tao, N.J., 2003. Measurements of single-molecule electromechanical properties. Journal of the American Chemical Society, 125(52), pp.16164-16165.

4. Martín, S., Haiss, W., Higgins, S., Cea, P., Lopez, M.C. and Nichols, R.J., 2008. A comprehensive study of the single molecule conductance of α, ω-dicarboxylic acid-terminated alkanes. *The Journal of Physical Chemistry C*,*112*(10), pp.3941-3948.

5. Chen, F., Li, X., Hihath, J., Huang, Z. and Tao, N., 2006. Effect of anchoring groups on single-molecule conductance: comparative study of thiol-, amine-, and carboxylic-acid-terminated molecules. *Journal of the American Chemical Society, 128*(49), pp.15874-15881.

6. Dappe, Y.J., González, C. and Cuevas, J.C., 2014. Carbon tips for all-carbon single-molecule electronics. *Nanoscale*, *6*(12), pp.6953-6958.

7. González, C., Abad, E., Dappe, Y.J. and Cuevas, J.C., 2016. Theoretical study of carbon-based tips for scanning tunnelling microscopy.*Nanotechnology*, *27*(10), p.105201.

8. Yan, H., Bergren, A.J. and McCreery, R.L., 2011. All-carbon molecular tunnel junctions. *Journal of the American Chemical Society, 133*(47), pp.19168-19177.

9. Wen, Y., Chen, J., Guo, Y., Wu, B., Yu, G. and Liu, Y., 2012. Multilayer Graphene-Coated Atomic Force Microscopy Tips for Molecular Junctions.*Advanced Materials, 24*(26), pp.3482-3485.

10. Soler, J.M., Artacho, E., Gale, J.D., García, A., Junquera, J., Ordejón, P. and Sánchez-Portal, D., 2002. The SIESTA method for ab initio Order-N materials simulation. *Journal of Physics: Condensed Matter, 14*(11), p.2745.

11. Ceperley, D.M. and Alder, B.J., 1980. Ground state of the electron gas by a stochastic method. *Physical Review Letters, 45*(7), p.566.

12. Ferrer, J., Lambert, C.J., García-Suárez, V.M., Manrique, D.Z., Visontai, D., Oroszlany, L., Rodríguez-Ferradás, R., Grace, I., Bailey, S.W.D., Gillemot,

K. and Sadeghi, H., 2014. GOLLUM: a next-generation simulation tool for electron, thermal and spin transport. *New Journal of Physics*, *16*(9), p.093029.

13. Lambert, C.J., 2015. Basic concepts of quantum interference and electron transport in single-molecule electronics. *Chemical Society Reviews*, *44*(4), pp.875-888.

14. a) Quek, S.Y., Kamenetska, M., Steigerwald, M.L., Choi, H.J., Louie, S.G., Hybertsen, M.S., Neaton, J.B. and Venkataraman, L., 2009. Mechanically controlled binary conductance switching of a single-molecule junction. *Nature nanotechnology*, *4*(4), pp.230-234. b) Wen, H.M., Yang, Y., Zhou, X.S., Liu, J.Y., Zhang, D.B., Chen, Z.B., Wang, J.Y., Chen, Z.N. and Tian, Z.Q., 2013. Electrical conductance study on 1, 3-butadiyne-linked dinuclear ruthenium (II) complexes within single molecule break junctions. *Chemical Science*, *4*(6), pp.2471-2477. c) Li, Y., Baghernejad, M., Qusiy, A.G., Zsolt Manrique, D., Zhang, G., Hamill, J., Fu, Y., Broekmann, P., Hong, W., Wandlowski, T. and Zhang, D., 2015. Three-State Single-Molecule Naphthalenediimide Switch: Integration of a Pendant Redox Unit for Conductance Tuning. *Angewandte Chemie International Edition*, *54*(46), pp.13586-13589.

15. Chen, F., Li, X., Hihath, J., Huang, Z. and Tao, N., 2006. Effect of anchoring groups on single-molecule conductance: comparative study of thiol-, amine-, and carboxylic-acid-terminated molecules. Journal of the American Chemical Society, 128(49), pp.15874-15881.

16. Wang, Y.H., Zhou, X.Y., Sun, Y.Y., Han, D., Zheng, J.F., Niu, Z.J. and Zhou, X.S., 2014. Conductance measurement of carboxylic acids binding to palladium nanoclusters by electrochemical jump-to-contact STM break junction. *Electrochimica Acta*, *123*, pp.205-210.

17. Martín, S., Haiss, W., Higgins, S., Cea, P., Lopez, M.C. and Nichols, R.J., 2008. A comprehensive study of the single molecule conductance of α, ω-dicarboxylic acid-terminated alkanes. *The Journal of Physical Chemistry C*,*112*(10), pp.3941-3948.

# Chapter 5

# 5. Charge Transport through alkane and oligoethylene glycol chains

## 5.1. Introduction

Understanding electronic transfer in single molecules is not enough, we need to control it, to develop new electronic devices. That requires the creation of various kinds of molecular bridges to connect functional groups of molecules together and produce effective electrical circuits. One of the most studied type of molecular bridges (chain) are saturated hydrocarbon chains. Alkanes are the simplest hydrocarbons (organic molecules), consisting of a series of methyl units connect by single bonds $(CH_2)$. There are a number of theoretical and experimental studies alkane chains in the literature. Therefore, I use it as benchmark to compare with an alternative kind of chain called oligoether chains.

In this chapter, I compare the conductance of alkane chains $(-CH_2-CH_2-CH_2)$and oligoethers $(-CH_2-CH_2-O-)$ of corresponding length. The DFT is used to calculate the single molecule conductance of two groups of length: short and long length chains. The short length alkanedithiols containing 5, 8, and 11 methylene units, and the three corresponding dithiolated oligoethers, where the third methylene unit is replaced by an oxygen. Then, I investigated the second long length situation by increasing the molecular length in both types of chains to (5, 8, 11, 14, 17, and 20) units.

The aim of this project is understand the effect of exchanging carbon by oxygen in saturated chains on charge transfer through single molecule. In order to understand the conductance-length dependence, I will study the tunnelling constant decay $\beta_N$.

## 5.2. Theoretical calculations

To calculate electrical properties of the alkanedithiol chains (5-C, 8-C, and 11-C) and oligethylene chains in the same length (5-O, 8-O and 11-O) shown in Figure 5.1a-b, I use DFT code SIESTA [1] to relaxed the geometry of each isolated molecule, which employs Troullier-Martins pseudopotentials to represent the potentials of the atomic cores [2], and a local atomic-orbital basis set. I used a double-zeta polarized basis set for all atoms and the generalized gradient approximation (GGA-PBE) for the exchange and correlation (GGA) [3]. The Hamiltonian and overlap matrices are calculated on a real -space grid defined by a plane-wave cut off of 150 Ry. Each molecule was relaxed to the optimum geometry until the forces on the atoms are smaller than 0.02 eV/Å and in case of the isolated molecules, a sufficiently-large unit cell was used to avoid spurious steric effects.

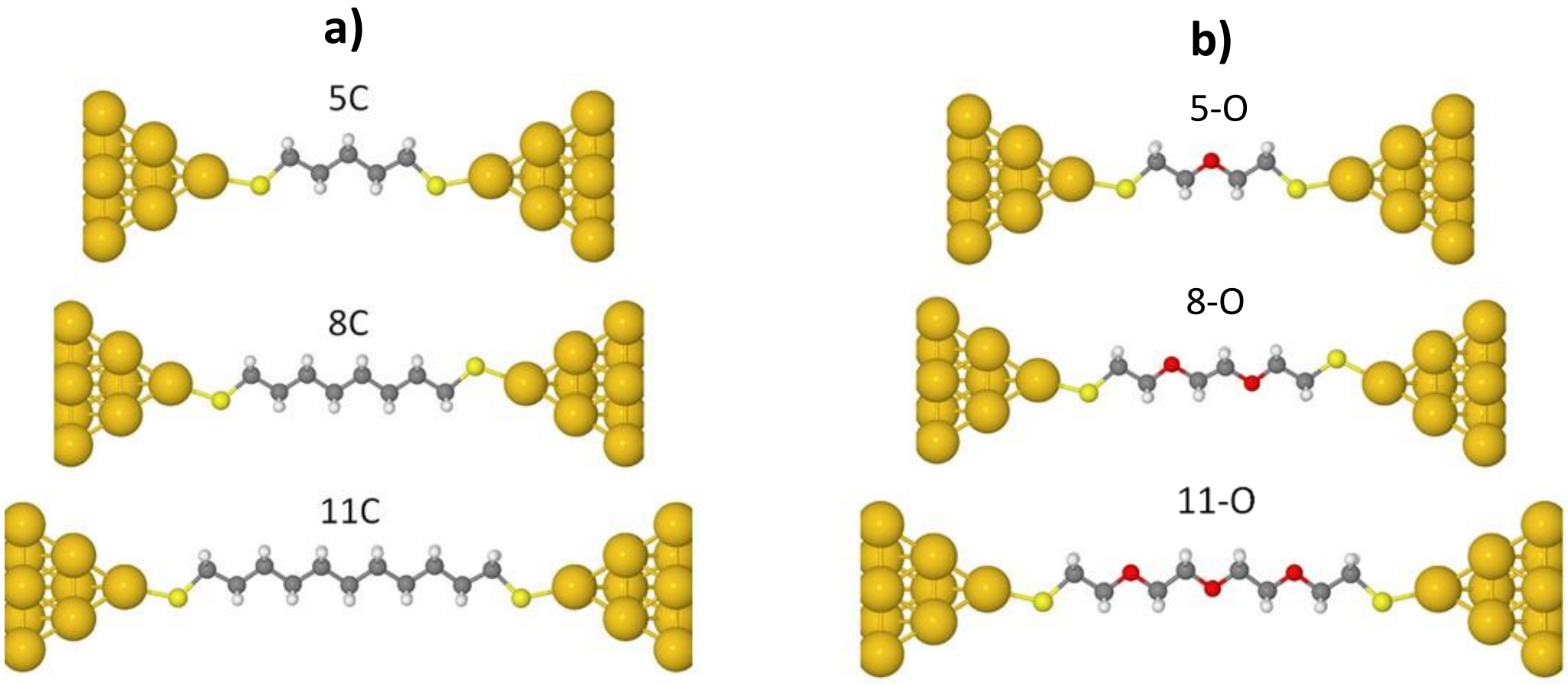

**Figure 5.1**. Optimized geometries of systems containing the alkanedithiol chains (5-C, 8-C, and 11-C) in a) and oligethylene chains in the same length (5-O, 8-O, and 11-O) in b) connected to the gold electrodes.

After obtaining the relaxed geometry of an isolated molecule, the molecule was then placed between (111) gold electrodes and the molecule plus electrodes allowed to further relax to yield the optimized structures shown in Figures 5.1a-b. The optimization was performed with the SIESTA code with the same parameters as used for the isolated molecule. During the relaxation the gold atoms were fixed and for the gold a double-zeta basis set was used. The initial distance between the S atom and the centre of the apex atom of each gold pyramid was initially 2.3 Å. After geometry optimization the distance changed to a final value of 2.5 Å. For each structure in Figures 5.1a-b, we use the GOLLUM method [4] to compute the transmission coefficient $T(E)$ for electrons of energy $E$ passing from the left gold electrode to the right electrode. GOLLUM is a next-generation code, born out of the SMEAGOL code [5] and uses to compute transport properties of a wide variety of nanostructures. Once the $T(E)$ is computed, we calculated the zero-bias electrical conductance G using the Landauer formula see Figure 5.3:

$$G = {}^{I}/_{V} = G_0 \int_{-\infty}^{\infty} dE \, T(E) \left(-\frac{df(E)}{dE}\right) \tag{5.1}$$

Where $G_0 = \left(\frac{2e^2}{h}\right)$ is the quantum of conductance, $f(E)$ is Fermi distribution function defined as $f(E) = \left[e^{(E-E_F)k_BT} + 1\right]^{-1}$ where $k_B$ is Boltzmann constant and $T$ is the temperature. Since the quantity $\left(-\frac{df(E)}{dE}\right)$ is a normalised probability distribution of width approximately equal to $k_BT$, centred on the Fermi energy $E_F$, the above integral represents a thermal average of the transmission function $T(E)$ over an energy window of the width $k_BT$ (=25 meV at room temperature) [4].

## 5.3. Conductance

For the structures in Figures 5.1(a-b), Figure 5.3(a-b) shows the room-temperature electrical conductance of three alkoxy dithiolated saturated linear molecules [2-mercaptoethyl ether (5-O), 2,2′-(ethylenedioxy)diethanethiol (8-O), and tetra(ethylene glycol)dithiol (11-O)] and three alkanedithiols of corresponding length [1,5-pentanedithiol (5-O), 1,8-octanedithiol (8-O), and 1,11-undecanedithiol (11-O)], over a range of Fermi energies $E_F$ in the proximity of the DFT-predicted Fermi energy $E_F^{DFT}$ . For a wide range of values of $E_F$, the conductance decreases when the molecular length increases in both molecules (alkane and oligoethers), and that is agreement with experimental measurements reported by Emil Wierzbinski [6], where he studied the same two groups of alkane and oligoether chains by the STM break junction method [7] show in Figure 5.2, Figure 5.4 shows a direct comparison between the conductance distributions constructed for alkanedithiol (5-C, 8-C, and 11-C) and oligoether (5-O, 8-O, and 11-O) molecules, while Table 5.1 shows comparison between our theoretical calculations and the experimental and other theory group results reported in same paper[6]. We can see from Table 5.1 that my calculations are closer to the experiment measurements than the theory calculations that reported in the same paper which are very low conductance.

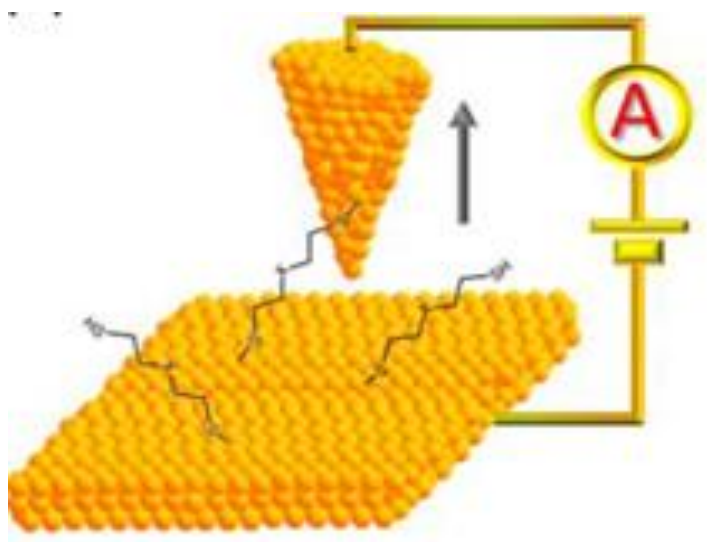

**Figure 5.2**. Shows a cartoon representation of the STM break junction, which used by Wierzbinski [6] to measure the conductance of three alkoxy dithiolated saturated linear molecules.

The conductances values of the alkanedithiol chains (5-C, 8-C, and 11-C) shown in Table 1, are very close to others values found it by others group and reported in [8, 9]. Furthermore, a comparison of the conductance value for the 5-O is greater than 5-C chains at EF = 0 eV which is disagreement with the results reported by in [6], that shows very similar conductances for these two compounds. But we can explain that by looking to Figure 5.5, which shows the room temperature electrical conductance of 5-C and 5-O chains in range of the Fermi Energy. The two carves are very similar in this range except in the two peaks in the right and left side of the main peak in $E_F$= 0.75 eV. This right peak is due to the anchor group which is the sulfur and the left peak appeared after replacing the ether $-CH_2$ by oxygen O. Therefore, the conductanse value for both alkane and oligoethers components indicates that, replacing a carbon atom in the hydrocarbon chain by an oxygen, reduces the conductance of the chain.

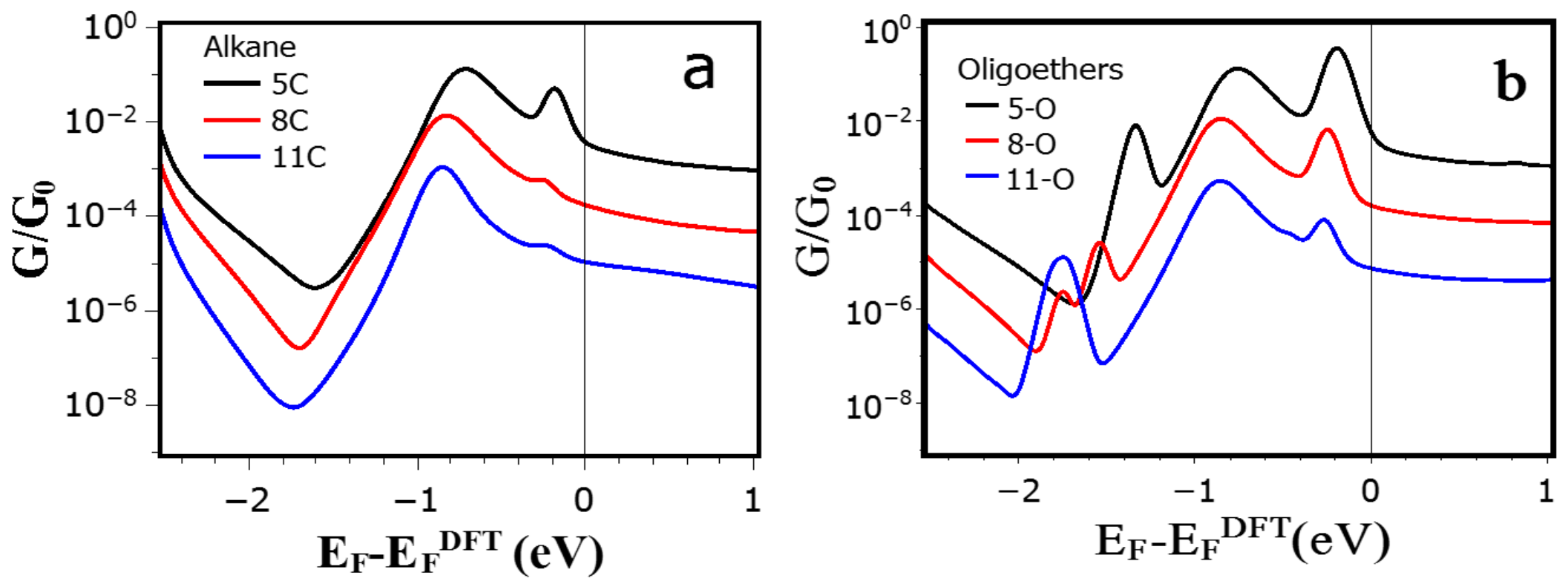

**Figure 5.3**. Shows the comparison between the room temperature electrical conductance of n-Alkane chain and n-oligoethers, where in is the $CH_2$ unit number. (5, 8 and 11).

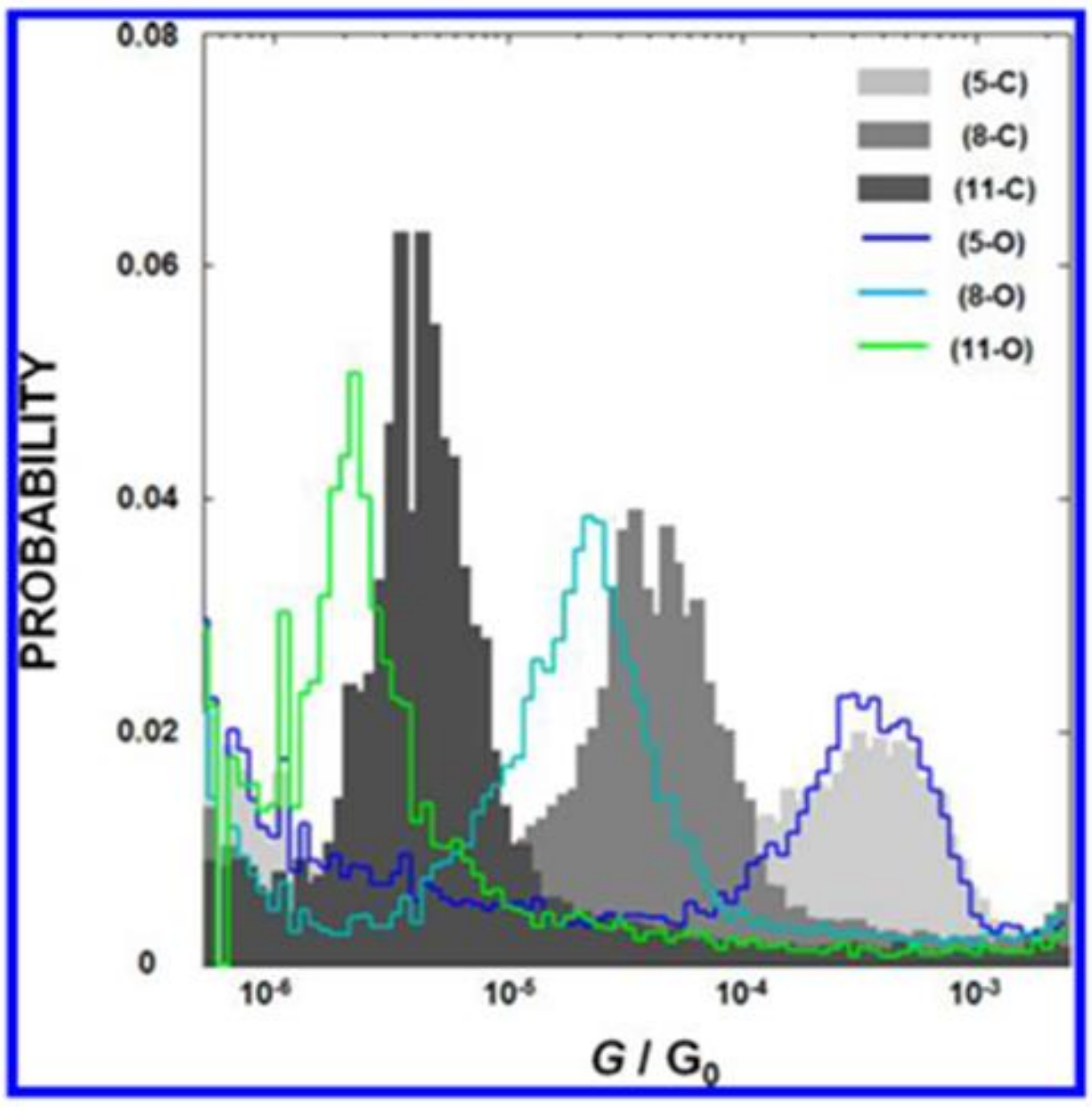

**Figure 5.4**. The plot compares the conductance distributions of dithiolated hydrocarbons (5-C, 8-C, 11-C) and oligoethers (5-O, 8-O, 11-O). The shaded distributions represent the hydrocarbon chains, and the colored curves sketch the distributions for the corresponding oligoethers (green is 11-O, aqua is 8-O, and blue is 5-O).

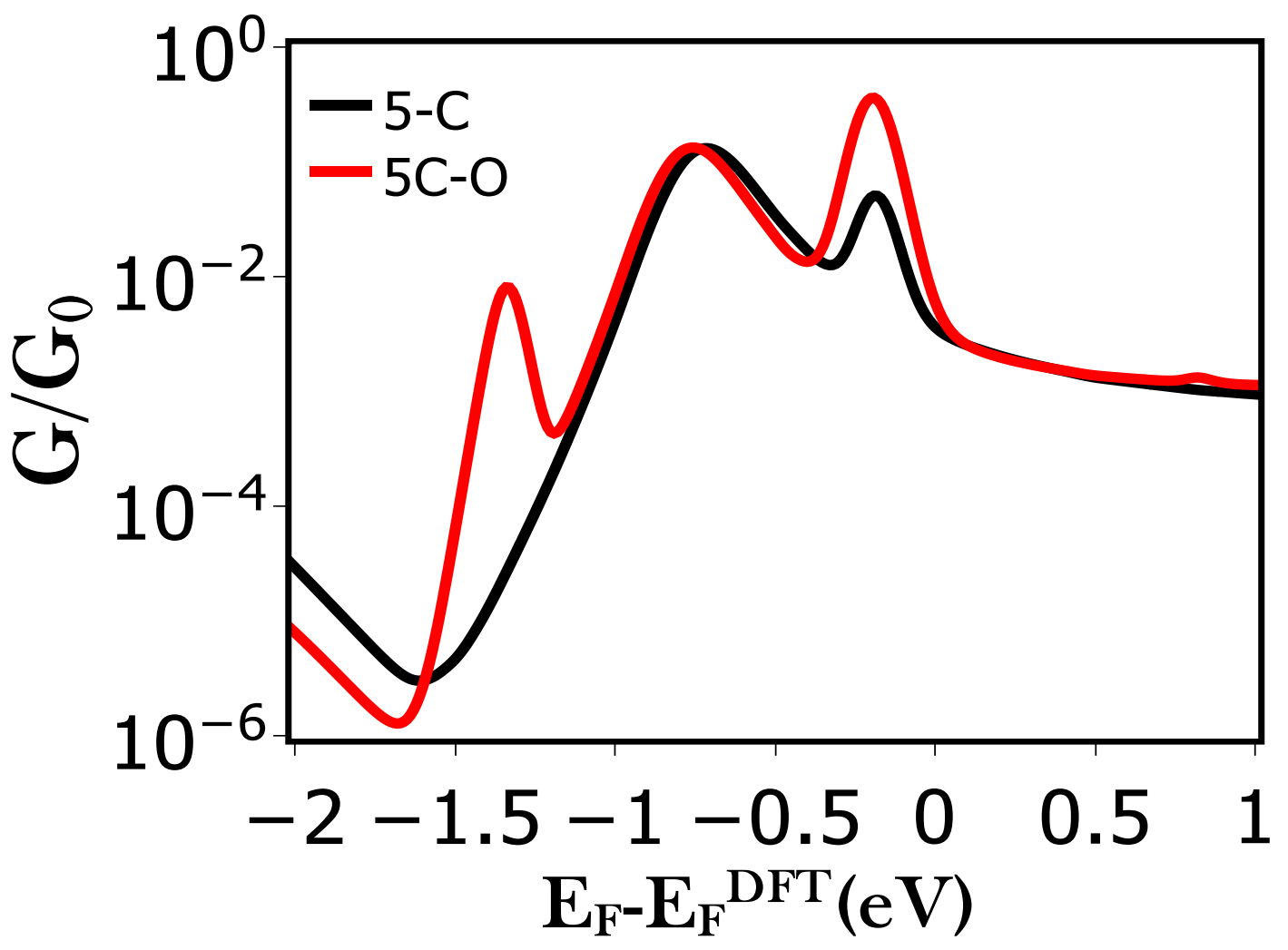

**Figure 5.5.** Comparison between the room temperature electrical conductance of alkane (5-C) in black line and oligoethers (5-O) in red line.

| *Molecule* | *Conductance, $G_0$* | | |
|---|---|---|---|
| | ***Calculated*** ***Our work*** | ***Measured[6]*** | ***Calculated[6]*** |
| **5-C** | $3.65 \times 10^{-3}$ | $(4.6 \pm 2.9) \times 10^{-4}$ | $2.6 \times 10^{-9}$ |
| **5-O** | $6.07 \times 10^{-3}$ | $(4.2 \pm 2.3) \times 10^{-4}$ | $1.9 \times 10^{-9}$ |
| **8-C** | $1.72 \times 10^{-4}$ | $(5.2 \pm 2.5) \times 10^{-5}$ | $7.0 \times 10^{-11}$ |
| **8-O** | $1.61 \times 10^{-4}$ | $(2.5 \pm 1.2) \times 10^{-5}$ | $4.1 \times 10^{-11}$ |
| **11-C** | $1.06 \times 10^{-5}$ | $(4.7 \pm 1.9) \times 10^{-6}$ | $2.2 \times 10^{-12}$ |
| **11-O** | $7.44 \times 10^{-6}$ | $(2.1 \pm 0.7) \times 10^{-6}$ | $1.3 \times 10^{-12}$ |

**Table 5.1**. Comparison between our theoretical calculations and the experimental results which reported recently [7] of the alkane and oligoethers series.

## 5.4. Beta tunnelling decay

The beta value increased with the Fermi energy in alkane and oligoether molecules as shown in Figure 5.6. At the same Fermi energy $E_F$ = -0.15 eV, the beta values of alkane reaches to the maximum value $\beta_N$= 1.3 per $CH_2$, but it is higher in alkane which is 1.6 per $CH_2$ at the same Fermi energy. After the maximum beta value decreases between -0.15 to 0 eV and I therefore predict that the actual value of the Fermi energy should be in this range between -0.15 to 0 eV.

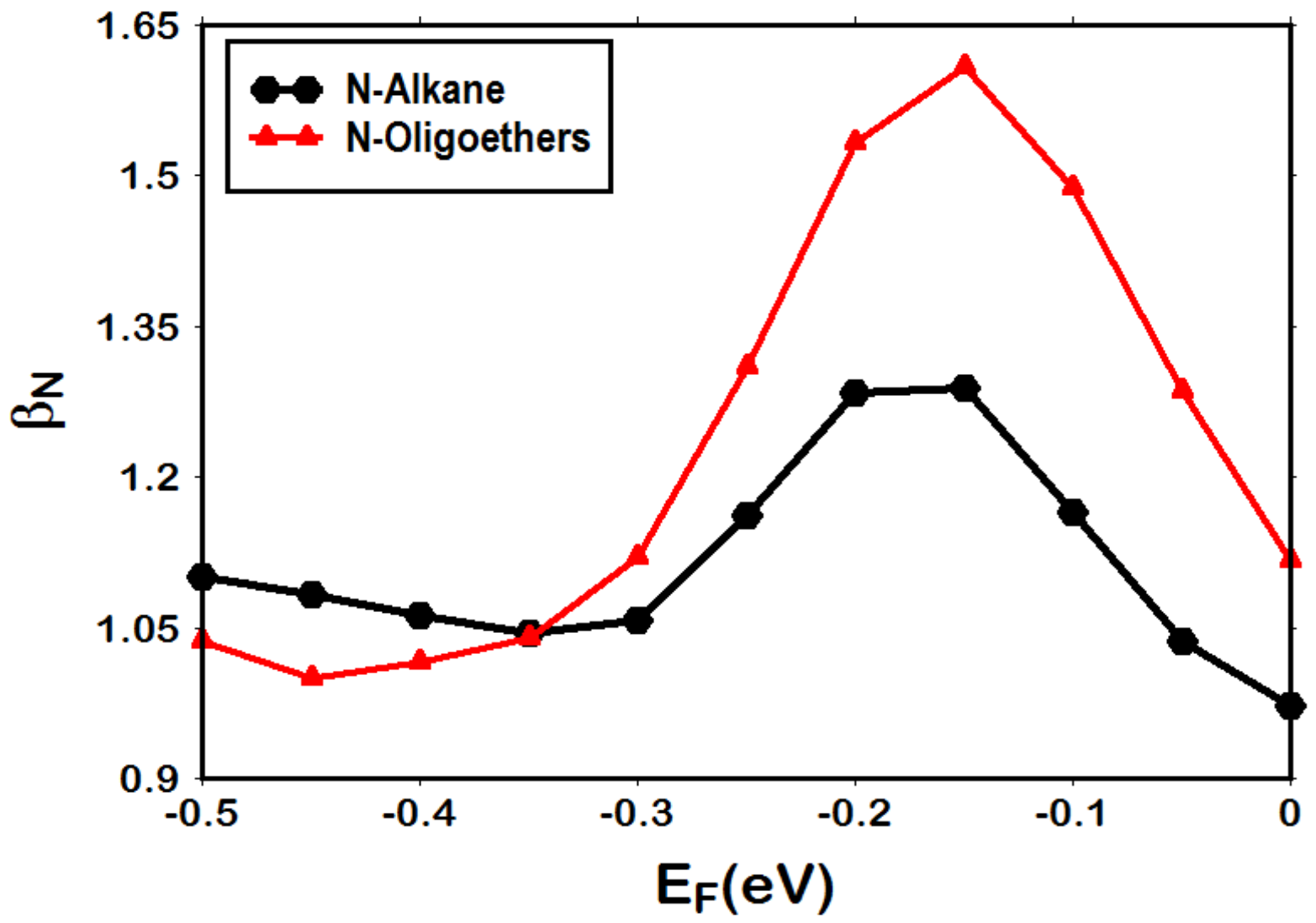

**Figure 5.6**. Tunnelling constant decay $\beta_N$ functional of range of Fermi energies for three alkanes (5-C, 8-C, and 11-C) in black line, and three oligethers in the same length (5-O, 8-O, and 11-O) in red line.

The DFT predicted the Fermi energy is zero and the $\beta_N$ for the alkane group at the DFT Fermi energy $(E_F = 0\ eV)$ is 0.97 per $CH_2$ unit. This value is in agreement with other theoretical and experiment groups [7,10, 11], where the $\beta_N$ for the oligoether group is found to be 1.12 per $CH_2$ unit. The higher value of $\beta_N$ in oligoether illustrates the effect of replacing the methyl by oxygen in the saturated chains such like alkane and that will reduce the charge transfer and the conductance.

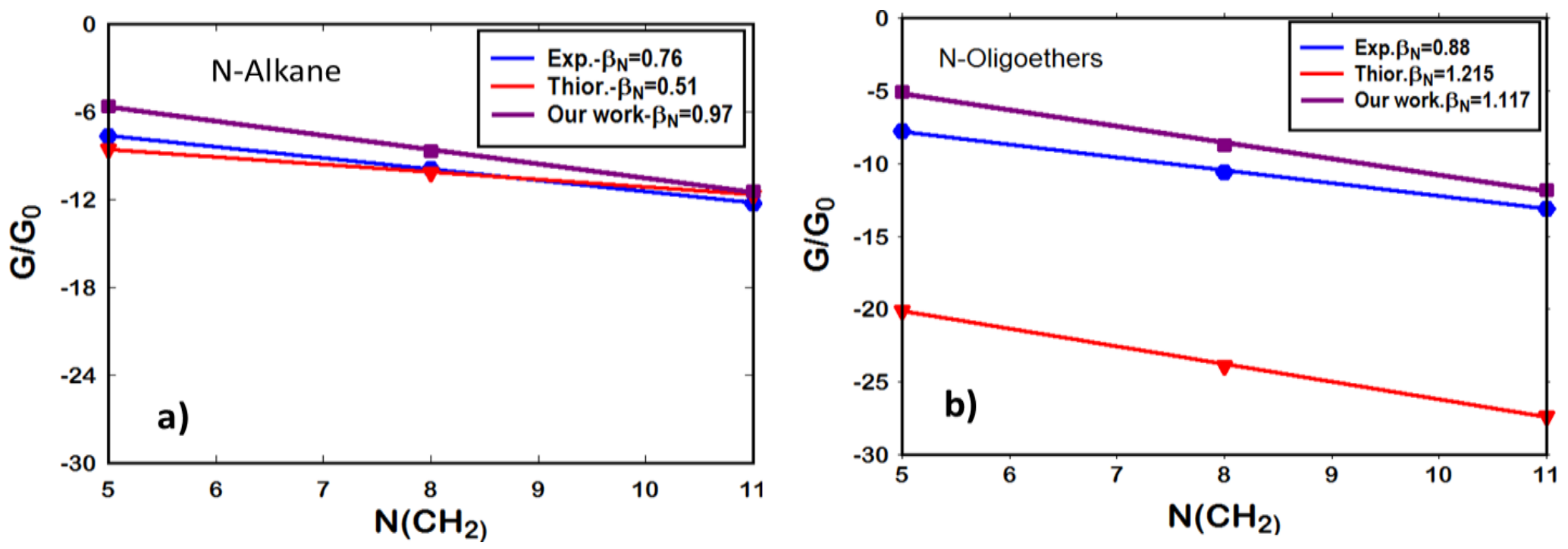

**Figure 5.7.** Comparison between theory and experiment of the logarithm of single – molecule conductance versus number of (-CH2) units in a) alkanes and b) oligoethers. The purple line represent our theoretical results obtained using gold electrodes, blue and red lines represent the experiment and theory results which reported recently [6].

Figure 5.7 shows a logarithmic plot of predicted single-molecule conductances versus the number of (-CH2) units in the alkane and oligoether chain, along with a comparison with experiment reported in [6]. Figure 5.7 also shows that, our theoretical conductance values are slightly higher than measured ones for all molecular lengths, which can be attributed to the tendency of GGA to underestimate the HOMO-LUMO gap, which results in an overestimated of the conductance [12].

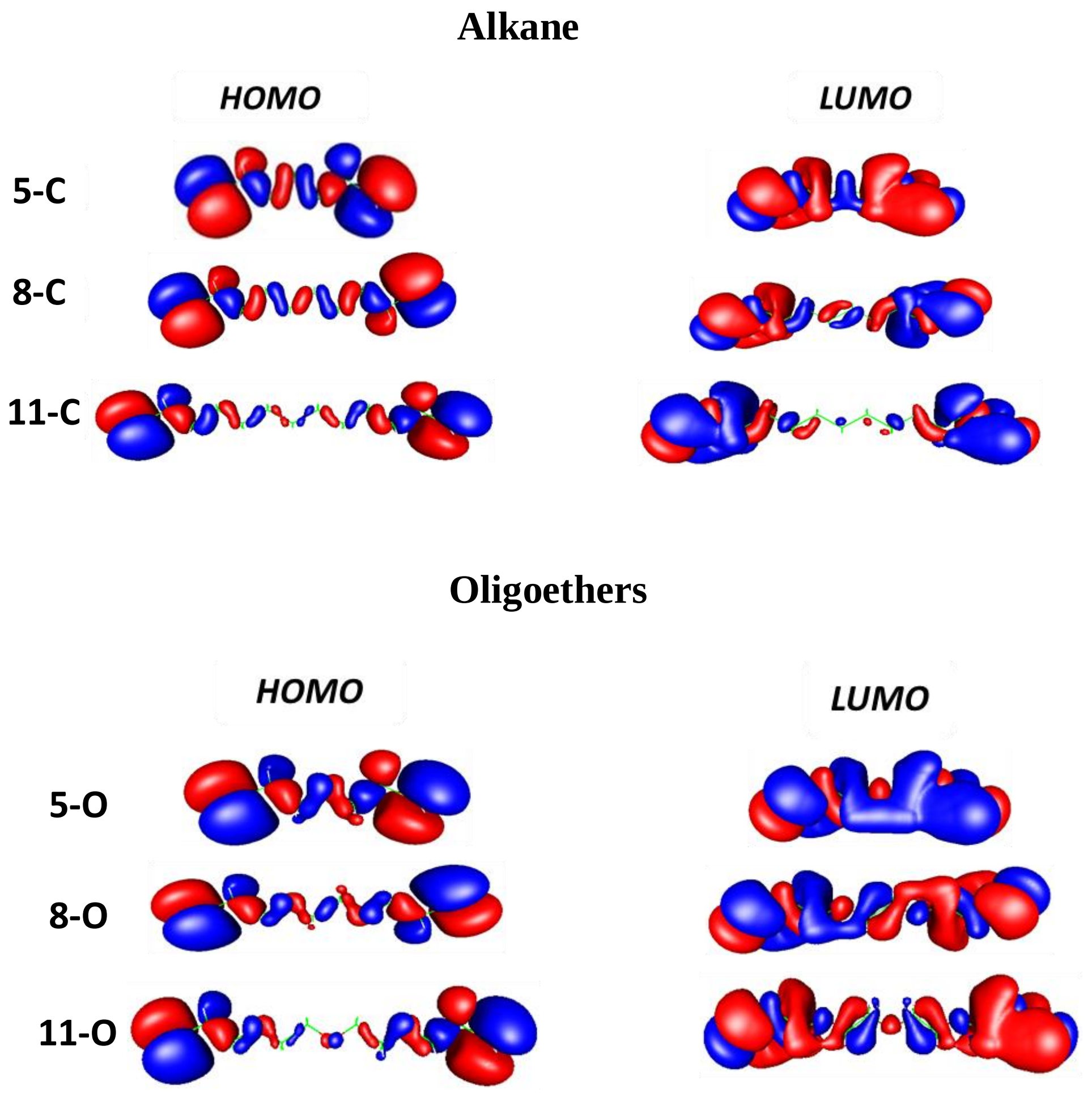

**Figure 5.8.** The distributions of the HOMO and LUMO orbital obtained using the DFT code SIESTA. Red corresponds to positive and blue to negative regions of the wave functions, shown for alkane (5-C, 8-C, and 11-C), and oligoethers (5-O, 8-O, and 11-O).

Figure 5.8, shows the distribution of molecular orbitals for all alkane and oligoethers components, in alkane group. Both the HOMO and LUMO are do localized across the molecules for short lengths. The wavefunctions are reduced in the bridge with increasing molecule length and more localized on the anchor groups. The density of states for both the HOMO and the LUMO have large amplitudes around the anchor groups, and a little density on the bridge specially for the long length. This means the bridge acts as a tunnel barrier between the thiol anchors. This barrier increased with the molecule length and that explains, the decreased of the conductance with the increased the molecule length.

Furthermore, the distribution of HOMO and LUMO in the bridge shws that, the alkane and oligoerthers have the odd-even effected.

## 5.5. Long length molecules

What happened if we increased the length of molecules more? And what will change in the conductance and beta? To answer these questions, I increased the alkane chains length to (14-C, 17-C and 20-C) and the oligoether (14-O, 17-O and 20-O).

The conductance decreases when the molecular length increases in both molecules alkane and oligoether as show in Figure 5.8a-b. Figure 5.8a shows the conductance of alkanes at different lengths. There is a small peak in the right side of HOMO peak in 5C (alkane) curve which becomes smaller until it disappears when the molecular length is increased. This peak is different in oligoether curve, it is higher than the one in alkane curve as we see in Figure 5.5 and also higher than the HOMO peak. There is another peak in the right side the HOMO peak, the height of this peak increased randomly but it is clear that the cause of this peak is the presence of oxygen.

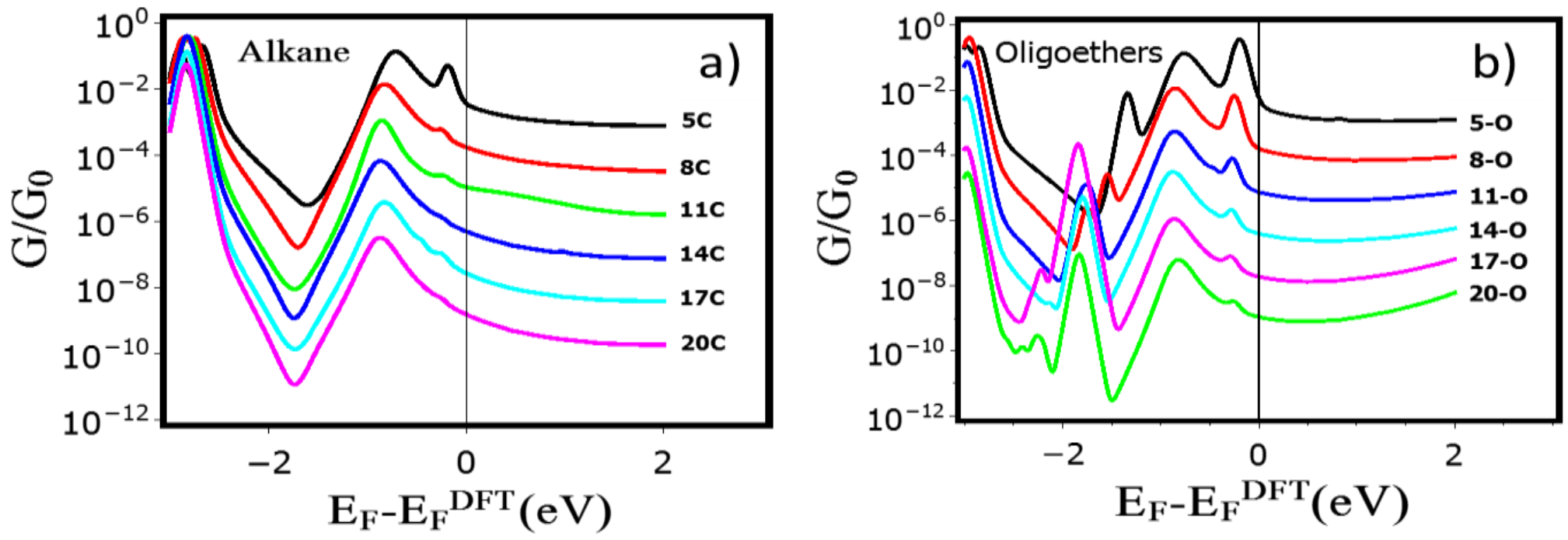

**Figure 5.9**. Shows the comparison between the room temperature electrical conductance of n-Alkane chain and n-oligoethers, where in is the $CH_2$ unit number. (5, 8, 11, 14, 17 and 20).

In the long length molecule, the beta decay for the oligoether is still higher than the beta decay for the alkane as I expected. At Fermi energy zero the $\beta_N$ for the long length alkane is 0.98 per $CH_2$ unit, and also the $\beta_N$ for the oligoether group is 1.03 per $CH_2$ unit. That means the increased length over the 11 units of $CH_2$ does not affected of the beta decay in both alkane and oligoether.

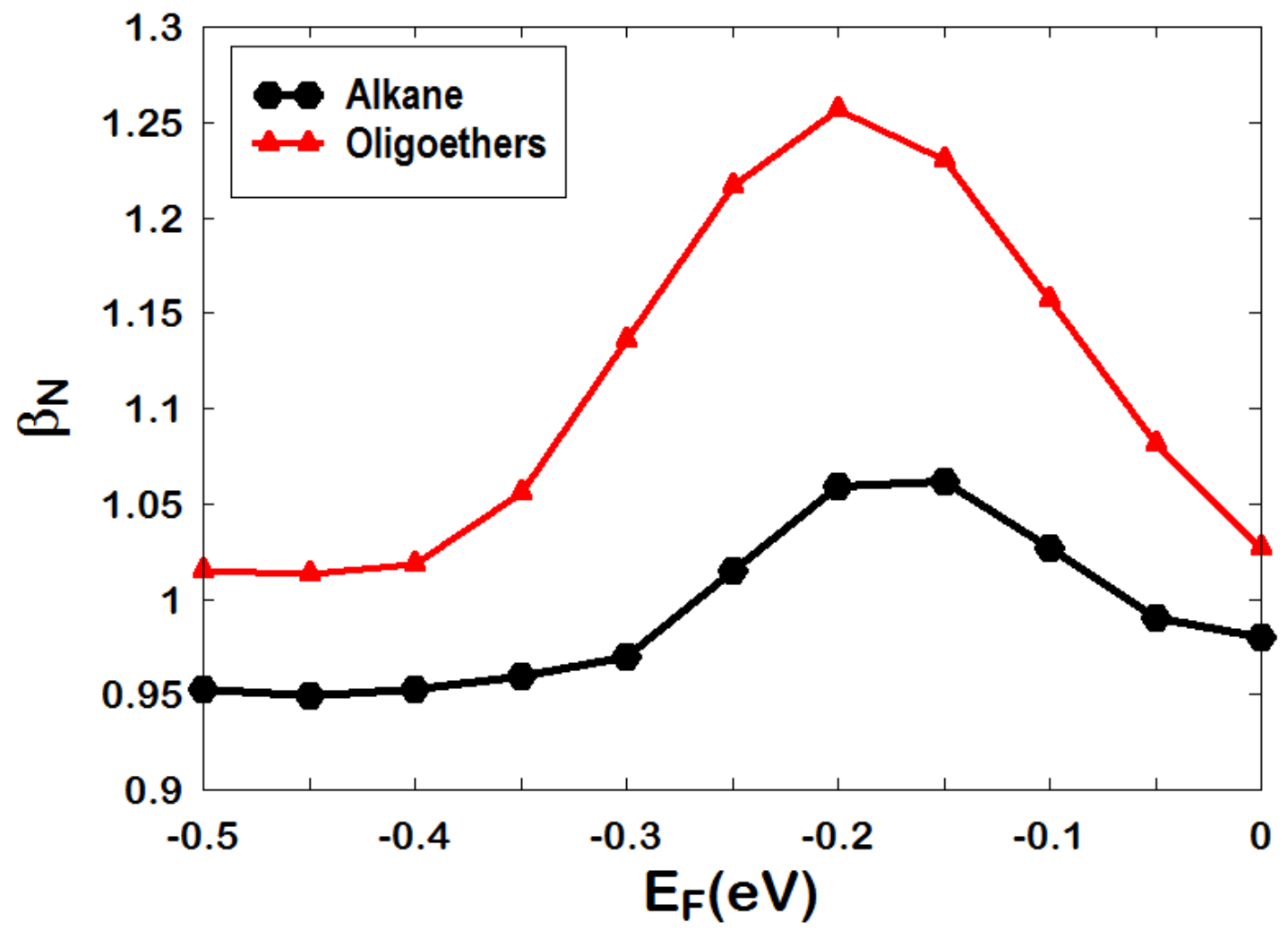

**Figure 5.10**. Tunnelling constant decay $\beta_N$ functional for the long length molecules of range of Fermi energies for N-alkanes in black line, and N-oligethers in red line, where N is the unit number (5, 8, 11, 14, 17 and 20).

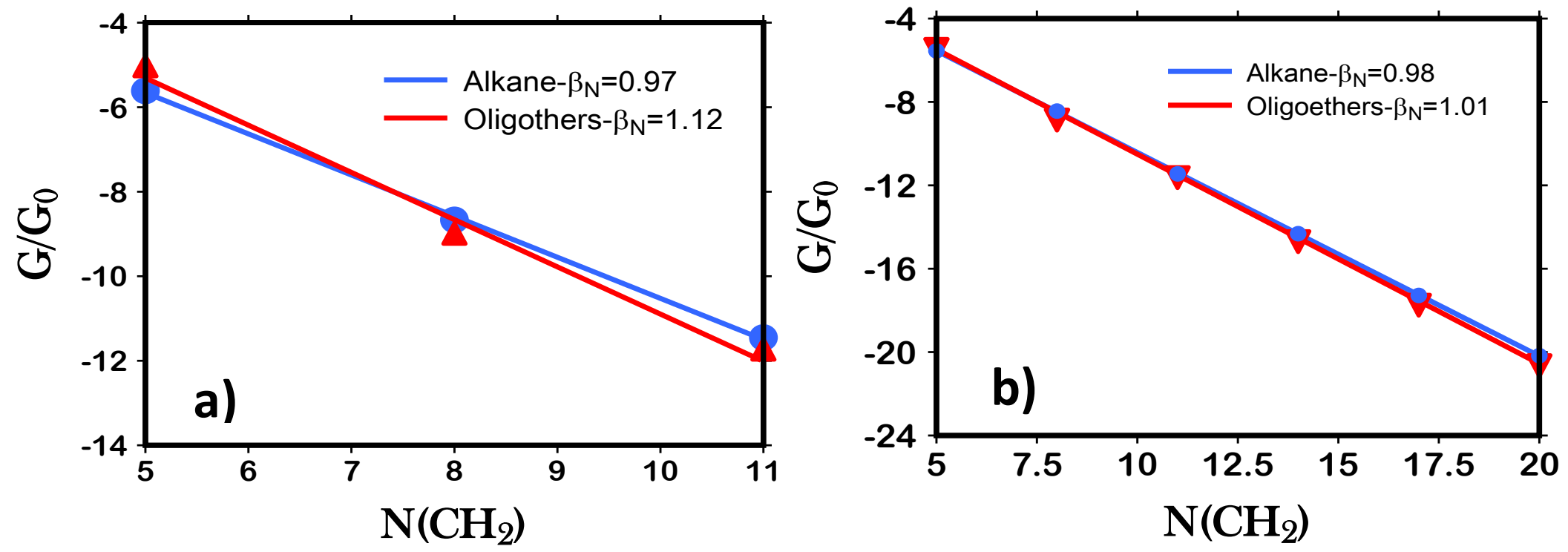

**Figure 5.11**. a) Comparison the logarithm of single –molecule conductance versus number of (-CH2) units of alkanes and oligoethers. The blue line represent alkane and red line represent oligethers, where a) Short length (5C, 8C and 11C), and B) long length (5C-20C).

As you see in Figure 5.12, the distribution of molecular orbitals for all alkane and oligoethers components for the length over 11 unites, in alkane group both HOMO and LUMO fully localized on the anchor group. The HOMO localization for oligoethers almost the same with the alkane but the LUMO still distribution across the molecules as in the short length in the previous section.

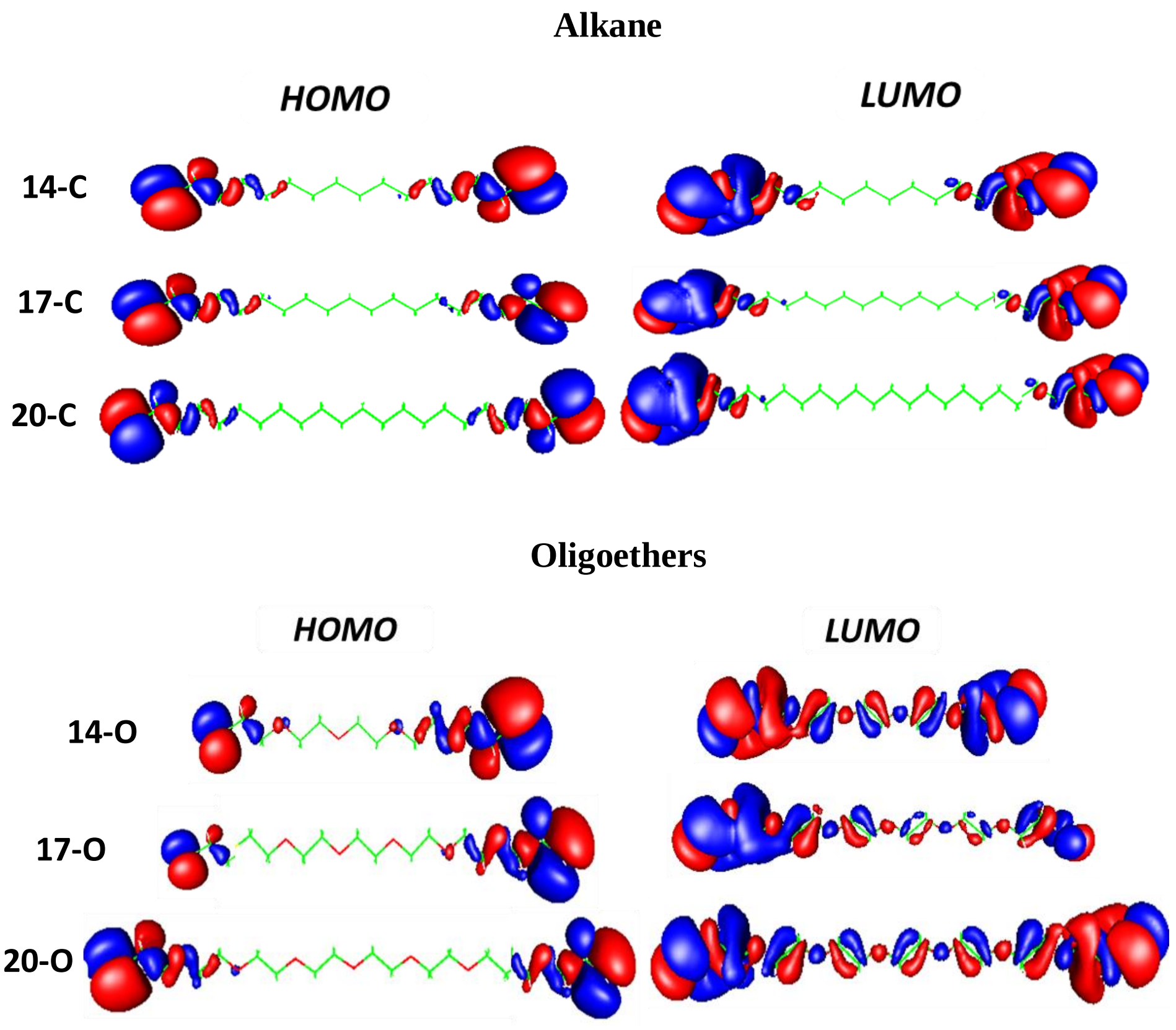

**Figure 5.12.** The distributions of the HOMO and LUMO orbital for the long chain, obtained using the DFT code SIESTA. Red corresponds to positive and blue to negative regions of the wave functions, shown for alkane (14-C, 17-C, and 20-C), and oligoethers (14-O, 17-O, and 20-O).

## Comparison the Decay Values $\beta_N$ for Alkane with different anchor group:

| Anchor group | | $\beta_N$ | Reference |
|---|---|---|---|
| **Thiol** | -SH | 0.98 | Our work(Theory) |
| | | 1.02 | Experiment, HC[10] |
| | | 1.08 | Experiment, LC[10] |
| | | 1.07 | Experiment H&L, [11] |
| | | 0.96 | Experiment L, [9] |
| | | 0.94 | Experiment M, [9] |
| | | 0.88 | Theory [9] |
| | | | |
| **Carboxylic acid** | -COOH | 0.81 | previous work in [11] |
| | | 0.81±0.01 | Experiment, HC[10] |
| | | 0.77 | Experiment, LC[10] |
| | | 0.78±0.07 | Experiment, [17] |
| | | | |
| **Amine** | $-NH_2$ | 0.71 | previous chapter 4 |
| | | 0.81 | Experiment, HC[10] |
| | | 0.88 | Experiment, LC[10] |
| | | 0.82 | Theory[14] |
| | | 0.98 | Theory[15] |
| | | 0.91 | Experiment [16] |

## Comparison the Decay Values $\beta_N$ for different molecules group:

| Molecule | | | |
|---|---|---|---|
| alkanedithiols | 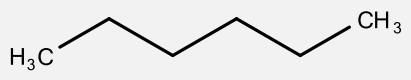  | 0.94±0.06 $A^{0-1}$ | Theory [20] |
| | | 0.73 $A^{0-1}$ | Experimet[18] |
| **phenylenedithiols** | 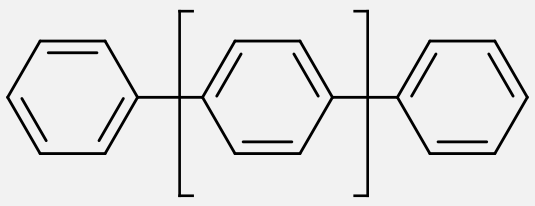 | 0.42±0.07 $A^{0-1}$ | Theory [20] |
| | | 0.42±0.07 $A^{0-1}$ | |
| **OPE** | 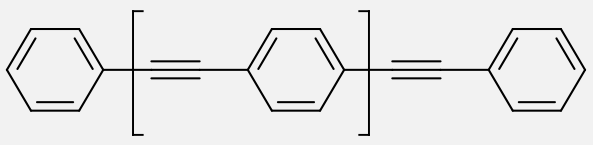 | 0.19 | Theory [18] |
| **OPV** | 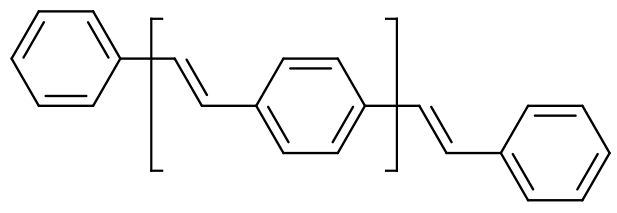 | 0.17 | Theory [18] |
| **PF** | 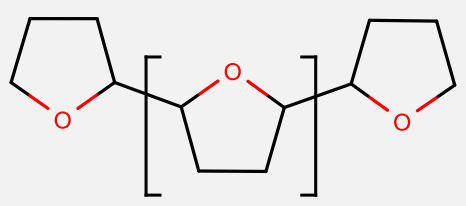 | 0.16 | Theory [18] |
| **ct-PA** | 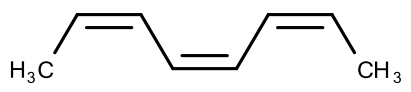 | 0.13 | Theory [18] |
| **t-PA** | 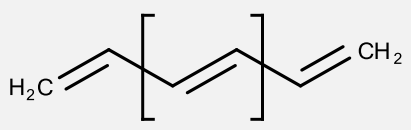 | 0.036 | Theory [18] |

## 5.6. Conclusion

The theoretical and experimental single molecule conductance of alkanes and their corresponding oligoether show that molecular conductance of alkane chains is greater than that of oligoether chains. The oligoether chains possess a higher $\beta$ factor than alkane chain.

## 5.7. Reference

1. J.M. Soler, E. Artacho, J.D. Gale, A. García, J. Junquera, P. Ordejón, D. Sánchez-Portal, The SIESTA method for ab inito order-N materials simulation, Journal of Physics: Condensed Matter, 2002, 14, 2745.

2. Troullier, N. and Martins, J.L., 1991. Efficient pseudopotentials for plane-wave calculations. *Physical review B*, *43*(3), p.1993.

b) Perdew, J.P., Burke, K. and Ernzerhof, M., 1996. Generalized gradient approximation made simple. *Physical review letters*, *77*(18), p.3865.

2. Ferrer, J., Lambert, C.J., García-Suárez, V.M., Manrique, D.Z., Visontai, D., Oroszlany, L., Rodríguez-Ferradás, R., Grace, I., Bailey, S.W.D., Gillemot,

K. and Sadeghi, H., 2014. GOLLUM: a next-generation simulation tool for electron, thermal and spin transport. *New Journal of Physics*, *16*(9), p.093029.

3. Lambert, C.J., 2015. Basic concepts of quantum interference and electron transport in single-molecule electronics. *Chemical Society Reviews*, *44*(4), pp.875-888.

4. Wierzbinski, E., Yin, X., Werling, K. and Waldeck, D.H., 2012. The Effect of Oxygen Heteroatoms on the Single Molecule Conductance of Saturated Chains. *The Journal of Physical Chemistry B*, *117*(16), pp.4431-4441.

5. Xu, B. and Tao, N.J., 2003. Measurement of single-molecule resistance by repeated formation of molecular junctions. *Science*, *301*(5637), pp.1221-1223.

6. Li, X., He, J., Hihath, J., Xu, B., Lindsay, S.M. and Tao, N., 2006. Conductance of single alkanedithiols: conduction mechanism and effect of molecule-electrode contacts. *Journal of the American Chemical Society*,*128*(6), pp.2135-2141.

7. Li, C., Pobelov, I., Wandlowski, T., Bagrets, A., Arnold, A. and Evers, F., 2008. Charge transport in single Au| alkanedithiol| Au junctions: coordination geometries and conformational degrees of freedom. *Journal of the American Chemical Society*, *130*(1), pp.318-326.

8. Chen, F., Li, X., Hihath, J., Huang, Z. and Tao, N., 2006. Effect of anchoring groups on single-molecule conductance: comparative study of thiol-, amine-, and carboxylic-acid-terminated molecules. *Journal of the American Chemical Society*, *128*(49), pp.15874-15881.

9. Li, X., He, J., Hihath, J., Xu, B., Lindsay, S.M. and Tao, N., 2006. Conductance of single alkanedithiols: conduction mechanism and effect of molecule-electrode contacts. *Journal of the American Chemical Society*,*128*(6), pp.2135-2141.

a) Quek, S.Y., Kamenetska, M., Steigerwald, M.L., Choi, H.J., Louie, S.G., Hybertsen, M.S., Neaton, J.B. and Venkataraman, L., 2009. Dependence of single-molecule junction conductance on molecular conformation. Nat. Nanotechnol, 4, p.230. b) Hong, W., Manrique, D.Z., Moreno-Garcia, P.,

Gulcur, M., Mishchenko, A., Lambert, C.J., Bryce, M.R. and Wandlowski, T., 2012. Single molecular conductance of tolanes: experimental and theoretical study on the junction evolution dependent on the anchoring group. Journal of the American Chemical Society, 134(4), pp.2292-2304. c) Li, Y., Baghernejad, M., Qusiy, A.G., Zsolt Manrique, D., Zhang, G., Hamill, J., Fu, Y., Broekmann, P., Hong, W., Wandlowski, T. and Zhang, D., 2015. Three-State Single-Molecule Naphthalenediimide Switch: Integration of a Pendant Redox Unit for Conductance Tuning. Angewandte Chemie International Edition, 54(46), pp.13586-13589.

10. Liu, L., Zhang, Q., Tao, S., Zhao, C., Almutib, E., Al-Galiby, Q., Bailey, S.W., Grace, I., Lambert, C.J., Du, J. and Yang, L., 2016. Charge transport through dicarboxylic-acid-terminated alkanes bound to graphene–gold nanogap electrodes. *Nanoscale*, *8*(30), pp.14507-14513.

11. Prodan, E. and Car, R., 2008. Tunneling conductance of amine-linked alkyl chains. *Nano letters*, *8*(6), pp.1771-1777.

12. Fagas, G. and Greer, J.C., 2007. Tunnelling in alkanes anchored to gold electrodes via amine end groups. *Nanotechnology*, *18*(42), p.424010.

13. Venkataraman, L., Klare, J.E., Tam, I.W., Nuckolls, C., Hybertsen, M.S. and Steigerwald, M.L., 2006. Single-molecule circuits with well-defined molecular conductance. *Nano Letters*, *6*(3), pp.458-462.

14. Martín, S., Haiss, W., Higgins, S., Cea, P., Lopez, M.C. and Nichols, R.J., 2008. A comprehensive study of the single molecule conductance of α, ω-dicarboxylic acid-terminated alkanes. *The Journal of Physical Chemistry C*,*112*(10), pp.3941-3948.

15. Liu, H., Wang, N., Zhao, J., Guo, Y., Yin, X., Boey, F.Y. and Zhang, H., 2008. Length-Dependent Conductance of Molecular Wires and Contact Resistance in Metal–Molecule–Metal Junctions. *ChemPhysChem*, *9*(10), pp.1416-1424.

16. Kaun, C.C. and Guo, H., 2003. Resistance of alkanethiol molecular wires.*Nano letters*, *3*(11), pp.1521-1525.

# Chapter 6

# 6. Toward transport self-assembly monolayer

## 6.1. Molecular dynamics simulation

Molecular dynamics (MD) is a computational simulation technique used to study the physical behaviour and interactions of atoms and molecules in a fixed period of time which gives an insight to the dynamical development of the system [1]. In this chapter, I will spotlight a new kind of a self-assembled monolayer (SAM) [2, 3], which will present a new direction to understand the transport through SAMs. For example: if you want to make a thermal electric device, you would like the total current to be large, but the current through one molecule is never large. However, if many molecules are placed in parallel to create a self-assembled monolayer, one can achieve a huge value of current. Therefore, as the first step to understanding how a SAM can form, I am going to perform molecular dynamics simulations (MD) to examine the molecular assembly of two amphiphilic candidate molecules for graphene based molecular electronics. The first has a single pyrene anchor, Pyrene-PEGn-exTTF (PPT) and the other has three pyrene anchors, tri-pyrene (TPPT). Both will be assembled on a disordered graphene surface. The adsorption packing of PPT and TPPT onto the graphene surface will be demonstrated both in aqueous solvent and vacuum condition.

For many years most researchers who studied classes of SAMs prefer to use metals such as gold [4], silver [5], copper [5], and mercury [6]. Graphene is hydrophobic with low solubility in aqueous solution [7] and has attracted enormous interest in many different fields due to its unique properties [8]. It is harder than diamond, and 200 times stronger than steel,1 million times thinner than a human hair, more conductive than copper and more flexible than rubber. This miracle material, promises to [9] transform the technology. Graphene has many potential applications, it could use to make better batteries, medical scanners, transistor, nanomaterials, sensors, and computers.

Functionalization of graphene [10, 11] can improve its dispersion in solvents and extend its applications. Graphene can be functionalized by using various conjugated compounds, such as polyethylene glycol (PEG) [12], aromatic compounds [13], polyacetylenes [14], and pyrene–polyethylene glycol (Py–PEG) polymer [15].

Before I discuss the methodology and the project result, I will introduce some important definitions I will use it in this chapter. A *micelle* is an aggregate (or assembly) of surfactant molecules dispersed in a liquid colloid. There are two types of micelle, the normal-phase micelle (oil-in-water micelle) which have the hydrophilic head contact with aqueous solution and hydrophobic single-tail regions in the micelle centre see Figure 6.1a. The other type is referred to as Inverse micelles which have the head groups at the centre with the tails extending out (water-in-oil micelle) Figure 6.1b.

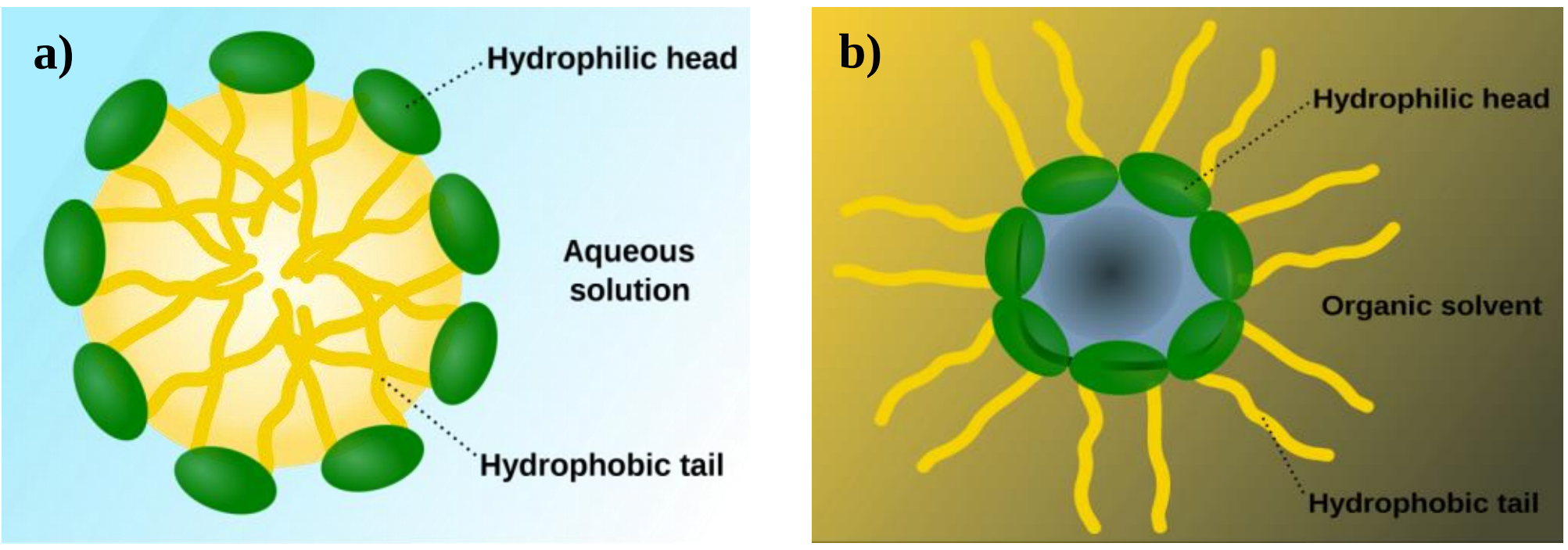

**Figre 6.1.** The tow type of micell: a) normal-phase micelle, and b) Inverse micelles. These two figure take it from Micelle_scheme-en.svg.

Dr. Beatriz M. Illescas y from Dept. de Quimica Organica Universidad de Computense, has designed a new type of amphiphilic molecule with two main groups, pyrene which is the hydrophobic (water hating) head and exTTF which is hydrophilic (water loving) tail. These two groups are connected by the PEG chain which has been used to create hydrophilic surfaces [16, 17] Figure 6.2a-b, shows our two molecules

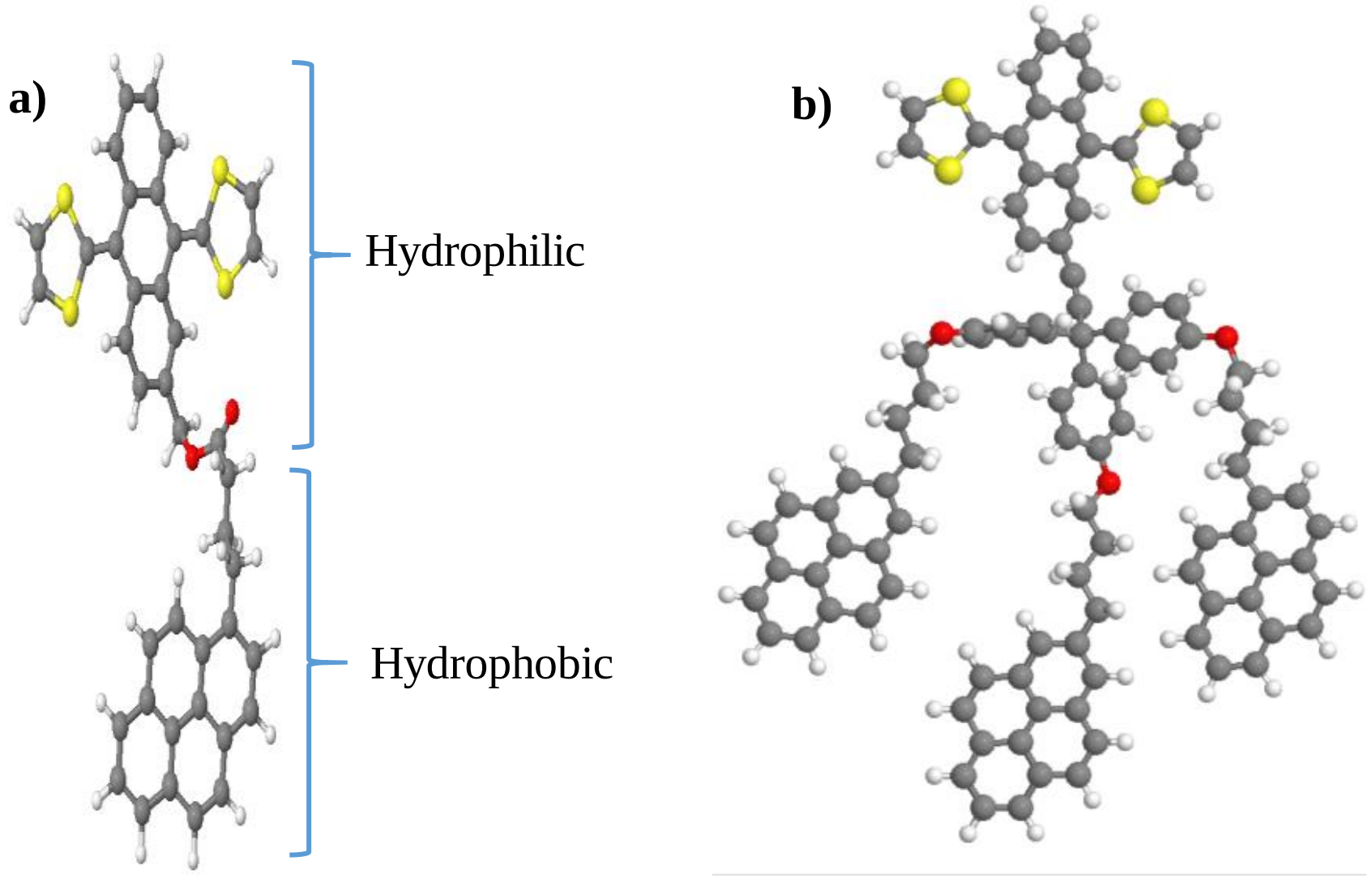

**Figuer 6.2.** amphiphilic a bifunctional molecule, pyrene-exTTF, a) with a unit of pyrene, and b) Three untis of pyrene.

### 6.1.1. Methodology

I used the molecular dynamics (MD) package DLPOLY_4 [18] to investigate the interaction of Py–PEG-exTTF(PPT) and Triple-Py–PEG-exTTF(TPPT) with a periodic graphene sheet, in vacuum and aqueous phases. The Dreiding force field was carefully adapted to model the system where the charge on each atom was obtained from ab-initio density functional theory (DFT) by using the SIESTA [19] package. The simulations have been carried out at room temperature (300 K) and atmospheric pressure using the DL_POLY package [18], putting the molecules and water molecules in a cubic box and applying periodic boundary conditions in three dimensions. In the aqueous phase the TIP3P water model was employed with an (NVT) ensemble for both phases, using the Nose-Hoover thermostat [20, 21].

Aggregation of six PPT and TPPT molecules on the graphene surface (GS) in vacuum and in water are shown below for in-plane and through plane views at 300K and 1 atmos. The snapshots show 4 periodic sheets and the water is omitted for clarity.

a) b)

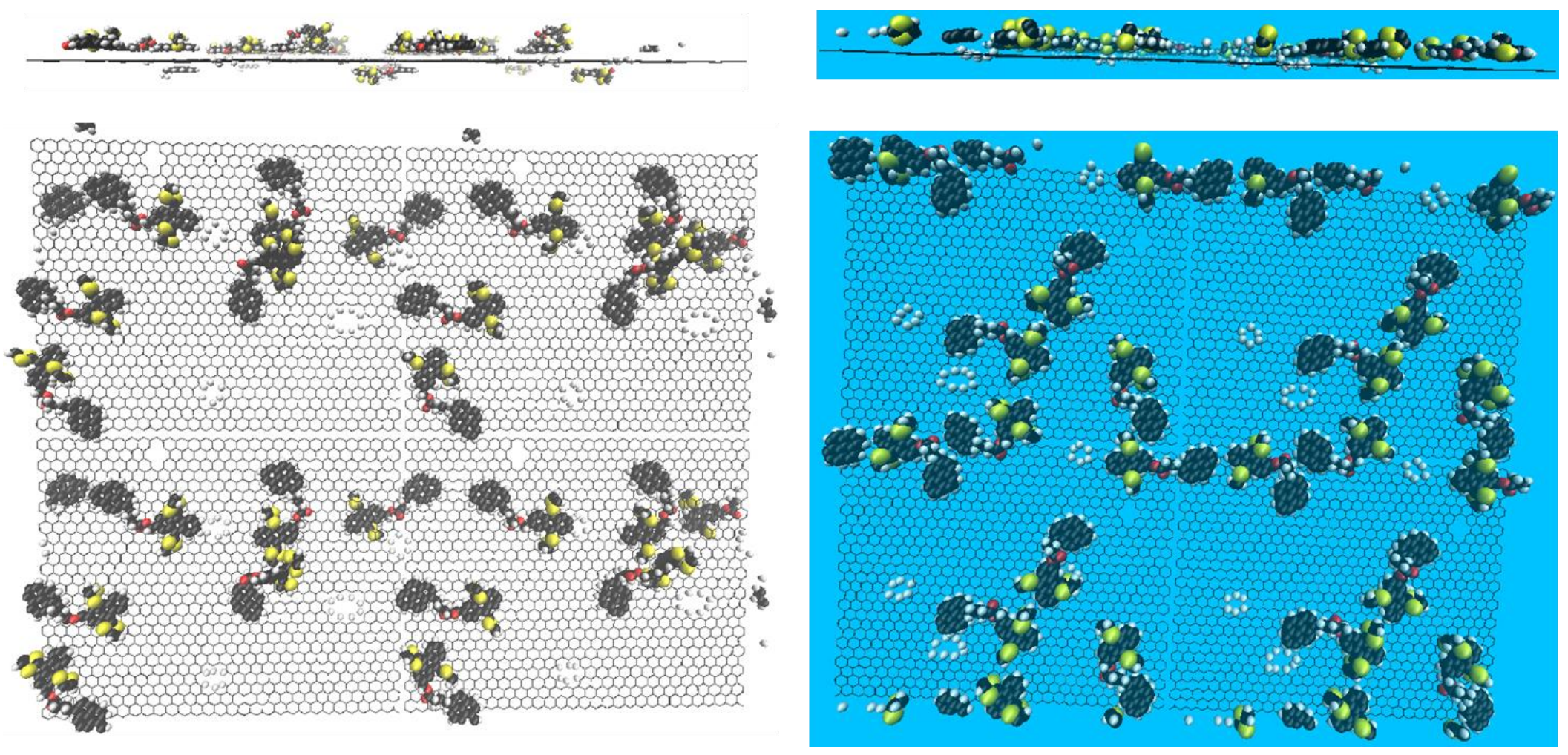

**Figure 6.4.** six molecules of pyrene-PEGn-exTTF (PPT) onto graphene sheet, a) in vacuum, and b) in aqueous solution.

The aggregation of six PPT molecules on the GS are compared in vacuum and aqueous phases. At this concentration in vacuum one PPT is ejected from the surface and in both phases the molecules are randomly distributed over the surface in a monolayer, Figure 6.4.

a) b)

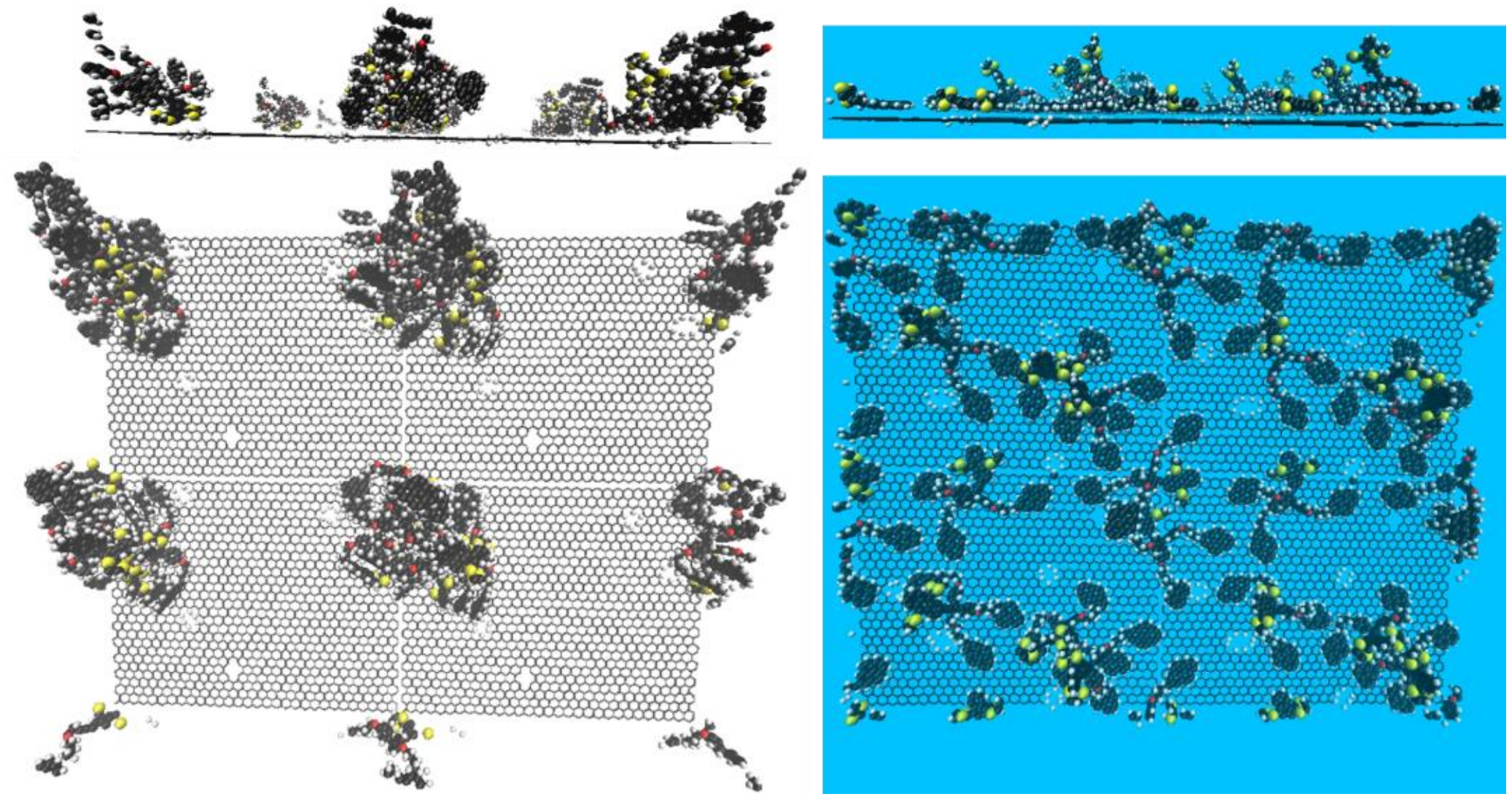

**Figure 6.5.** six molecules of triple-pyrene-PEGn-exTTF (TPPT) onto graphene sheet, a) in vacuum, and b) in aqueous solution.

The aggregation of six TPPT molecules on the GS are compared in vacuum and aqueous phases. The six TPPT molecules in vacuum form semicircular micelles on the surface, which after adding water are, randomly dispersed as the water breaks the micelle apart, Figure 6.5.

Molecular dynamics simulations were used to investigate the aggregation of PPT and TPPT molecules on a disordered GS. The pyrene anchors are found to be attached to the GS for both molecules but can be displaced. The PPT are randomly distributed over the GS but the TPPT form conical surface micelle structures. In the absence of water, the RDF plots show that raising the temperature changes the surface configurations for both PPT and TPPT. For example, at 300K in the case of PPT both head group and pyrene lie flat on the surface whereas for the TPPT case surface bound 3-D micelles form.

## 6.2. The electrical properties of a single molecules pyrene-exTTF

The π-extended tetrathiafulvalene (exTTF) has been reported as strong donor character in thousands of papers such as [24, 25, 26]. It has remarkable applications in several interesting fields, including covalent and supramolecular ensembles, such as molecular wires [26], artificial photosynthetic systems as well as photovoltaic devices. The exTTF terminated by pyrene will be the accepter group of our amphiphilic molecules PPT and TPPT. The pyrene anchor will provide strong p-interaction with carbon surfaces [16, 27]. The two heads (exTTF and pyrene) are connected by the hydrophilic PEG chain which has been used to create hydrophilic surfaces [16, 17], and it will be useful to improve the solubility of pyrene-labeled polymer [28]. PEG can also use for exfoliation of graphite oxide to produce graphene sheets [29].

In this section, I study the electrical properties of a single molecule of pyrene-exTTF (PPT) and TPPT with graphene based electrodes. First, I use the MD simulation to study the movement or (the best orientation) of the molecules. As shown in Figure 6.6a-b, PPT adsorbs onto the surface of the graphene and lays horizontally on the surface. This means there are strong interactions between the molecule and graphene and that could be different if we change the PEGn chain length. In the other molecules of TPPT, we found there are interactions between pyrenes and the molecules coil themselves.

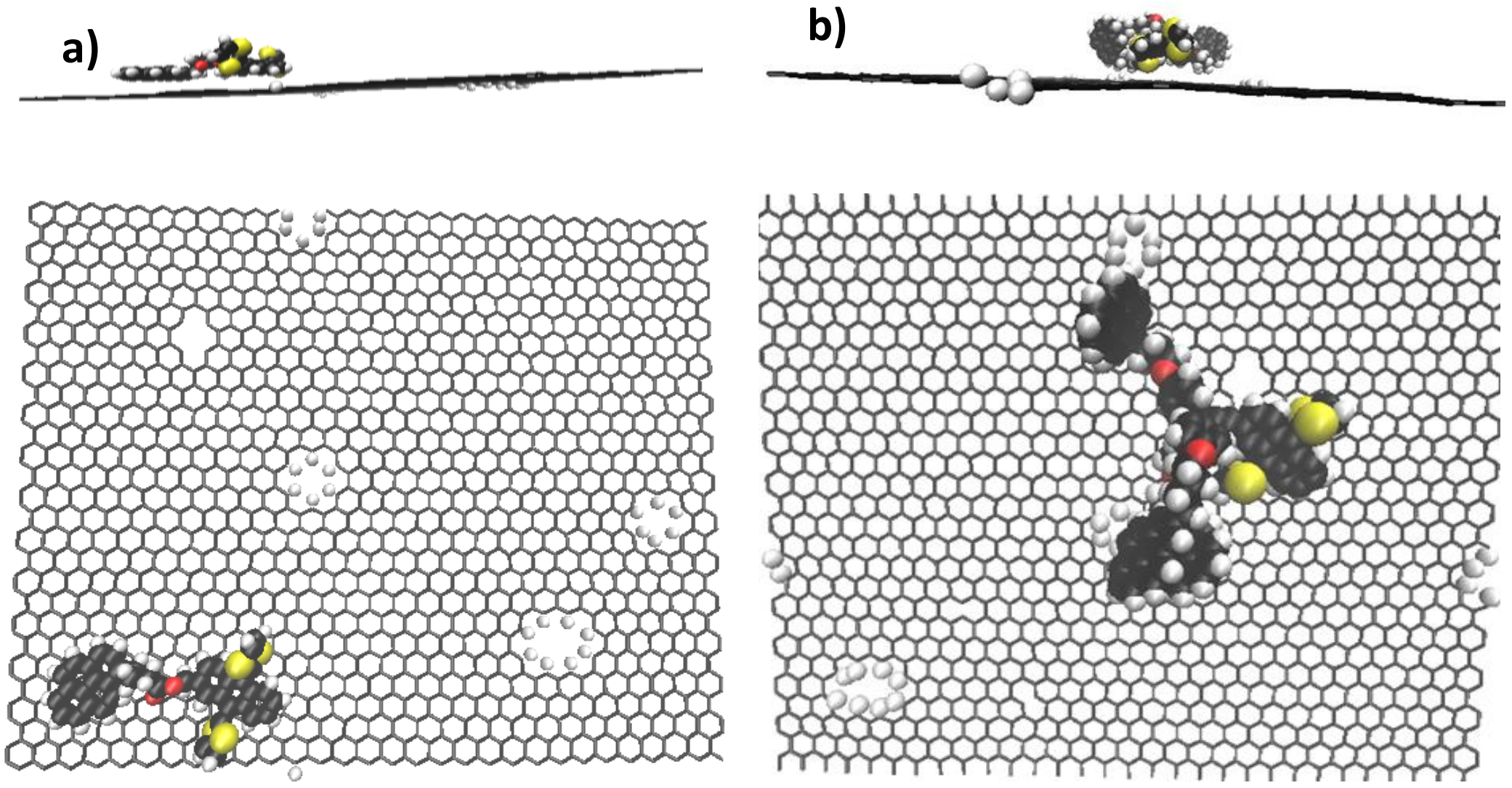

**Figure 6.6.** A snapshot taking by MD(dl-ploly) [18] showing the single molecule on a graphene sheet simulated in vacuum at room temperature a) shows how the PPT which has a unit pyrene lies flat on the GS, and in b) The TPPT with triple-pyrenes form conical surface micelles onto GS.

After obtaining a clear idea about how the molecules interact with the GS, I calculated the electrical properties of the molecules in Figure 6.7a, with two flat graphene electrodes. The density functional theory (DFT) code SIESTA [19] used to relaxed geometry of each isolated molecule which employs Troullier-Martins pseudopotentials to represent the potentials of the atomic cores and a local atomic-orbital basis set. I used a double-zeta polarized basis set for all atoms and the local density functional approximation (LDA-CA) by Ceperley and Adler [31]. The Hamiltonian and overlap matrices are calculated on a real -space grid defined by a plane-wave cutoff of 150 Ry.

After obtaining the relaxed geometry of an isolated molecule, I use the π-π bonding method to connect the molecule with the graphene electrodes as shown in Figure 6.7-b. The molecules plus electrodes were allowed to further relax to yield the optimized structures shown in Figures 6.7b-c. The GOLLUM method [32] was used to compute the transmission coefficient $T(E)$ for electrons of energy $E$ passing from the left electrode to the right graphene electrode. Once the $T(E)$ is computed, I calculated the zero-bias electrical conductance $G$ using the Landauer formula:

$$G = \frac{I}{V} = G_0 \int_{-\infty}^{\infty} dE T(E) \left( -\frac{df(E)}{dE} \right) \tag{6.2}$$

where $G_0 = \left(\frac{2e^2}{h}\right)$ is the quantum of conductance, $f(E)$ is Fermi distribution function defined as $f(E) = e^{(E-E_F)k_BT}$ where $k_B$ is Boltzmann constant and $T$ is the temperature. Since the quantity $-\frac{df(E)}{dE}$ is a normalised probability distribution of width approximately equal to $k_BT$, centred on the Fermi energy $E_F$

, the above integral represents a thermal average of the transmission function $T(E)$ over an energy window of the width $k_B T$ (=25 meV at room temperature) [33].

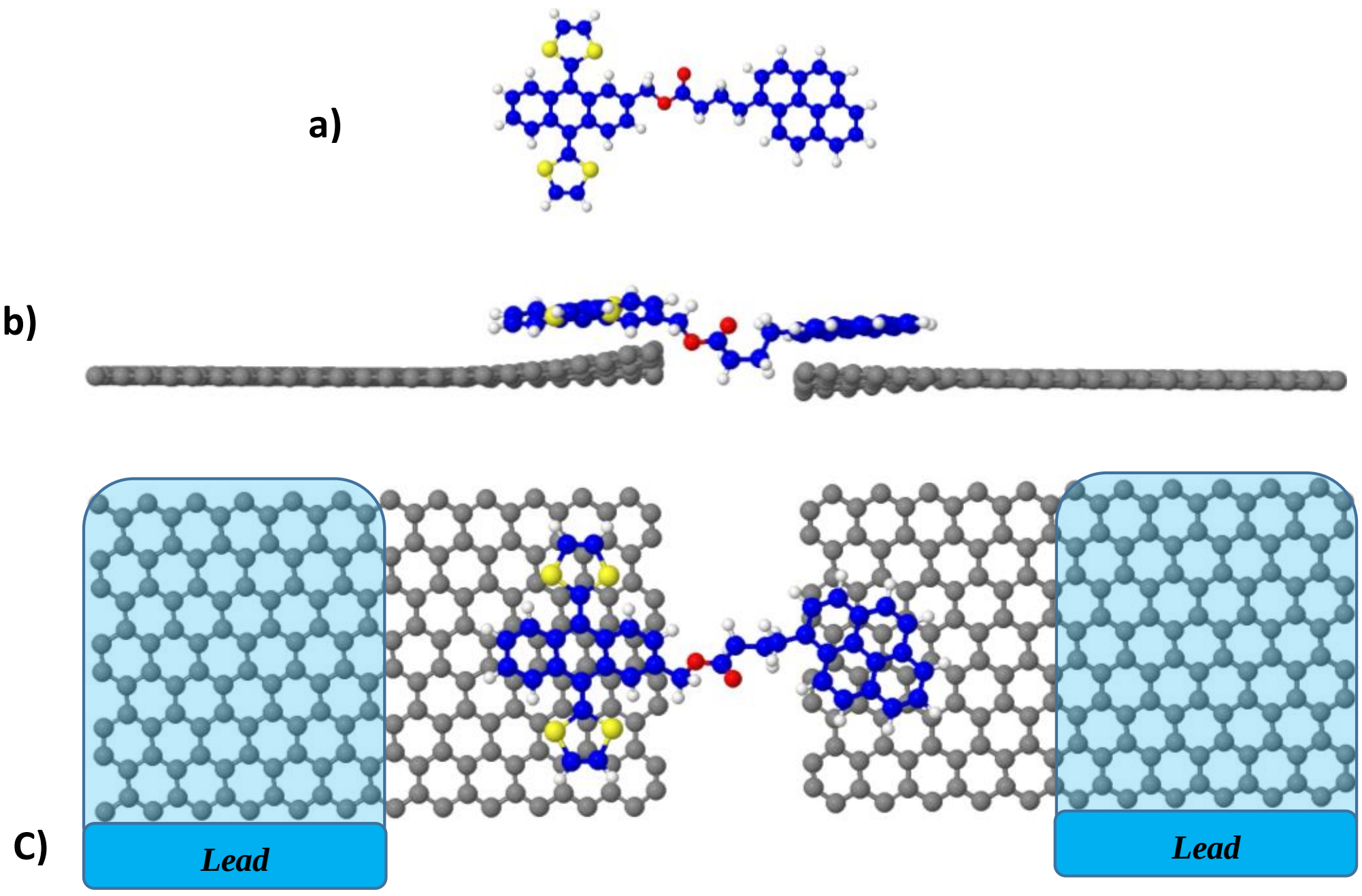

**Figure 6.7.** a) The PPT molecular structure, gray balls are carbon atoms in the graphene sheet, blue balls are carbon atoms in the molecule, red is oxygen, yellow is sulfur, and white is hydrogen, b and c) show the optimized geometries of systems containing the PPT molecule connected to the two graphene electrodes by π-π stacking.

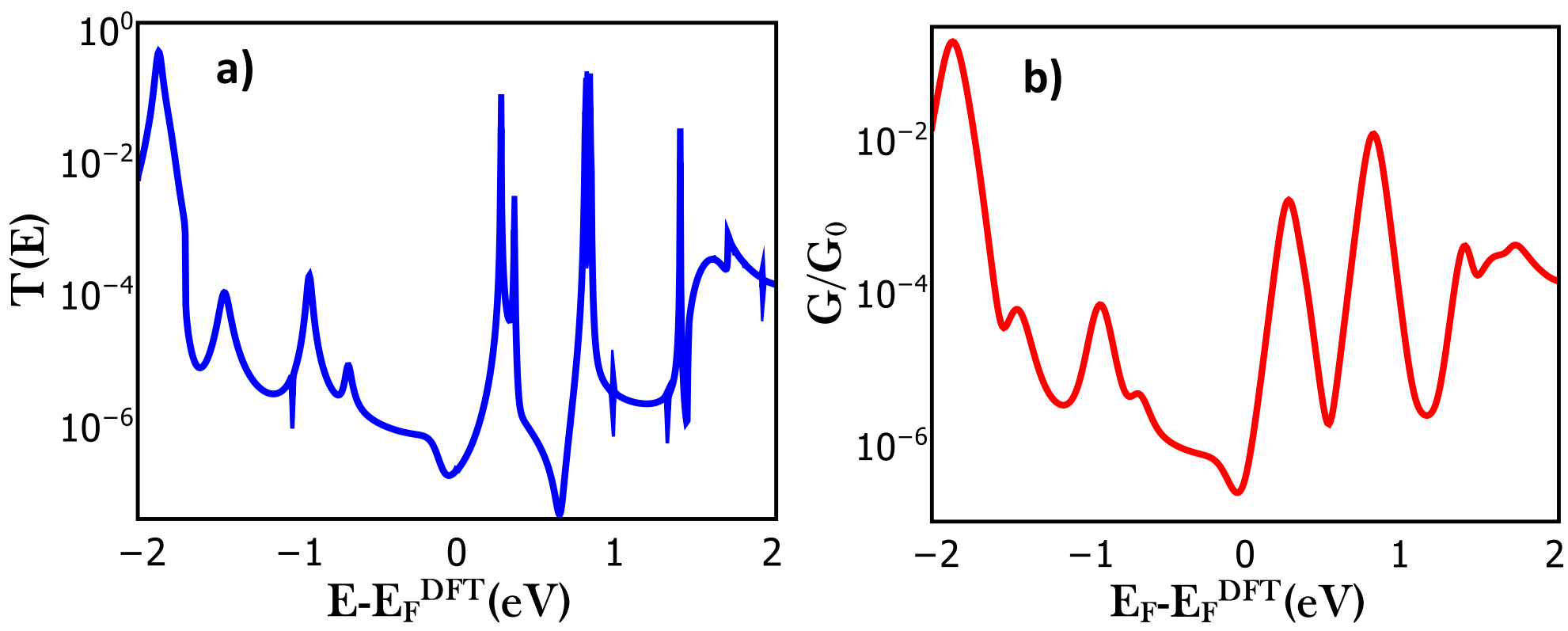

**Figure 6.8.** For the structures in Figure 6.7, a) the transmission coefficients as a function of energy of the systems containing PPT molecule attached to the two graphene electrodes, b) demonstrates the room temperature electrical conductances over a range of Fermi energies.

Figure 6.8a shows the transmission coefficients as a function of energy of the single molecule PPT connected to the two graphene electrodes, see Figure 6.7a-c, exhibit LUMO-dominated transitions at the DFT Fermi energy. While in Figure 6.8b, the room-temperature electrical conductance over a range of Fermi energies $E_F$ in the vicinity of the DFT-predicted Fermi energy $E_F^{DFT}$, which show the low conductance $3.36 \times 10^{-7}$ at $E_F = 0\ eV$.

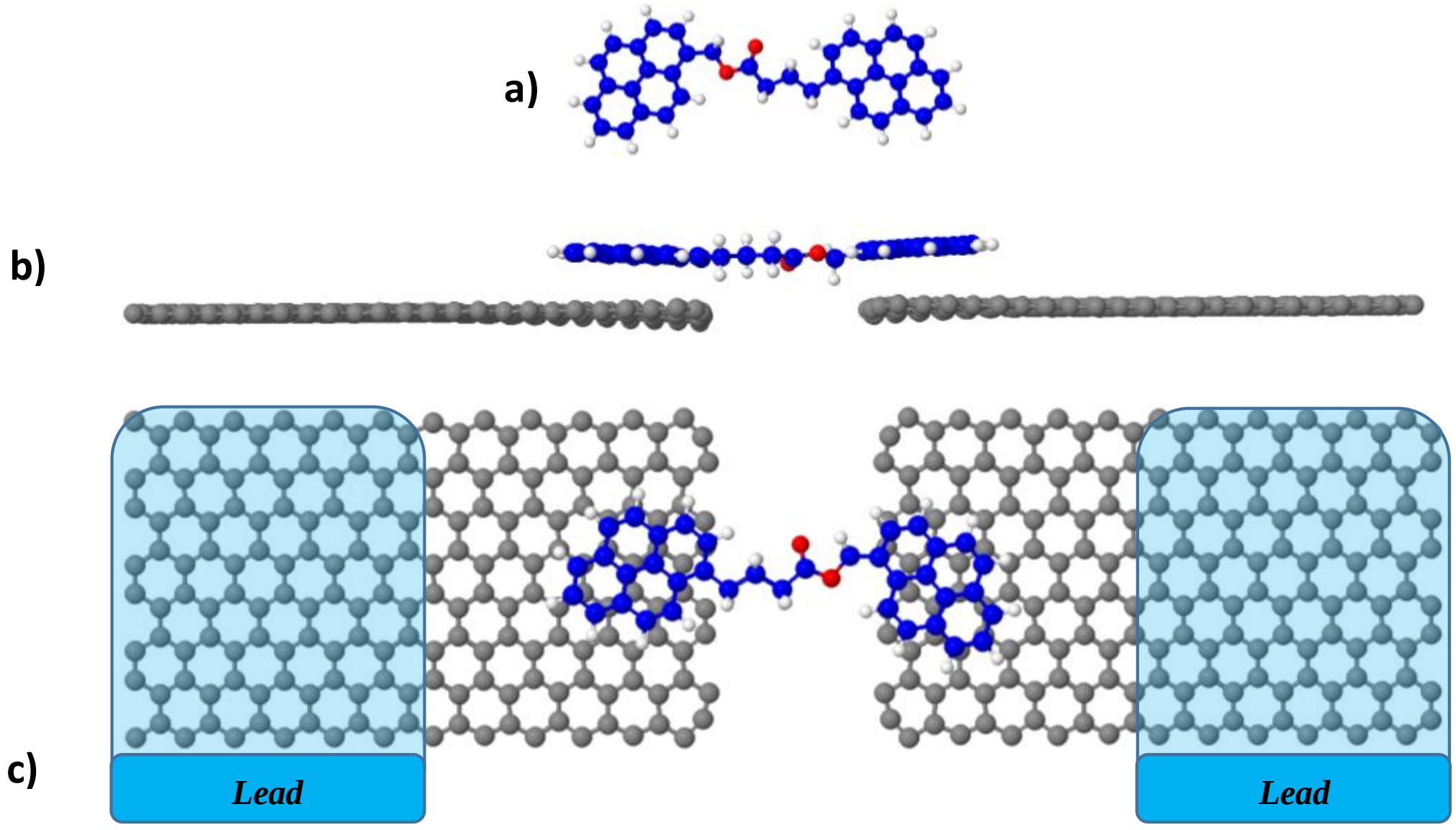

**Figure 6.9.** a) show the PP molecular structure, the gray balls are carbon atoms in the graphene sheet, blue balls are carbon atoms in the molecule, red is oxygen, yellow is sulfur, and white is hydrogen, b and c) show the optimized geometries of systems containing the PP molecule connected to the two graphene electrodes by π-π stacking.

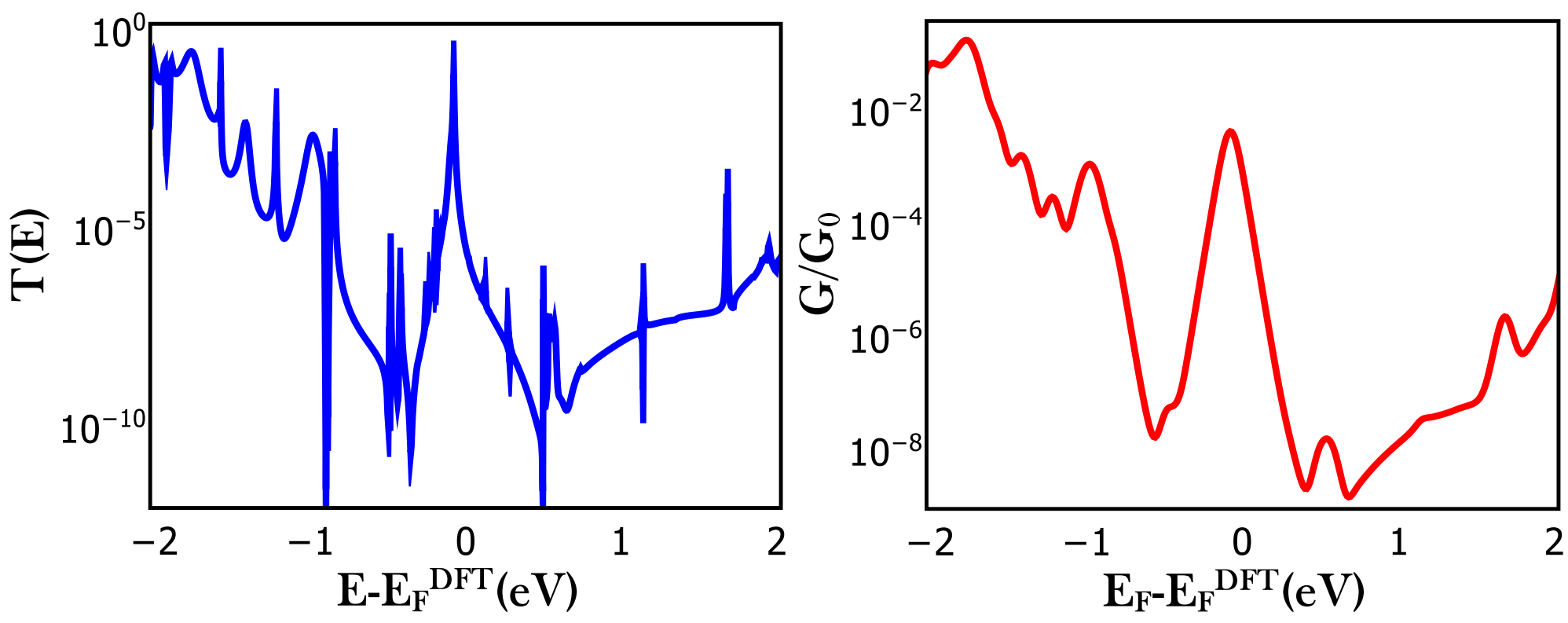

**Figure 6.10.** For the structures in Figure 6.9, a) the transmission coefficients as a function of energy of the systems containing PPT molecule attached to the two graphene electrodes, b) demonstrates the room temperature electrical conductances over a range of Fermi energies.

In order to make systematic comparison, I compared the transmission coefficient and conductance of PPT which is an asymmetric molecule, with the symmetric molecules pyrene-pyrene (PP) and exTTF-exTTF (TT). Figure 6.10a shows the transmission coefficients as a function of energy of the single molecule PPT connected to the two graphene electrodes, exhibit on resonant transmission at the the DFT-predicted Fermi energy $E_F^{DFT}$ and conductance value (G=7.42×10$^{-04}$ G$_0$). Furthermore, the TT conductance curve is on resonance also but, the conductance value is a higher than that for PP (G=3.23×10$^{-03}$G$_0$). To clarify this comparison between the asymmetric molecule PPT and asymmetric molecules PP, TT, I plot the three curves with together in Figure 6.13.

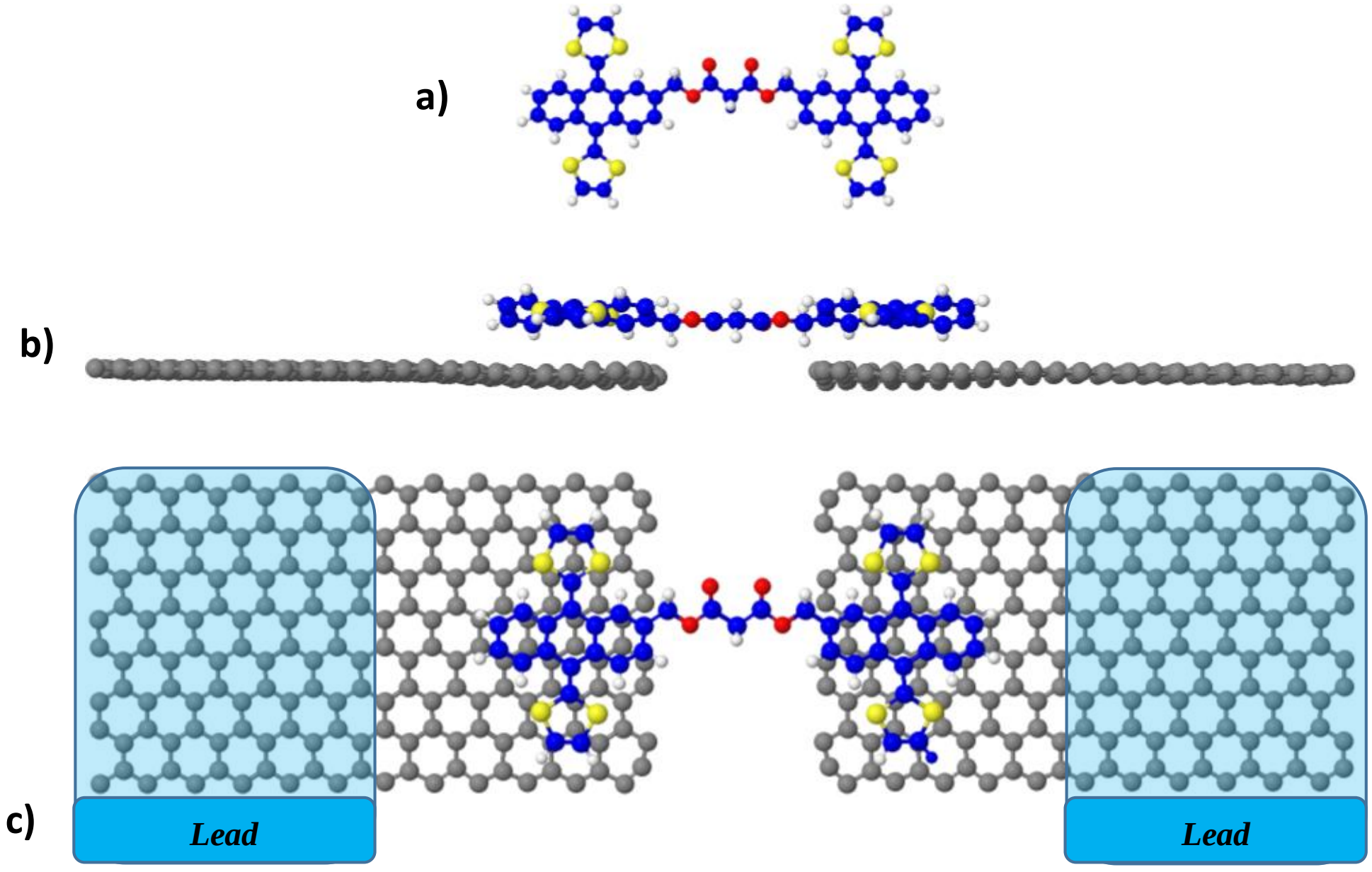

**Figure 6.11.** a) show the TT molecular structure, gray balls are carbon atoms in the graphene sheet, blue balls are carbon atoms in the molecule, red is oxygen, yellow is sulfur, and white is hydrogen, b and c) show the optimized geometries of systems containing the TT molecule connected to the two graphene electrodes by π-π stacking.

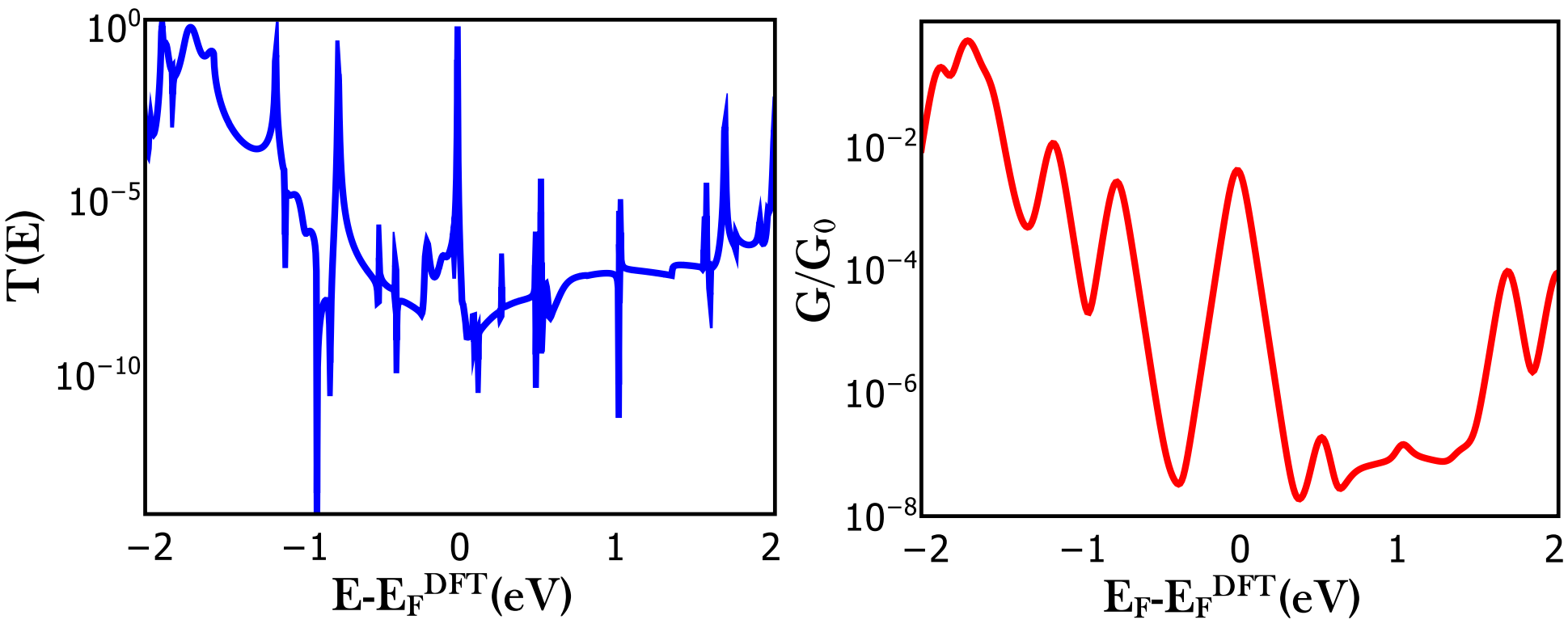

**Figure 6.12.** For the structures in Figure 6.11, a) the transmission coefficients as a function of energy of the systems containing TT molecule attached to the two graphene electrodes, b) demonstrates the room temperature electrical conductances over a range of Fermi energies.

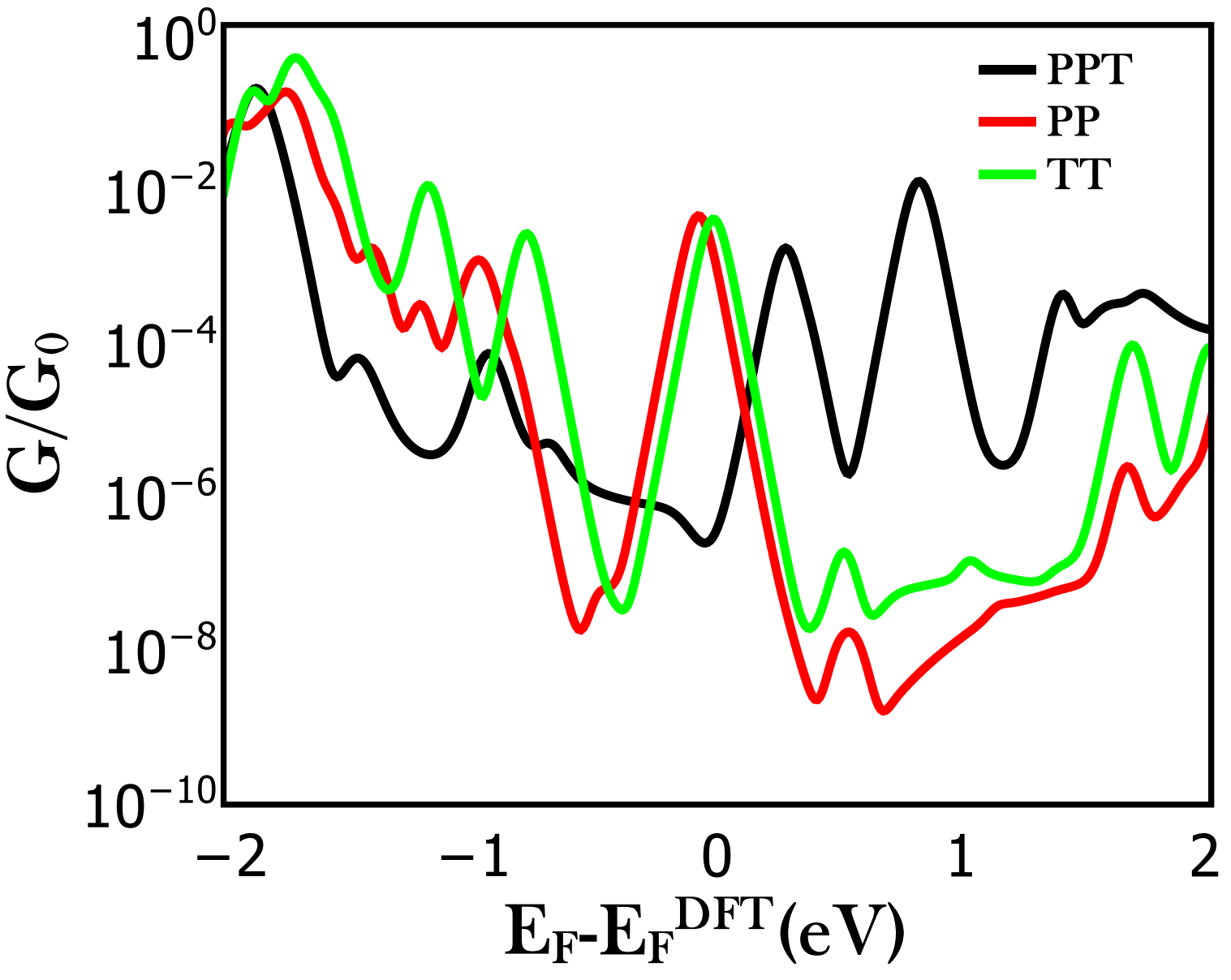

**Figure 6.13.** Comparison between the room temperature electrical conductances over a range of Fermi energies for three situations, PPT is the amphiphilic molecules which has two heads exTTF and pyrene, PP has pyrene-pyrene and TT has exTTF-exTTF.

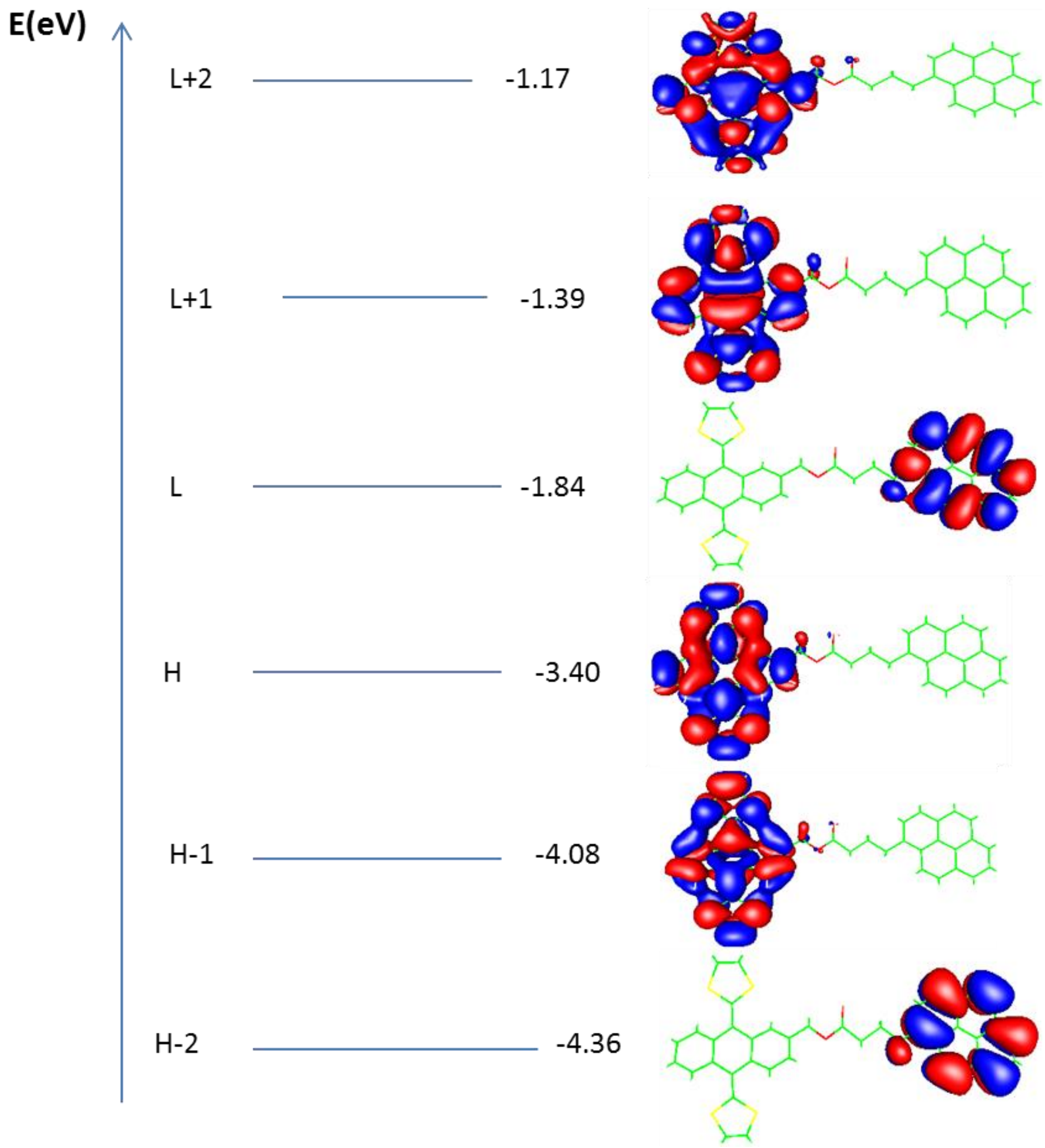

**Figure 6.14.** shows the highest-occupied (HOMOs) and lowest unoccupied (LUMOs), obtained using the DFT code SIESTA, red corresponds to positive and blue to negative regions of the wave functions. The HOMO (-3.39940) is localized on the electron-donor exTTF unit. In contrast, the LUMO (-1.83582) is localized on the electron-acceptor Pyrene unit.

Figure 6.14, shows the distribution of molecular orbitals for PPT for the length. The HOMO is fully localized on the exTTF and the LUMO is fully localized on the pyrene. This means that the pyrene-exTTF molecule is a donor-acceptor (D-A) molecule.

## 6.3. Conclusion

Molecular dynamic simulations are used to examine the molecular assembly of two candidate molecules for graphene based molecular electronics, one with one pyrene anchor, Pyrene-PEGn-exTTF (PPT) and the other with three pyrene anchors, tri-pyrene derivative (TPPD) on a disordered graphene surface. PPT is seen to form flat structures whilst TPPT is seen to form semi-circular cone-like micelle structures on the graphene surface. In the presence of water, the PPT tends to aggregate whereas the TPPT micelle expands. The hydrophobic pyrene anchors are firmly attached to the graphene surface in both cases and it is the hydrophyllic dithiol heads groups which allow the water to disperse the micelles.
The electrical properties for a single molecule PPT is demonstrated, and we found that this molecule has low conductance and it is LUMO-dominated transitions at the DFT Fermi energy.

## 6.4. Acknowledgment

Molecule design from Dr. Beatriz M. Illescas y Dept. de Quimica Organica

Universidad de Computense.

## 6.5. Reference

1. Frenkel, D. and Smit, B., 2002. Understanding molecular simulation: from algorithms to applications. *Computational sciences series*, *1*, pp.1-638.
2. Love, J.C., Estroff, L.A., Kriebel, J.K., Nuzzo, R.G. and Whitesides, G.M., 2005. Self-assembled monolayers of thiolates on metals as a form of nanotechnology. Chemical reviews, 105(4), pp.1103-1170.
3. Vericat, C., Vela, M.E., Corthey, G., Pensa, E., Cortés, E., Fonticelli, M.H., Ibanez, F., Benitez, G.E., Carro, P. and Salvarezza, R.C., 2014. Self-assembled monolayers of thiolates on metals: a review article on sulfur-metal chemistry and surface structures. RSC Advances, 4(53), pp.27730-27754.
4. Nakamura, T., Kondoh, H., Matsumoto, M. and Nozoye, H., 1996. Scanning tunneling microscopy observations of α, ω-bis (mercaptomethylthienyl) alkane derivatives self-assembled on Au (111). Langmuir, 12(25), pp.5977-5979.
5. Laibinis, P.E., Whitesides, G.M., Allara, D.L., Tao, Y.T., Parikh, A.N. and Nuzzo, R.G., 1991. Comparison of the structures and wetting properties of self-assembled monolayers of n-alkanethiols on the coinage metal surfaces, copper, silver, and gold. Journal of the American Chemical Society, 113(19), pp.7152-7167.
6. Muskal, N., Turyan, I. and Mandler, D., 1996. Self-assembled monolayers on mercury surfaces. Journal of Electroanalytical Chemistry, 409(1), pp.131-136.
7. Yan, L., Zheng, Y.B., Zhao, F., Li, S., Gao, X., Xu, B., Weiss, P.S. and Zhao, Y., 2012. Chemistry and physics of a single atomic layer: strategies and challenges for functionalization of graphene and graphene-based materials. Chemical Society Reviews, 41(1), pp.97-114.

8. Geim, A.K., 2009. Graphene: status and prospects. science, 324(5934), pp.1530-1534.
9. Novoselov, K.S., Geim, A.K., Morozov, S.V., Jiang, D., Zhang, Y., Dubonos, S.V., Grigorieva, I.V. and Firsov, A.A., 2004. Electric field effect in atomically thin carbon films. science, 306(5696), pp.666-669.
10. Kuila, T., Bose, S., Mishra, A.K., Khanra, P., Kim, N.H. and Lee, J.H., 2012. Chemical functionalization of graphene and its applications. Progress in Materials Science, 57(7), pp.1061-1105.
11. Huang, Y., Yan, W., Xu, Y., Huang, L. and Chen, Y., 2012. Functionalization of Graphene Oxide by Two-Step Alkylation. Macromolecular Chemistry and Physics, 213(10-11), pp.1101-1106.
12. Zhang, S., Xiong, P., Yang, X. and Wang, X., 2011. Novel PEG functionalized graphene nanosheets: enhancement of dispersibility and thermal stability. Nanoscale, 3(5), pp.2169-2174.
13. Das, S., Irin, F., Ahmed, H.T., Cortinas, A.B., Wajid, A.S., Parviz, D., Jankowski, A.F., Kato, M. and Green, M.J., 2012. Non-covalent functionalization of pristine few-layer graphene using triphenylene derivatives for conductive poly (vinyl alcohol) composites. Polymer, 53(12), pp.2485-2494.
14. Xu, X., Ou, D., Luo, X., Chen, J., Lu, J., Zhan, H., Dong, Y., Qin, J. and Li, Z., 2012. Water-soluble graphene sheets with large optical limiting response via non-covalent functionalization with polyacetylenes. Journal of Materials Chemistry, 22(42), pp.22624-22630.

15. Xu, L. and Yang, X., 2014. Molecular dynamics simulation of adsorption of pyrene–polyethylene glycol onto graphene. Journal of colloid and interface science, 418, pp.66-73.
16. Liu, J., Bibari, O., Mailley, P., Dijon, J., Rouviere, E., Sauter-Starace, F., Caillat, P., Vinet, F. and Marchand, G., 2009. Stable non-covalent functionalisation of multi-walled carbon nanotubes by pyrene–polyethylene glycol through π–π stacking. New Journal of Chemistry, 33(5), pp.1017-1024.
17. Zu, S.Z. and Han, B.H., 2009. Aqueous dispersion of graphene sheets stabilized by pluronic copolymers: formation of supramolecular hydrogel. The Journal of Physical Chemistry C, 113(31), pp.13651-13657.
18. Todorov, I.T., Smith, W., Trachenko, K. and Dove, M.T., 2006. DL_POLY_3: new dimensions in molecular dynamics simulations via massive parallelism. Journal of Materials Chemistry, 16(20), pp.1911-1918.
19. Soler, J.M., Artacho, E., Gale, J.D., García, A., Junquera, J., Ordejón, P. and Sánchez-Portal, D., 2002. The SIESTA method for ab initio order-N materials simulation. Journal of Physics: Condensed Matter, 14(11), p.2745.
20. Nosé, S., 1984. A unified formulation of the constant temperature molecular dynamics methods. The Journal of chemical physics, 81(1), pp.511-519.
21. Evans, D.J. and Holian, B.L., 1985. The nose–hoover thermostat. The Journal of chemical physics, 83(8), pp.4069-4074.
22. D'Angelo, P., Serva, A., Aquilanti, G., Pascarelli, S. and Migliorati, V., 2015. Structural Properties and Aggregation Behavior of 1-Hexyl-3-methylimidazolium Iodide in Aqueous Solutions. The Journal of Physical Chemistry B, 119(45), pp.14515-14526.

23. Mayo, S.L., Olafson, B.D. and Goddard, W.A., 1990. DREIDING: a generic force field for molecular simulations. Journal of Physical chemistry, 94(26), pp.8897-8909.
24. Sánchez, L., Pérez, I., Martín, N. and Guldi, D.M., 2003. Controlling Short-and Long-Range Electron Transfer Processes in Molecular Dyads and Triads. Chemistry–A European Journal, 9(11), pp.2457-2468.
25. Calbo, J., Viruela, P.M. and Ortí, E., 2014. Theoretical insight on novel donor-acceptor exTTF-based dyes for dye-sensitized solar cells. Journal of molecular modeling, 20(4), pp.1-10.
26. Illescas, B.M., Santos, J., Wielopolski, M., Atienza, C.M., Martín, N. and Guldi, D.M., 2009. Electron transfer through exTTF bridges in electron donor–acceptor conjugates. Chemical Communications, (36), pp.5374-5376.
27. Leng, Y., Chen, J., Zhou, B. and Gräter, F., 2010. Rupture Mechanism of Aromatic Systems from Graphite Probed with Molecular Dynamics Simulations. Langmuir, 26(13), pp.10791-10795.
28. Siu, H., Prazeres, T.J., Duhamel, J., Olesen, K. and Shay, G., 2005. Characterization of the aggregates made by short poly (ethylene oxide) chains labeled at one end with pyrene. Macromolecules, 38(7), pp.2865-2875.
29. Barroso-Bujans, F., Fernandez-Alonso, F., Pomposo, J.A., Cerveny, S., Alegría, A. and Colmenero, J., 2012. Macromolecular structure and vibrational dynamics of confined poly (ethylene oxide): From subnanometer 2d-intercalation into graphite oxide to surface adsorption onto graphene sheets. ACS Macro Letters, 1(5), pp.550-554.

30. Jiang, P., Morales, G.M., You, W. and Yu, L., 2004. Synthesis of diode molecules and their sequential assembly to control electron transport.*Angewandte Chemie*, *116*(34), pp.4571-4575.

31. Ceperley, D.M. and Alder, B.J., 1980. Ground state of the electron gas by a stochastic method. Physical Review Letters, 45(7), p.566.

32. Ferrer, J., Lambert, C.J., García-Suárez, V.M., Manrique, D.Z., Visontai, D., Oroszlany, L., Rodríguez-Ferradás, R., Grace, I., Bailey, S.W.D., Gillemot, K. and Sadeghi, H., 2014. GOLLUM: a next-generation simulation tool for electron, thermal and spin transport. New Journal of Physics, 16(9), p.093029.

33. Troullier, N. and Martins, J.L., 1991. Efficient pseudopotentials for plane-wave calculations. *Physical review B, 43*(3), p.1993.

# Chapter 7

## 7. Conclusion

For the past few decades, thiol terminated molecules attached to gold electrodes have been a major focus for research into single-molecule electronics. However, with the recent discovery of methods for processing graphene, there is now an opportunity to explore new combinations of molecules and electrode materials.

In this thesis, I have taken two steps in this direction. First examined quantum transport through single molecules formed from alkyl chains or ethylene glycol chains and compared their properties with the same molecules attached to gold electrodes. Graphene can be engineered to possess defects [1] or most-likely is created with defects and grain boundaries. Therefore, I also examined the role of defects in determining transport properties through single molecules.

Nowadays one of the drivers of molecular-electronic research is the search for high-performance thermoelectric materials. Since the thermal conductance of such materials should be low, and since phonon scattering at boundaries is one approach to reducing phonon transport [2], one strategy is to create massively parallel arrays of single molecules sandwiched between graphene electrodes, such that the current passes through the molecules, perpendicular to the graphene sheets. Translating single-molecule functionality into such self-assembled monolayers is highly non-trivial and require fundamental understanding of the nature of self-assembly onto graphene. In this thesis I therefore performed molecular dynamics simulations of such self-assembly and combined these with quantum transport calculations of the electrical conductance of the resulting structures.

For the future it will be of interest to examine the thermoelectric properties of these assemblies and to examine their stability. For example, it is known that pyrene anchor groups bind strongly to graphene, but the energy barrier to sliding is rather low [4]. This means that such groups may not be able to resist current-induced forces [5] and therefore it may be necessary to engineer defects into the graphene to pin such anchors. All of these issues will be interesting topics for future studies.

## References.

1. XH Zheng, GR Zhang, Z Zeng, VM García-Suárez, CJ Lambert, 2009, Effects of antidots on the transport properties of graphene nanoribbons, Physical Review B 80 (7), 075413 (2009)
2. G. Fagas, A.G. Kozorezov, C.J. Lambert, J.K. Wigmore, A. Peacock,A. Poelaert, R. den Hartog, 1999, Lattice dynamics of a disordered solid-solid interface, *Phys. Rev*. B60, 6459 (1999)
3. Haoxue Han, Yong Zhang, Zainelabideen Y Mijbil, Hatef Sadeghi, Yuxiang Ni, Shiyun Xiong, Kimmo Saaskilahti, Steven Bailey, Yuriy A Kosevich, Johan Liu, Colin J Lambert, Sebastian Volz, 2016, Functionalization mediates heat transport in graphene nanoflakes, *Nature Communications* 7 11281 (2016)
4. Bailey, S.; Visontai, D.; Lambert, C. J.; Bryce, M. R.; Frampton, H.; Chappell, D., 2014, A study of planar anchor groups for graphene-based single-molecule electronics, *Journal of Chemical Physics 140,* 054708

5. I. Amanatidis, S.W. Bailey and C.J. Lambert, 2008, Carbon nanotube electron windmills: a novel design for nano-motors, Phys. *Rev. Lett.* 100 256802